

\def\unredoffs{\hoffset-.14truein\voffset-.21truein
                     \vsize=9.5truein}
\newbox\leftpage \newdimen\fullhsize \newdimen\hstitle \newdimen\hsbody
\tolerance=1000\hfuzz=2pt
\catcode`\@=11 
\def\bigans{b }
\magnification=1100\unredoffs\baselineskip=16pt plus 2pt minus 1pt
\hsbody=\hsize\hstitle=\hsize
\def\almostshipout#1{
\count1=1 \message{[\the\count0.\the\count1]}
      \global\setbox\leftpage=#1 \global\let\l@r=R}
\newcount\yearltd\yearltd=\year\advance\yearltd by -1900

%
%

\def\draftmode{\message{ DRAFTMODE }\def\draftdate{{\rm preliminary draft:
\number\month/\number\day/\number\yearltd\ \ \hourmin}}%
\headline={\hfil\draftdate}\writelabels\baselineskip=10pt plus 2pt minus 2pt
 {\count255=\time\divide\count255 by 60 \xdef\hourmin{\number\count255}
  \multiply\count255 by-60\advance\count255 by\time
  \xdef\hourmin{\hourmin:\ifnum\count255<10 0\fi\the\count255}}}
\def\nolabels{\def\wrlabeL##1{}\def\eqlabeL##1{}\def\reflabeL##1{}}
\def\writelabels{\def\wrlabeL##1{\leavevmode\vadjust{\rlap{\smash%
{\line{{\escapechar=` \hfill\rlap{\sevenrm\hskip.03in\string##1}}}}}}}%
\def\eqlabeL##1{{\escapechar-1\rlap{\sevenrm\hskip.05in\string##1}}}%
\def\reflabeL##1{\noexpand\llap{\noexpand\sevenrm\string\string\string##1}}}
\nolabels
%
\global\newcount\secno \global\secno=0
\global\newcount\meqno \global\meqno=1
\def\newsec#1{\global\advance\secno by1\message{(\the\secno. #1)}
\global\subsecno=0\eqnres@t\noindent{\bf\the\secno. #1}
\writetoca{{\secsym} {#1}}\par\nobreak\medskip\nobreak}
\def\eqnres@t{\xdef\secsym{\the\secno.}\global\meqno=1\bigbreak\bigskip}
\def\sequentialequations{\def\eqnres@t{\bigbreak}}\xdef\secsym{}
\global\newcount\subsecno \global\subsecno=0
\def\subsec#1{\global\advance\subsecno by1\message{(\secsym\the\subsecno. #1)}
\ifnum\lastpenalty>9000\else\bigbreak\fi
\noindent{\it\secsym\the\subsecno. #1}\writetoca{\string\quad
{\secsym\the\subsecno.} {#1}}\par\nobreak\medskip\nobreak}
\def\appendix#1#2{\global\meqno=1\global\subsecno=0\xdef\secsym{\hbox{#1.}}
\bigbreak\bigskip\noindent{\bf Appendix #1. #2}\message{(#1. #2)}
\writetoca{Appendix {#1.} {#2}}\par\nobreak\medskip\nobreak}
%
%
\def\eqnn#1{\xdef #1{(\secsym\the\meqno)}\writedef{#1\leftbracket#1}%
\global\advance\meqno by1\wrlabeL#1}
\def\eqna#1{\xdef #1##1{\hbox{$(\secsym\the\meqno##1)$}}
\writedef{#1\numbersign1\leftbracket#1{\numbersign1}}%
\global\advance\meqno by1\wrlabeL{#1$\{\}$}}
\def\eqn#1#2{\xdef #1{(\secsym\the\meqno)}\writedef{#1\leftbracket#1}%
\global\advance\meqno by1$$#2\eqno#1\eqlabeL#1$$}
%
\newskip\footskip\footskip14pt plus 1pt minus 1pt 
\def\footnotefont{\ninepoint}\def\f@t#1{\footnotefont #1\@foot}
\def\f@@t{\baselineskip\footskip\bgroup\footnotefont\aftergroup\@foot\let\next}
\setbox\strutbox=\hbox{\vrule height9.5pt depth4.5pt width0pt}
\global\newcount\ftno \global\ftno=0
\def\foot{\global\advance\ftno by1\nobreak\footnote{$^{\the\ftno}$}}
%
\newwrite\ftfile
\def\footend{\def\foot{\global\advance\ftno by1\chardef\wfile=\ftfile
$^{\the\ftno}$\ifnum\ftno=1\immediate\openout\ftfile=foots.tmp\fi%
\immediate\write\ftfile{\noexpand\smallskip%
\noexpand\item{f\the\ftno:\ }\pctsign}\findarg}%
\def\footatend{\vfill\eject\immediate\closeout\ftfile{\parindent=20pt
\centerline{\bf Footnotes}\nobreak\bigskip\input foots.tmp }}}
\def\footatend{}
%
%
\global\newcount\refno \global\refno=1
\newwrite\rfile
\def\ref{[\the\refno]\nref}
\def\nref#1{\xdef#1{[\the\refno]}\writedef{#1\leftbracket#1}%
\ifnum\refno=1\immediate\openout\rfile=refs.tmp\fi
\global\advance\refno by1\chardef\wfile=\rfile\immediate
\write\rfile{\noexpand\item{#1\ }\reflabeL{#1\hskip.31in}\pctsign}\findarg}
\def\findarg#1#{\begingroup\obeylines\newlinechar=`\^^M\pass@rg}
{\obeylines\gdef\pass@rg#1{\writ@line\relax #1^^M\hbox{}^^M}%
\gdef\writ@line#1^^M{\expandafter\toks0\expandafter{\striprel@x #1}%
\edef\next{\the\toks0}\ifx\next\em@rk\let\next=\endgroup\else\ifx\next\empty%
\else\immediate\write\wfile{\the\toks0}\fi\let\next=\writ@line\fi\next\relax}}
\def\striprel@x#1{} \def\em@rk{\hbox{}}
\def\lref{\begingroup\obeylines\lr@f}
\def\lr@f#1#2{\gdef#1{\ref#1{#2}}\endgroup\unskip}
\def\semi{;\hfil\break}
\def\addref#1{\immediate\write\rfile{\noexpand\item{}#1}} 
\def\startrefs#1{\immediate\openout\rfile=refs.tmp\refno=#1}
\def\xref{\expandafter\xr@f}\def\xr@f[#1]{#1}
\def\refs#1{\count255=1[\r@fs #1{\hbox{}}]}
\def\r@fs#1{\ifx\und@fined#1\message{reflabel \string#1 is undefined.}%
\nref#1{need to supply reference \string#1.}\fi%
\vphantom{\hphantom{#1}}\edef\next{#1}\ifx\next\em@rk\def\next{}%
\else\ifx\next#1\ifodd\count255\relax\xref#1\count255=0\fi%
\else#1\count255=1\fi\let\next=\r@fs\fi\next}
%

%
\newwrite\ffile\global\newcount\figno \global\figno=1
\def\fig{fig.~\the\figno\nfig}
\def\nfig#1{\xdef#1{fig.~\the\figno}%
\writedef{#1\leftbracket fig.\noexpand~\the\figno}%
\ifnum\figno=1\immediate\openout\ffile=figs.tmp\fi\chardef\wfile=\ffile%
\immediate\write\ffile{\noexpand\medskip\noexpand\item{Fig.\ \the\figno. }
\reflabeL{#1\hskip.55in}\pctsign}\global\advance\figno by1\findarg}
\def\xfig{\expandafter\xf@g}\def\xf@g fig.\penalty\@M\ {}
\def\figs#1{figs.~\f@gs #1{\hbox{}}}
\def\f@gs#1{\edef\next{#1}\ifx\next\em@rk\def\next{}\else
\ifx\next#1\xfig #1\else#1\fi\let\next=\f@gs\fi\next}
\newwrite\lfile
{\escapechar-1\xdef\pctsign{\string\%}\xdef\leftbracket{\string\{}
\xdef\rightbracket{\string\}}\xdef\numbersign{\string\#}}

\def\writestop{\def\writestoppt{\immediate\write\lfile{\string\pageno%
\the\pageno\string\startrefs\leftbracket\the\refno\rightbracket%
\string\def\string\secsym\leftbracket\secsym\rightbracket%
\string\secno\the\secno\string\meqno\the\meqno}\immediate\closeout\lfile}}
\def\writestoppt{}\def\writedef#1{}
\def\seclab#1{\xdef #1{\the\secno}\writedef{#1\leftbracket#1}\wrlabeL{#1=#1}}
\def\subseclab#1{\xdef #1{\secsym\the\subsecno}%
\writedef{#1\leftbracket#1}\wrlabeL{#1=#1}}
\newwrite\tfile \def\writetoca#1{}
\def\leaderfill{\leaders\hbox to 1em{\hss.\hss}\hfill}
\def\writetoc{\immediate\openout\tfile=toc.tmp
   \def\writetoca##1{{\edef\next{\write\tfile{\noindent ##1
   \string\leaderfill {\noexpand\number\pageno} \par}}\next}}}

%
\catcode`\@=12 
%
\edef\tfontsize{scaled\magstep3}
 \tfontsize  \tfontsize
 \tfontsize \font\titlei=cmmi10 \tfontsize
\font\titleis=cmmi7 \tfontsize \font\titleiss=cmmi5 \tfontsize
\font\titlesy=cmsy10 \tfontsize \font\titlesys=cmsy7 \tfontsize
\font\titlesyss=cmsy5 \tfontsize  \tfontsize
\skewchar\titlei='177 \skewchar\titleis='177 \skewchar\titleiss='177
\skewchar\titlesy='60 \skewchar\titlesys='60 \skewchar\titlesyss='60
\font\ninerm=cmr9 \font\sixrm=cmr6 \font\ninei=cmmi9 \font\sixi=cmmi6
\font\ninesy=cmsy9 \font\sixsy=cmsy6 \font\ninebf=cmbx9
\font\nineit=cmti9 \font\ninesl=cmsl9 \skewchar\ninei='177
\skewchar\sixi='177 \skewchar\ninesy='60 \skewchar\sixsy='60
\def\ninepoint{\def\rm{\fam0\ninerm}
\textfont0=\ninerm \scriptfont0=\sixrm \scriptscriptfont0=\fiverm
\textfont1=\ninei \scriptfont1=\sixi \scriptscriptfont1=\fivei
\textfont2=\ninesy \scriptfont2=\sixsy \scriptscriptfont2=\fivesy
\textfont\itfam=\ninei \def\it{\fam\itfam\nineit}\def\sl{\fam\slfam\ninesl}%
\textfont\bffam=\ninebf \def\bf{\fam\bffam\ninebf}\rm}
%
%
\def\noblackbox{\overfullrule=0pt}
\hyphenation{anom-aly anom-alies coun-ter-term coun-ter-terms}
\def\inv{^{\raise.15ex\hbox{${\scriptscriptstyle -}$}\kern-.05em 1}}

\def\Dsl{\,\raise.15ex\hbox{/}\mkern-13.5mu D} 
\def\dsl{\raise.15ex\hbox{/}\kern-.57em\partial}

\def\tr{{\rm tr}} \def\Tr{{\rm Tr}}

\def\lspace{\ifx\answ\bigans{}\else\qquad\fi}
\def\lbspace{\ifx\answ\bigans{}\else\hskip-.2in\fi} 
\def\boxeqn#1{\vcenter{\vbox{\hrule\hbox{\vrule\kern3pt\vbox{\kern3pt
	\hbox{${\displaystyle #1}$}\kern3pt}\kern3pt\vrule}\hrule}}}
\def\mbox#1#2{\vcenter{\hrule \hbox{\vrule height#2in
		\kern#1in \vrule} \hrule}}  
\def\tilde{\widetilde} \def\bar{\overline} 
%

\def\darr#1{\raise1.5ex\hbox{$\leftrightarrow$}\mkern-16.5mu #1}

\def\half{{\textstyle{1\over2}}} 
\def\roughly#1{\raise.3ex\hbox{$#1$\kern-.75em\lower1ex\hbox{$\sim$}}}

\def\footsy{\foot}

\def\pslash{p\!\!/}

\def\p2inf{\mathrel{\mathop{\sim}\limits_{\scriptscriptstyle
{p^2 \rightarrow \infty }}}}
\def\kap2inf{\mathrel{\mathop{\sim}\limits_{\scriptscriptstyle
{\kappa \rightarrow \infty }}}}
\def\x2inf{\mathrel{\mathop{\sim}\limits_{\scriptscriptstyle
{x \rightarrow \infty }}}}
\def\Lam2inf{\mathrel{\mathop{\sim}\limits_{\scriptscriptstyle
{\Lambda \rightarrow \infty }}}}
\def\frac#1#2{{{#1}\over {#2}}}
\def\half{\hbox{${1\over 2}$}}
\def\quarter{\hbox{${1\over 4}$}}
\def\smallfrac#1#2{\hbox{${{#1}\over {#2}}$}}

\def\tr{{\rm tr}}\def\Tr{{\rm Tr}}

\def\blackbox{\vrule height7pt width5pt depth2pt}

\def\VertL{\Vert_{\Lambda}}\def\VertR{\Vert_{\Lambda_R}}
\def\Real{\Re e}\def\Imag{\Im m}
\def\bp{\bar{p}}\def\bq{\bar{q}}\def\br{\bar{r}}
\catcode`@=11 
\def\slash#1{\mathord{\mathpalette\c@ncel#1}}
 \def\c@ncel#1#2{\ooalign{$\hfil#1\mkern1mu/\hfil$\crcr$#1#2$}}
\def\lsim{\mathrel{\mathpalette\@versim<}}
\def\gsim{\mathrel{\mathpalette\@versim>}}
 \def\@versim#1#2{\lower0.2ex\vbox{\baselineskip\z@skip\lineskip\z@skip
       \lineskiplimit\z@\ialign{$\m@th#1\hfil##$\crcr#2\crcr\sim\crcr}}}
\catcode`@=12 

\def\PR{{\it Phys.~Rev.~}}
\def\PRL{{\it Phys.~Rev.~Lett.~}}
\def\NP{{\it Nucl.~Phys.~}}
\def\PL{{\it Phys.~Lett.~}}
\def\PRep{{\it Phys.~Rep.~}}
\def\AP{{\it Ann.~Phys.~}}
\def\CMP{{\it Comm.~Math.~Phys.~}}
\def\JMP{{\it Jour.~Math.~Phys.~}}
\def\NC{{\it Nuov.~Cim.~}}
\def\SJNP{{\it Sov.~Jour.~Nucl.~Phys.~}}
\def\ZP{{\it Zeit.~Phys.~}}
\def\JP{{\it Jour.~Phys.~}}
\def\vol#1{{\bf #1}}
\def\vyp#1#2#3{\vol{#1} (#2) #3}


\noblackbox
\pageno=0
\nopagenumbers
\tolerance=10000
\hfuzz=5pt
\line{\hfill CERN-TH.7067/93}
\line{\hfill OUTP-93-23P}
\vskip 36pt
\centerline{\bf RENORMALIZABILITY OF EFFECTIVE SCALAR FIELD THEORY}
\vskip 36pt\centerline{Richard~D.~Ball$^{a,b}$ and Robert~S.~Thorne$^{a}$}
\vskip 12pt
\centerline{\it Theoretical Physics, 1 Keble Road,}
\centerline{\it  Oxford OX1 3NP, U.K.~$^{a}$}
\vskip 10pt
\centerline{\it and }
\vskip 10pt
\centerline{\it Theory Division, CERN,}
\centerline{\it CH-1211 Gen\`eve 23, Switzerland.$^{b}$}
\vskip 36pt
{\narrower\baselineskip 10pt
\centerline{\bf Abstract}
\medskip
We present a comprehensive discussion of the consistency of the
effective quantum field theory of a single $Z_2$ symmetric scalar field.
The theory is constructed from a bare Euclidean action which at a
scale much greater than the particle's mass is constrained only
by the most basic requirements;
stability, finiteness, analyticity, naturalness, and global symmetry.
We prove to all orders in perturbation theory the boundedness, convergence,
and universality of the theory at low energy scales, and thus that the
theory is perturbatively renormalizable in the sense that to a certain
precision over a range of such scales it depends only on a finite
number of parameters. We then
demonstrate that the effective theory has a well
defined unitary and causal analytic S--matrix at all energy scales.
We also show that redundant terms in the Lagrangian may be
systematically eliminated by field redefinitions without changing the
S--matrix, and discuss the extent to which effective
field theory and analytic S--matrix theory are actually equivalent.
All this is achieved by a systematic
exploitation of Wilson's exact renormalization group flow equation, as
used by Polchinski in his original proof of the renormalizability of
conventional $\varphi^4$-theory.
}

\vskip 36pt
\line{CERN-TH.7067/93\hfill}
\line{OUTP-93-23P\hfill}
\line{November 1993\hfill}
\vfill
\eject
\footline={\hss\tenrm\folio\hss}



An effective quantum field theory is any quantum field theory designed
to give a description of the physics of a particular set
of particles over a limited range of scales, without requiring
detailed knowledge of any further physics outside of this range.
Consider the following familiar examples: very soft photons scattering
electromagnetically at energies below the electron mass; soft pions
scattering strongly at energies below the scale of chiral symmetry
breaking; electroweak processes involving leptons, pions and kaons;
deep inelastic scattering at energies below
the mass of a heavy quark; weak processes involving heavy quarks with masses
below that of the intermediate vector bosons; electroweak processes
involving intermediate
vector bosons at energies below the mass of the Higgs boson (or
the supersymmetry scale, or the technicolour scale, or whatever);
the supersymmetric standard model below the scale of supersymmetry
breaking; the
evolution of standard model couplings below the scale of grand
unification; quantum gravity below the Planck scale. Most of the
physics we know, or even hope to know, seems to be described by
effective quantum field theories, which are useful only until
the scale of some new physical process is reached.

Despite the wide variety of increasingly useful applications,
from a formal point of view effective quantum field theories are
still relatively poorly understood. Most of the original work on
the consistency of quantum field theory dealt only with idealized
theories, supposedly fundamental in the sense that they attempted
to describe physics at arbitrarily high energies. In
particular the S--matrix of the theory had to exist, be manifestly
finite and well defined, and then satisfy the usual physical requirements
of Lorentz invariance, unitarity, cluster decomposition and
causality; in short, the theory
had to be `renormalizable'. Only a rather small
class of such theories was discovered; this was regarded as a virtue,
since it helped guide the physicist to the `correct' theory --- the
standard model. Now however we have become more discerning;
from the examples given above it is clear that in many situations an effective
theory can in practice be more useful than the underlying more
fundamental theory which gives rise to it, and conversely that any theory
we at present like to think of as fundamental might eventually
turn out to be itself an effective theory. It thus becomes important
to find out just how consistent an effective quantum field theory can
be made to be; does it have a finite, well defined S--matrix, to what
extent does this S--matrix depend on the unknown (and thus in
principle arbitrary) physics at higher scales, and what about
unitarity and causality? In other words, to what extent
is effective field theory `renormalizable'?

The purpose of this article, and a number of succeeding articles, is
to address this question, using the exact renormalization group
pioneered by Wilson\ref\rii{
K.~G.~Wilson and J.~G.~Kogut, \PRep\vyp{12C}{1974}{75}.}.
The techniques necessary to prove conventional perturbative renormalizability
of conventional scalar field theories using renormalization group flow
were developed by Polchinski\ref\rix
{J. Polchinski, \NP\vyp{B231}{1984}{269}.}; we
will show how to extend his ideas to effective field theories,
in such a way as to permit a thorough investigation of their consistency.
In this article we treat only the special case
of a $Z_2$ symmetric scalar field propagating in $3+1$ dimensions,
using this as a theoretical
model with which to develop the necessary techniques. In a number
of subsequent articles we hope to consider more interesting (and indeed
physically relevant) theories.\vfill\eject

We begin with a brief review of renormalization theory,
renormalization group flow, and effective field theory. These subjects
are very closely
tied together since they are all concerned with the interplay
between physics at different energy scales. However, this way of
understanding renormalization is a relatively recent development
in the history of quantum field theory.

When they were first devised, quantum field theories were expected to
be valid for all energies, i.e fundamental.
Renormalization was originally introduced
simply as a means of subtracting divergences encountered when integrating
the four--momenta of internal virtual loops over all
scales\ref\rfd{F.~J.~Dyson, \PR\vyp{75}{1949}{486,~1736}.}. The theory was
defined by a classical Lagrangian containing a finite number of masses
and coupling
constants, and some sort of regularization scheme to render all loop
integrals finite. It was then said to be perturbatively renormalizable
if these `bare' masses and coupling constants could be fine-tuned in
such a way that all S--matrix elements, calculated in perturbation
theory, are finite (bounded) and tend towards a well-defined limit
(convergent) as the regularization was removed.  It
was found in general that this is possible if the classical
Lagrangian contains all possible terms with canonical dimension less than
or equal to the dimension of space--time, consistent with
the symmetries (for example Lorentz invariance or internal global
symmetries) of the theory. Furthermore provided
the classical field equations supported only stable, causal solutions
(which in practice meant that the classical Lagrangian could contain
terms with at most two derivatives of the
fields\ref\rpu{A.~Pais and G.~E.~Uhlenbeck, \PR\vyp{79}{1950}{145}.}),
it was possible to show that not only did the S--matrix exist, but
also that it was unitary and causal, with the appropriate analyticity
properties\nref\rlandau{L.~D.~Landau, \NP\vyp{13}{1959}{181}.}
\nref\rui{R.~E.~Cutkovsky, \JMP\vyp{1}{1960}{429}.}\refs{\rlandau,\rui}.

\nref\ri{N.~N.~Bogoliubov and D.~V.~Shirkov, {\it
Fort.~Phys.~}\vyp{4}{1956}{438}\semi N.~N.~Bogoliubov and
O.S. Parasiuk, {\it Acta Math.~}\vyp{97}{1957}{227}\semi
K.~Hepp, \CMP\vyp{15}{1969}{301}\semi W. Zimmermann,
\CMP\vyp{15}{1969}{208}\semi P.~Breitenlohner and D.~Maison,
\CMP\vyp{52}{1977}{11,~39,~55}\semi for a review see
J.~C.~Collins, ``Renormalization'', C.U.P.,
Cambridge, 1984.}
\nref\rdiagrammar{G.~'t~Hooft and M.~Veltman,
``Diagrammar'', CERN Report 73-9.}
Proofs of renormalizability to all orders in perturbation
theory\refs{\ri,\rdiagrammar} were notoriously long and
complicated (involving notions of graph topologies, skeleton
expansions, overlapping divergences, the forest theorem,
Weinberg's theorem,
etc.), and left the abiding impression that
renormalization was nothing but an artificial trick in which
the divergences were, in Dirac's words, being ``swept under the carpet''.
The question of why physics should be described by a theory which could be
renormalized in this way was not directly addressed; it was simply
found to be impossible to do calculations with a non--renormalizable
theory without encountering divergences that
could only be removed by introducing an infinite number of
counterterms into the theory, and thus seemingly losing all predictive power.

A new, much more intuitive way of looking at the renormalization of a
quantum field theory was developed in the 1970s when Wilson introduced
the flow of effective Lagrangians as generated by a renormalization group
transformation\rii . Taking seriously the idea that a particular
quantum field theory may not be valid for arbitrarily high energies a
smooth regularization is introduced at some scale $\Lambda_0$. To maintain
explicit Lorentz invariance, it is
necessary to define the theory in Euclidean space, where the Lorentz group is
compact; the regularization then restricts the length $p^2$ of all
four--momenta.
The physics below the scale $\Lambda_0$ is then described by a very general
`non--local' bare Lagrangian\foot{We describe a Lagrangian as local if its
inverse propagator and vertices are polynomials in momentum space;
otherwise it is conventionally (though perhaps misleadingly) described
as `non--local'. We will see later however that a Lagrangian only
really deserves to be labeled non--local if its inverse propagator
and vertices have singularities; if they are regular the term
`quasi--local' is more appropriate since a truncation of their
Taylor expansion renders them local.} which is constrained only
by the most basic requirements\ref
\riii{S.~Weinberg, {\it Physica~}\vyp{96A}{1979}{327}.}:
rotational invariance; analyticity in momentum
(so that a Taylor expansion in powers of momentum converges);
stability, meaning that the equations of motion have a unique stable
solution; analyticity in field space at this solution so that
perturbations about it are well defined; small couplings such that
perturbative expansions makes sense; and possibly consistency with some
internal symmetry.\nobreak\foot{As the Lagrangian of an effective theory is
of necessity not local, the treatment of local symmetries is
inevitably nontrivial; this technical difficulty will be explored
in a later article.}
Thus, in general the Lagrangian will consist of an infinite series of
local terms, together with an infinite set of coupling constants,
one for each independent term in the Lagrangian. Using the
criterion of `naturalness' \ref\rxx{G.~'t~Hooft, in ``Recent Developments
in Gauge Theories'' (Cargese Summer School, 1979), eds. G.~'t~Hooft et
al. (Plenum, New York, 1980).}, these couplings are chosen to be of order
unity in units of scale $\Lambda_0$. Weinberg conjectures\riii\ that
the perturbative S--matrix for such a theory will simply be the most general
possible consistent with basic principles; Lorentz invariance,
perturbative unitarity, cluster decomposition, analyticity and
causality, and the internal symmetry.

For this effective quantum field theory, ultraviolet finiteness is no
longer an issue; all loop integrals are
cut off. Rather we must ask whether, since the theory now contains an
infinite number of coupling constants, it still has any
predictive capability.

We can address this question by considering some energy $E$,
far below $\Lambda_0$. If we were to compute perturbatively S--matrix
elements for processes occurring at these energies using the Lagrangian
described above, then we would eventually have to bring into play
all of the coupling constants, and the theory would appear
superficially to have no real
predictive power. However using Wilson's exact
renormalization group \rii\ we can consider smoothly lowering the
regularization scale
to some value $\Lambda_R$ say, of order $E$. To
keep the S--matrix fixed, the coupling constants must change as the
regularization scale does. Hence we have a `running' or `effective'
Lagrangian,
which `flows' with $\Lambda$. We may then use the Lagrangian
defined at $\Lambda_R$ to calculate S--matrix elements at the scale $E$.
It is then not the coupling constants at $\Lambda_0$ that are
important, but those at the scale $\Lambda_R$.  So we say that
an effective field theory is `renormalizable' if we can calculate
all the S--matrix elements for processes with energy scale
$E$, up to small errors which vanish as powers of $E/\Lambda_0$, once
we have determined a finite number of coupling constants at some
renormalization scale $\Lambda_R \sim E$. These coupling constants
are called `relevant'; all others are `irrelevant'. We can then reliably
describe low energy processes without knowing any of the details of
the physics at much higher energies, or more
specifically the values of the infinite set of irrelevant
coupling constants at $\Lambda_0$.

Denoting all relevant couplings by $\lambda(\Lambda)$ and all irrelevant
couplings by $\eta(\Lambda)$,  this coupling constant flow may be
depicted schematically as in \fig\flow{Schematic picture of the
renormalization group flow. $\lambda$ and $\eta$ are the relevant and
irrelevant couplings respectively, and each solid line is a line of
constant physics; the arrow shows how the couplings flow as $\Lambda$
is decreased. The dotted line gives the value of the couplings at the
naturalness scale $\Lambda_0$, the dotted line their value at the
renormalization scale $\Lambda_R$.}. Each trajectory corresponds to a
different choice of `bare' couplings
$\{\lambda(\Lambda_0),\eta(\Lambda_0)\}$, which we assume to be natural.
As we flow down to $\Lambda_R \ll \Lambda_0$, then the flow in coupling
constant space converges towards a submanifold of dimension $N_{\lambda}$,
the number of independent relevant coupling constants in the theory.
This submanifold has a thickness of order $\Lambda_R /\Lambda_0$ or
less --- the renormalized irrelevant couplings
$\eta(\Lambda_R)$ are thus determined in terms of the renormalized relevant
couplings $\lambda(\Lambda_R)$ up to corrections of order
$\Lambda_R/\Lambda_0$.

This picture is in accord with naive power counting arguments, and
indeed it is usual to classify the terms in the Lagrangian according to their
canonical (or `engineering' or `classical') dimension; those with
canonical dimension $n\leq d$,
the dimension of spacetime, are called `relevant',
those with $n=d$ `marginal', while those with $n>d$ are
`irrelevant'.\footsy{From the conventional
viewpoint terms which are classically relevant
or marginal are called `renormalizable', while classically
irrelevant terms are called `nonrenormalizable'; we will not use this
terminology for obvious reasons.} The coupling constants
associated with these terms are classified correspondingly.
The dimension which determines the true relevance or irrelevance of a
coupling constant is in general however not its canonical dimension,
but the sum of the canonical dimension and an `anomalous'
dimension due to quantum corrections. Thus a coupling constant which
is classically irrelevant may turn out to be relevant in the quantum
theory. Perturbative
renormalizability amounts to the simple assertion that perturbatively quantum
corrections to classical scaling are at most logarithmic; classically
relevant(irrelevant) couplings are then indeed truly relevant(irrelevant),
provided the irrelevant couplings are not fine--tuned to unnaturally
large values.

It is not difficult to see that this more general notion of
renormalizability includes the conventional one as a special case;
 we may
remove the regularization by taking the formal limit
$\Lambda_0/\Lambda_R \rightarrow\infty$.
This would then give an S--matrix which depended only on the relevant
renormalized couplings $\lambda(\Lambda_R)$, or in other words on
a renormalized classical Lagrangian.\foot{
Although some of these renormalized couplings may turn
out to be necessarily zero in the infinite cut--off limit; this is
probably the case for the coupling constant of $\phi^4$-theory
in four dimensions, and may also be true of QED.} Since the
conventional bare Lagrangian
contains only relevant bare couplings, the conventional formulation
corresponds to the subset of flows which cross the surface
$\eta(\Lambda_0)=0$, in the limit $\Lambda_0/\Lambda_R \to\infty$, with
$\lambda(\Lambda_R)$ held fixed.

However it will also be true in general that wherever we choose to
define $\Lambda_0$ (as long as it is large enough), and whatever
we choose the irrelevant bare couplings $\eta(\Lambda_0)$ to be (as
long as they are natural enough), then for a particular choice of
$\lambda(\Lambda_R)$ the $\eta(\Lambda_R)$ will be the same up
to small corrections, of the order of $(\Lambda_R/\Lambda_0)$.
In other words, for any point on the submanifold at $\Lambda_R$ there
is a flow towards it from a wide variety of initial Lagrangians at
$\Lambda_0$, all of these being equivalent as far as
the values of S--matrix elements for processes with energies of order
$E\sim\Lambda_R$ are concerned. This more general aspect of
renormalizability we will call `universality'.

By now it should be clear that the word `renormalizability' is
rather overworked; to avoid confusion we will try to avoid using
it at all except in the following specific sense. If the
S--matrix is made $\Lambda_0$-finite by determining a finite number
$N_{\lambda}$ of renormalized coupling constants then the theory
will be said to be bounded. If it also has a well defined limit as
$\Lambda_0/\Lambda_R \rightarrow \infty$, with the renormalized couplings held
fixed, the theory will be described as convergent. If in this limit the
perturbative S--matrix is independent of the form of the bare Lagrangian,
depending only on the $N_{\lambda}$ renormalized couplings, the theory will
be said to be universal. A theory is then `renormalizable' only if it is
bounded, convergent and universal; processes at scales far below
$\Lambda_0$ may then be determined to a given precision in terms of
only a finite number of physically relevant coupling constants.

An effective quantum theory thus gives us a much more general notion
of renormalizability than we had in conventional quantum field theory:
the regularization need no longer be removed, and the irrelevant bare
couplings need not be zero.\footsy{In the language of statistical
mechanics, an effective quantum field theory is constructed near, but
not precisely at, an ultraviolet fixed point; the closer the theory is to
such a fixed point, the fewer the number of physically relevant
couplings needed to describe it, and the more renormalizable it becomes.}
However, boundedness, convergence and universality are not the
only requirements for a meaningful quantum field theory ---
it is also necessary that the equations of motion have a unique
stable solution at all scales $\Lambda$ (so that a Fock space of
asymptotic states may be constructed), and that furthermore the
S--matrix of transition amplitudes between these asymptotic states
is then Lorentz invariant, unitary and causal. Now because an
effective quantum field theory has some form of
momentum cut--off the Lagrangian is necessarily
non--local (or at best quasi--local if the cut--off is sufficiently
smooth\rii). This should
not really come as any surprise; in an effective
field theory we are absorbing the effects of new, high energy physics into our
Lagrangian, and the particles we describe may even be composite.
However, it is well known that such theories are in general
inconsistent (\nref\rbs{N.N. Bogoliubov and D.V. Shirkov,
``Introduction to the Theory of Quantized Fields'',
Wiley-Interscience, New York, 1959.}\refs{\rpu,\rbs}
 --- for a review from a modern perspective see
\ref\riv{D.A. Eliezer and R.P. Woodard, \NP\vyp{B325}{1989}{389}.}).
If explicit Lorentz invariance is maintained the equations of motion
of local theories containing terms in the Lagrangian with more
than two time derivatives necessarily have negative energy solutions. When
the Lagrangian is not local there are in general an infinite number
of such solutions.
The theory is then nonperturbatively unstable; the S--matrix does not
exist, except in perturbation theory about a stable solution, and the
perturbative S--matrix is rendered nonunitary and acausal by the
unstable solutions.\foot{Although as explained in ref.\riv\ this
only obscures the fundamental nature of the problem.}
In open string field theory this disease seems at present to be
incurable\riv.

Clearly instability cannot be a property of all theories with
Lagrangians containing terms with more than two time derivatives however since
counter--examples are rather easy to come by; consider for example a
conventional field theory containing one stable and one unstable scalar
field, with the unstable field integrated out to give a non--local
theory of a single scalar field which is manifestly stable, unitary
and causal\ref\rvel{M.~Veltman, {\it Physica~}\vyp{29}{1963}{186}.}.
Indeed by considering an auxiliary field which has no pole in its
propagator one may construct a quasi--local
effective Lagrangian with finite regularization scale from a local one
in such a way that the number and nature of the solutions to the
equations of motion is unaltered\ref\rv{G.~Kleppe and R.~P.~Woodard,
\NP\vyp{B388}{1993}{81};\AP\vyp{221}{1993}{206}.}. A simple
generalization of this construction using the classical
renormalization group flow equation will allow us to find a whole
class of effective field theories which are not only demonstrably
bounded, convergent and universal (at least within perturbation
theory), but also stable, with a unitary causal analytic S--matrix ---
in short consistent. Indeed we can show explicitly that there is
a one--to--one correspondence between such theories and all possible
analytic S--matrices with the same particle content, thus confirming
Weinberg's conjecture\riii.
\medskip

 The contents of the paper are organized as follows: in
\S 1 we discuss regularization and the exact renormalization
group, modifying the discussion in \rix\ so as to make it
possible to have a consistent effective theory at all energy scales; in
\S 2 we prove that in its conventional formulation $\phi^4$
scalar field theory is bounded and convergent to all orders in
perturbation theory, with a unitary and causal perturbative S--matrix;
in \S 3 we extend this proof of
convergence to the more general effective theory in order to prove
universality, which also allows us to discuss
the systematic improvement of the predictive power of the theory
achieved by
specifying further renormalized couplings;
in \S 4 we prove the equivalence theorem, and then use it
to eliminate redundant terms by field redefinitions;
\S 5 deals with the construction of a stable theory, the
proof of the unitarity and causality of the S--matrix, triviality, and
Weinberg's equivalence conjecture; in \S 6 we give a brief
outline of the how the methods developed here can be applied to other
effective field
theories; and finally in the concluding section we summarize
the fundamental principles on which we have based our construction of a
renormalizable effective quantum field theory, and consider some of
the general implications of our results.

\newsec{Renormalization Group Flow.}

In this article we will define an effective quantum field theory in
terms of a regularized classical action (\S 1.1), and a renormalization group
equation (\S 1.2) which then allows us to smoothly evolve the classical action
until its vertices become equal to the n-point connected amputated
Green's functions, which are themselves directly related to S--matrix
amplitudes for processes involving $n$ external particles.
Although this definition does not of itself rely on perturbation
theory, it is equivalent order by order to the usual
diagrammatic definition (as
adopted for example in \rdiagrammar) when the interaction is expanded
perturbatively in some small parameter. However as shown by
Polchinski\rix\ defining the
theory through its evolution permits a particularly direct approach to
the proof of its renormalizability, which we will exploit is \S 2 and
\S 3 below.

\subsec{Regularization.}

Unless one is satisfied by purely formal arguments, before beginning
to study the renormalization of any non--trivial quantum field theory it is
first necessary to choose a regulator, such that the regulated
theory is manifestly finite. Many such regularizations have been
proposed, all with their own peculiar advantages and difficulties.
Here however we wish to discuss effective field theories, which means
in effect that our regulator is chosen for us as an intrinsic part of
the theory; Euclidean four--momenta are smoothly cut off when their lengths
exceed some scale $\Lambda_0$. In this way all momentum
integrals are made manifestly convergent, and no
infinities are ever encountered. In fact we will eventually be
able to show that if the regularization is chosen with sufficient
care it is, though possible in principle, quite unnecessary in
in practice to remove the regulator by
sending the regularization scale to infinity.

The most elegant way to implement a regularization is to
include it in the definition of the classical Euclidean action\rii .
In particular, we write
\eqn\classact
{S[\phi;\Lambda_0] = \half\big(\phi,P\inv
_{\Lambda_0}\phi\big)
+ S_{\rm int}[\phi;\Lambda_0],}
where the inner product $\big(\psi ,\phi\big)=\big(\phi ,\psi\big)\equiv
\int\frac{d^4p}{(2\pi )^4}\psi_p\phi_{-p}$ for real fields $\phi_p$
and $\psi_p$. The function $P_{\Lambda_0}(p)$ is the
free--particle propagator, while $S_{\rm int}[\phi;\Lambda_0]$
contains all the
interactions, which we assume may be expanded as an infinite series of
terms each containing a finite number of fields and
derivatives, with couplings ordered according to a power of
some coupling constant; the precise form of this expansion will be
specified later.
All terms in the action are explicitly invariant under Lorentz rotations.

In order that the Green's functions
calculated using this action are finite, $P_{\Lambda_0}(p)$ must
vanish sufficiently rapidly (more quickly than any polynomial because
in general there will be terms in the interaction
$S_{\rm int}[\phi;\Lambda_0]$ with arbitrarily high powers of $p$)
as $p^2\rightarrow\infty$. Also, as a necessary condition for a
unitary S--matrix,
$P_{\Lambda_0}(p)$, when analytically continued into the complex $p^2$
plane, must be analytic apart from simple poles on the negative real
axis with positive residue.\foot{Although not local, the free bare
Lagrangian may then still be `quasi--local' in the sense described in the
introduction. The necessity for a `smooth' cut-off, to avoid the
introduction of non--local interactions is emphasized in ref.\rii.}
We conventionally choose
it so that there is only one pole; the single field may then
correspond asymptotically to a single species of free particle.

In keeping with these general requirements we write
\eqn\exxix{P_{\Lambda_0}(p) = {K_{\Lambda_0}(p) \over (p^2 +
m^2)},}
where $m^2 > 0$, and the regulating function $K_{\Lambda}(p)$ is of the form
\eqn\kdef{K_{\Lambda}(p)\equiv K\big((p^2 + m^2)/\Lambda^2\big),}
with the real function $K(x)$ satisfying the following conditions:
\item{(1)} $x^n K(x) \rightarrow 0$ as $x\rightarrow\infty$ for any
finite value of $n$; this is so that $K(x)$ acts as a regulator, in
the sense that all Feynman integrals are rendered finite in Euclidean
space.\foot{This condition will be relaxed a little at the end of \S
4.3.} We also assume (for simplicity) that $K(x)$ falls monotonically.
\item{(2)} When analytically continued into the complex plane, $K(z)$
is a regular\foot{Analytic throughout the complex
plane, except at the point at infinity.} function of $z$; this
guarantees that $P_{\Lambda}(p)$ is meromorphic, with only a
single, simple pole. This is, as we show below, sufficient for an
analytic, unitarity and causal perturbative S--matrix. By the Schwarz
reflection principle \hbox{$K(z^*)=K(z)^*$.} We will further
assume that $1/K(z)$ is also regular, so $K(z)$ has no zeros.
\item{(3)} $K(0) =1$ so that the residue of the physical pole is
normalized to unity; this condition also means that as
$\Lambda /m\rightarrow\infty$, $P_{\Lambda}(p)\rightarrow
P_{\infty}(p)=(p^2+m^2)\inv$, the unregularized propagator. Since
$1/K(z)$ is regular, we then also have $K(x)>0$ for all real finite $x$.

\medskip
In fact \exxix\ is nothing but Schwinger's proper--time regularization
\ref\rxiv{J.~Schwinger, \PR\vyp{82}{1951}{664}.}, in which the
regulated propagator is given by
\eqn\ecci{P_{\Lambda}(p) = \int^{\infty}_{0}
\!d \tau \, \rho(\tau,\Lambda) e^{-\tau (p^2 + m^2)},}
where $\rho(\tau,\Lambda)$ is such as to cut off the proper--time
integration at small $\tau$ (see ref.\ref
\rdbphysrep{R.~D.~Ball, \PRep\vyp{182}{1989}{1}.} for a more
complete discussion, and various examples). The simplest
choice is the proper--time cut--off
\hbox{$\rho(\tau,\Lambda) =\Theta(\tau - 1/\Lambda^2)$;} this
corresponds to the simple form $K(x)=e^{-x}$, or
\eqn\exxx{K_{\Lambda}(p) = \exp \biggl[
{-(p^2 + m^2)\over\Lambda^2}\biggr].}
This example is not quite so arbitrary as it looks; since both $K(z)$ and
$1/K(z)$ are regular, we may use Hadamard's factorization theorem and the
fundamental theorem of algebra to show that $K(z)$ is proportional to
$e^{-{\cal Q}(z)}$, where ${\cal Q}(z)$ is a polynomial. Conditions
1) and 3) above
then mean that ${\cal Q}(z)$ is at least of order one, with real
coefficients, positive for large real $z$ and with a zero at the
origin. $K(z)$ thus necessarily has an essential singularity at the
point at infinity; choosing the order $\sigma$ of this singularity to be one
gives \exxx\ uniquely. Regulating functions of higher order may of
course involve more parameters.

\nref\rfs{L.~D.~Faddeev and A.~A.~Slavnov, ``Gauge Fields:
Introduction to Quantum Theory'', Benjamin, New York, 1980.}
\nref\riix{B.~J.~Warr, \AP\vyp{183}{1988}{1,~59}.}
The class of regularizations \exxix\ is nothing but a
systematized form, implemented at Lagrangian level in the propagator,
of the crude momentum cut--offs or (in position space) point
splittings used in the early days of quantum field theory.\foot{Of
which lattice regularizations are a special case, if we allow cut-offs
which explicitly break rotational invariance.} However it differs
from such schemes by maintaining analyticity; strictly speaking high
momenta are never actually `cut off', only suppressed. It is
equivalent to the introduction of form factors into the vertices, as
may be readily appreciated by the simple nonsingular field
redefinition\foot{We  show in \S 4.1 below that this does not
change the S--matrix.}
\hbox{$\phi_p\rightarrow K_{\Lambda_0}(p)^{1/2}\phi_p$.}
Pauli--Villars regularization, although also formally equivalent if
there is an infinite set of regulator fields,
requires a pre--regulator if the subtraction of the divergences in the
unregularized loops by the divergences in regulator loops is to be
made well defined. If there are only a finite number of regulators the
theory is unstable, and unitarity is necessarily violated due
to negative metric poles at the regularization scale. Higher
derivative regularization (which in any case would only work if the
number of derivatives in the vertices were bounded) may also take the
form \exxix , but since $K(x)$ is then a polynomial conditions (1) and (2)
are necessarily violated, and again there is trouble with negative
metric poles.\foot{Furthermore, when using covariant higher
derivatives one loop divergences are still not
regulated since the number of derivatives in the vertices increases
with the number in the propagator; one is then drawn into a hybrid scheme,
as used for example
in refs.\refs{\rfs,\riix}.} Dimensional regularization, while again formally
a special case of \ecci\ (indeed the connection is made very explicit in
the second of ref.\rv ), will not be very useful here since it only
makes sense physically in the limit in which it is removed; it also
suffers from automatic subtraction, in the sense that it only picks up
logarithmic divergences, powerlike divergences being ignored when
divergent integrals are defined by analytic continuation.

The propagator used by Polchinski\rix\ is of the form \exxix ,
but with $K$ a function
of $p^2/\Lambda^2$, such that it is unity for $p^2<\Lambda^2$ and zero
for $p^2 > 4\Lambda^2$; then although $K(x)$ is still differentiable,
$K(z)$ is no longer regular (since a regular function
whose first derivative vanishes on some finite interval is identically
zero), and condition (2) is violated. This is undesirable because if
the regulating function is not regular, it is very difficult to
construct an analytic S--matrix. The reason for such an
awkward choice was that $K$ had to be held fixed for $0<p^2<\Lambda^2$
so that $\partial P_{\Lambda}/\partial \Lambda$ had no overlap
with the external source $J(p)$, the latter being set to zero
for $p^2>\Lambda^2$. Such a
technical restriction is fortunately no longer required if we use the
formulation of the exact renormalization group described
in the next section.

\subsec{The Exact Renormalization Group.}

The original idea of Wilson's exact renormalization group is to lower
the regularization scale $\Lambda$ below $\Lambda_0$ smoothly, integrating out
more and more of the high momentum modes,
while keeping the physics of the theory unchanged by simultaneously
adjusting the form of the action to compensate for the lost modes.
Although it is then only
strictly necessary to keep the S--matrix elements fixed as $\Lambda$ is
lowered, in fact it is possible, just as in statistical mechanics,
to keep the full off--shell connected Green's functions (or
`correlation functions') invariant,
and thus construct a (non--perturbative) scheme for the construction
of such Green's functions. Indeed if we
wish to use the regularized classical action \classact\ to
parameterize the evolution we must necessarily
work in Euclidean space;
it is then necessary to analytically continue the Euclidean Green's
functions back to Minkowski space in order to put them on--shell and
evaluate S--matrix elements by adding external line wave functions in
the usual way\ref
\rlsz{H.~Lehmann, K.~Symanzik and W.~Zimmermann, {\it
Nuov. Cim.~}\vyp{1}{1955}{205}.}.

We begin with the conventional generating functional for $n$-point
connected Euclidean Green's functions, expressed formally (though only for
convenience) as a functional integral:
\eqn\eiv{
\exp\,-W[J;\Lambda ]
\equiv \int\!{\cal D} \phi \exp \bigl[-S\big[\phi ; \Lambda \big]
-\big(J,\phi\big) \bigr]}
where the regularized classical action $S[\phi ;\Lambda]$
is of the form \classact\ described above.\foot{There is no need for a
measure factor in the path integral, because the bare regularized
action $S[\phi;\Lambda_0]$ is already entirely general; any reasonable measure
factor could be exponentiated and absorbed into it. Thus in an
effective quantum field theory there is nothing to be gained
from consideration of the path integral in the
Hamiltonian formulation.} We may assume that the propagator
has the form \exxix\ stipulated above, since we still have the freedom
to choose the evolution of the interaction
$S_{\rm int}[\phi;\Lambda]$. We would like to be able to do this in such a
way as to make $W[J;\Lambda]$ independent of $\Lambda$. In fact as we
will see in a moment this is not possible without restricting the
support of $J(p)$ and spoiling the analytic structure
of the propagator, so instead we consider the slightly
more general generating functional\rii
\eqn\exiii{\exp\,-\tilde{W}[J] = \int\!{\cal D} \phi \exp \bigl[ -\half (\phi,
P\inv_{\Lambda}\phi) - S_{\rm int}[\phi;\Lambda] -
\big(J,Q\inv_{\Lambda}\phi\big) - S_0[J;\Lambda] \bigr],}
where the function $Q_{\Lambda}(p)$ and the field independent
functional $S_0[J;\Lambda]$ are to be determined. Differentiating with
respect to $\Lambda$, we see that $\tilde{W}[J]$ will be independent of
$\Lambda$ provided
\eqn\ev{{\partial\tilde{W}\over\partial\Lambda}
= \Biggl\langle {1\over 2}\biggl( \phi,
{\partial P\inv_{\Lambda} \over\partial \Lambda} \phi
\biggr) + {\partial S_{\rm int} \over
\partial \Lambda} + \biggl(J,
{\partial Q\inv_{\Lambda} \over\partial \Lambda} \phi
\biggr) + {\partial S_0 \over
\partial \Lambda}\Biggr\rangle_{\!\Lambda} = 0.}
where $\langle X \rangle_{\Lambda}$ is the expectation value of $X$,
normalized so that $\langle 1\rangle_{\Lambda}=1$;
$$\langle X \rangle_{\Lambda}\equiv e^{\tilde{W}}\int\!{\cal D} \phi \, X\,
e^{-S[\phi,\Lambda]-(J,Q_\Lambda\inv \phi)-S_0}.$$

In order to find some condition on $S_{\rm int}[\phi;\Lambda]$ so that
\ev\ is satisfied we use the fact that, assuming suitable
boundary conditions on $\phi$, the functional integral of a total
functional derivative is zero.\foot{This may be proven
rigorously if we treat the path integral as the generator of
Feynman diagrams, i.e. as a perturbative object\rfs ; it may be
assumed to be true of any reasonable definition of functional
integration provided the action increases sufficiently rapidly for
large values of the field and its gradient.} This yields the
identity\refs{\rix,\riix}
\eqn\evi{ 0 \equiv \Biggl\langle{\delta \over \delta \phi_p}
\biggl( {1 \over 2} {\delta \over \delta \phi_{-p}} + P\inv_{\Lambda}
\phi_p + Q\inv_{\Lambda}J_p \biggr)\Biggr\rangle_{\!\Lambda},}
or, expanding out the functional derivatives, multiplying by
${\partial P_{\Lambda} \over \partial \Lambda}$, and
integrating over $p$,
\eqn\evii
{\eqalign{\Biggl\langle\biggl(\phi,
{\partial P\inv_\Lambda \over \partial \Lambda} \phi \biggr) \Biggr
\rangle_{\!\Lambda} &\equiv - \int\!{d^4 p \over (2 \pi)^4}
{\partial P_{\Lambda} \over \partial \Lambda }
\Biggl \langle \biggl[{\delta S_{\rm int}\over \delta \phi_{-p}}
{\delta S_{\rm int} \over \delta \phi_{p}}
-{\delta^2 S_{\rm int} \over \delta \phi_{-p} \delta \phi_p} \biggr]
\Biggr \rangle_{\!\Lambda}\cr
& -\Biggl\langle\biggl(J,2{\partial P\inv_\Lambda \over \partial \Lambda}
P_{\Lambda} Q\inv_{\Lambda}\phi \biggr) \Biggr
\rangle_{\!\Lambda}- \int\!{d^4 p \over (2 \pi)^4}\biggl[\delta^4(0)
{\partial\ln P_{\Lambda} \over \partial \Lambda }
-{\partial P_{\Lambda}\over\partial\Lambda}Q_{\Lambda}^{-2}J_pJ_{-p}\biggr].
\cr}}
Substituting this expression into \ev\ we find that $\tilde{W}[J]$ may be made
independent of $\Lambda$ if we choose
\eqn\eiix
{{\partial S_{\rm int} \over \partial \Lambda} = {1 \over
2} \int {d^4 p \over (2 \pi )^4}{\partial P_{\Lambda} \over
\partial \Lambda} \biggl[{\delta S_{\rm int} \over \delta \phi_p} {\delta
S_{\rm int} \over \delta \phi_{-p}} - { \delta^2 S_{\rm int} \over \delta
\phi_p
\delta \phi_{-p}} \biggr],}
\eqn\exiv{ Q_{\Lambda}(p) = Q(p) P_{\Lambda}(p),}
and
\eqn\exv{ S_0[J;\Lambda] = {1 \over 2} \int {d^4 p \over (2 \pi)^4} \biggl[
\delta^4(0)\ln P_{\Lambda} + Q^{-2}P\inv_{\Lambda}J_pJ_{-p}
\biggr]+{\cal E}_0(\Lambda).}
where $Q(p)$ is independent of $\Lambda$, and ${\cal E}_0(\Lambda)$ is
badly divergent but physically irrelevant.

Alternatively we may assume as Polchinski does\rix\ that
$P_{\Lambda}(p^2)$ and $J(p)$ have been
chosen such that the support of $ \partial P_{\Lambda} / \partial \Lambda$
has no overlap with that of $J$; in place of \exiv\ we may then
choose $Q_{\Lambda}=Q$, while since $S_0$ is independent of $J$ it
may be dropped without ceremony. The result \eiix\ is then unchanged.
The real problem with this approach is that the regularizing function
$K$ must necessarily be nonanalytic, and we will thus pursue it no further.

We consider the implications of the three equations \eiix --\exv\ in turn.
The first, eqn.\eiix\ is Wilson's exact renormalization group
equation, or `flow equation' \rii --- it tells us precisely how the
interaction evolves as the regularization scale is reduced, in order
to compensate for the modes which have been integrated out.\footsy{A
similar equation, formulated only for discrete momenta, may be found
in \ref\rWH{F.~J.~Wegner and A.~Houghton, \PR\vyp{A8}{1973}{401}}; a
continuum version of this equation was derived by
Weinberg\ref\rWerice{S.~Weinberg, in ``Understanding the Fundamental
Constituents of Matter'', Eric\'e~1976, ed.~A.~Zichichi
(Plenum,~1978).}. Unfortunately the sharp momentum cut--off used by
these authors renders the first term on the right hand side of the
equation singular, and seems to have discouraged further applications.}
The content of the equation is most readily appreciated by
depicting it graphically; graphs where the derivative of the
propagator connects two different vertices, as in
\nfig\frge{Schematic depiction of the effect of the two terms in the
exact renormalization group equation \eiix: a) two vertices with $n$
and $m$ legs combine to give a single vertex with $n+m-2$ legs, and b)
two legs of a single vertex with $n$ legs form a closed loop, to give
a vertex with $n-2$ legs. The solid lines denote the derivative with
respect to $\Lambda$ of the propagator.}
\frge a, produce
the first term on the right hand side of \eiix, while graphs where
the propagator connects two legs of
the same vertex, as in \frge b, produce the second term. The full power
of Wilson's equation should not be underestimated however; it is
formally exact since it follows directly from the
identity\evi. The graphical interpretation is so simple
because the infinitesimal change $dP_{\Lambda}$ has been exploited as
an expansion parameter. We find it nonetheless remarkable that the
full range of complexities of the quantized theory (both perturbative
and nonperturbative) may be built up by iteration of these two basic
ingredients.\foot{Indeed \eiix\ and the bare action
$S[\phi;\Lambda_0]$ could serve as a useful nonperturbative definition of the
theory. Such a definition, while less artificial than the lattice
definition, may prove more amenable to suitable truncations than the
Schwinger--Dyson equations (which can also be derived formally using
the same assumption on the boundary conditions of the
functional integral).}

It is not difficult to check that the flow equation is integrable;
indeed, the formal solution to \eiix\ with an appropriate boundary condition
at some scale $\Lambda_0$ is
\ref\rvii{G.~Keller, C.~Kopper and M.~Salmhofer,
{\it Helv.~Phys.~Acta.~}\vyp{65}{1992}{32}.},
\eqn\eix{\exp\bigl[-S_{\rm int}[\phi;\Lambda]-{\cal E}(\Lambda)\bigr]
=\exp ({\cal P}_{\Lambda_0}-{\cal P}_{\Lambda})
\exp -S_{\rm int}[\phi;\Lambda_0],}
where,
\eqn\exi{{\cal P}_\Lambda = {1 \over 2} \int {d^4 p \over
(2\pi)^4} P_{\Lambda}(p){\delta \over \delta \phi_p}
{\delta \over \delta \phi_{-p}}}
and ${\cal E}(\Lambda)$ collects the field-independent pieces, and is
again irrelevant since it makes no contribution to Green's functions.
{}From the form of this solution we can see that a vertex in
$S_{\rm int}[\phi;\Lambda]$ is constructed from connected diagrams
involving the vertices defined at $\Lambda_0$ and the regular `propagator'
${P}_{\Lambda_0}-P_{\Lambda}$.

The equation \exiv\ for $Q_\Lambda$ tells us that the we should couple
the source $J_p$ to $Q\inv(p)P\inv_{\Lambda}(p)\phi_p$
rather than just to $\phi_p$. It is simplest to take $Q(p)=1$. Then
comparing the two generating functionals \eiv\ and \exiii\ it is not
difficult to see that for $n>2$, $\Lambda$-independent n--point
connected Green's
functions obtained by functionally differentiating $\tilde{W}[J]$ with
respect to $J$, with $J$ then set to zero, are just the
$\Lambda$-dependent ones obtained similarly from $W[J;\Lambda]$, but
with the $\Lambda$-dependent free--particle propagator $P_\Lambda$
amputated from the external legs:
\eqn\exvi{\tilde{G}^{c}_n(p_1,\ldots ,p_n)=
P\inv_{\Lambda}(p_1)\cdots P\inv_{\Lambda}(p_n)
{G}^{c}_n(p_1,\ldots ,p_n;\Lambda ),}
where
\eqn\exxiii
{\eqalign{\prod_{i=1}^n\Big(-\frac{\delta}{\delta J_{p_i}}\Big)
W[J;\Lambda]\Big\vert_{J=0}&\equiv {G}^{c}_n(p_1,\ldots ,p_n;\Lambda ),\cr
\prod_{i=1}^n\Big(-\frac{\delta}{\delta J_{p_i}}\Big)
\tilde{W}[J]\Big\vert_{J=0}&\equiv \tilde{G}^{c}_n(p_1,\ldots ,p_n).\cr}}
This is particularly convenient since the $\Lambda$-independent
S--matrix elements may be obtained simply by analytically continuing
the Green's functions $\tilde{G}^c_n$ to the physical region, putting
them on shell, and adding external line wave functions in the usual
way\rlsz. If the more usual unamputated Green's
functions are preferred however, they may also be generated directly by
taking $Q(p)=P\inv_{\infty}(p)$; $\tilde{W}[J]$ then generates Green's
functions which are rendered $\Lambda$-independent by the removal of
the regulating factors $K_\Lambda(p)$ from their external legs (cf. \exxix ).

The field independent
term \exv\ is simply related to the connected generating functionals
$W_F[J;\Lambda]$ for the free theory obtained by
setting $S_{\rm int}$ to zero in \eiv ;
\eqn\exvii{\eqalign
{-W_F[P_{\Lambda}\inv J;\Lambda]
&\equiv\ln\bigg(\int {\cal D}\phi\exp\bigl[-\half(\phi,P\inv_{\Lambda}\phi)
-(J,Q P\inv_{\Lambda}\phi)\bigr]\bigg)\cr
&=\half (J,QP\inv_{\Lambda}J) +
\half\delta^4(0) \int\!{d^4 p \over (2 \pi)^4}\ln P_{\Lambda}
+{\cal E}_F(\Lambda ),\cr}}
Indeed for the particularly simple case of $Q=1$ the combined effect of \exiv\
and \exv\ may
be conveniently summarized by writing
\eqn\exxii{ \tilde{W}[J] = W[P_{\Lambda}\inv J;\Lambda]
- W_F[P_{\Lambda}\inv J;\Lambda].}
The $\Lambda$-independent connected two--point function is thus
given by \exxiii\ and \exxii\ as
\eqn\exxiv{\tilde{G}^{c}_2(p,-p) =
P\inv_{\Lambda}(p)G^c_2(p,-p;\Lambda)P\inv_{\Lambda}(p)
+P\inv_{\Lambda}(p).}

Using this formulation of the exact renormalization group, the
effective action $S[\phi,\Lambda]$ can be used to calculate Green's
functions for any external energy scale $E$ ($E > \Lambda$ no longer
providing any obstacle). When working at energy scale $E$ it is
often acceptable to set a renormalization scale $\Lambda_R\sim E$,
so this would not at first sight seem to be any great advantage.
However, it is more useful to set renormalization conditions
which may be directly related to physical observables, which means in
practice that we must take $\Lambda$ to zero, but with the external
momenta held fixed; it is then crucial to be able to treat $E>\Lambda$.

To see why this is so, note that if we let $\Lambda \rightarrow 0$, the
propagator is cut off for all
momenta ($P_0(p)=0$ for all Euclidean $p$) and the only Feynman
diagrams which give non--vanishing contributions are the interaction vertices
themselves.\footsy{
When taking the limit $\Lambda \rightarrow 0$ for a theory containing
massless particles more care is needed, since then $P_0(p)$ vanishes
only when $p^2>0$. For exceptional external
momenta the Green's functions may then develop infra-red
divergences. In this article we will always assume that $m>0$; the
massless theory is considered in ref.\ref\rbtir{R.~D.~Ball
and R.~S.~Thorne, CERN-TH.????/93, OUTP-93-??P.}.}
Now in perturbation theory the functional integral
\exiii\ may be performed to yield
\eqn\ewpert{\exp -\tilde{W}[J]=\exp {\cal P}_{\Lambda}
\exp\big[ -S_{\rm int}[\phi;\Lambda]-(J,P_\Lambda\inv\phi)
-\half(J,P_\Lambda\inv J)-{\cal E}_W(\Lambda)\big];}
the conventional perturbation theory results on expanding the
exponentials with $\Lambda=\Lambda_0$, the validity of the result for
all $\Lambda$ being guaranteed by the evolution \eix. In the limit
$\Lambda\to 0$ all reducible diagrams vanish, and the
effect of the first factor is merely to exchange $\phi$ for $J$ in the
interaction; the two quadratic terms cancel each other. We thus find
that
\eqn\exxiix{\tilde{W}[J]=S_{\rm int}[J;0]+{\cal E}_W(0)}
where ${\cal E}_W(0)$ is a constant, independent of $J$.
In fact we will show in \S 5.1 that in this limit
the effective action is also equal to the complete generating functional
for proper vertices. Here we merely remark that at $\Lambda=0$ the
parameters in the effective action may be directly related to physical
observables (by which we mean S--matrix elements); this is thus the
best place to set renormalization conditions.

\newsec{Perturbative Renormalizability of Conventional $\phi^4$
Theory.}

In this section we give a pedagogical proof of the boundedness
and convergence (or conventional renormalizability) of massive $\phi^4$
scalar field theory to all orders in perturbation theory. This proof
will use Polchinski's idea \rix\ of exploiting the simple form of
Wilson's flow equation \eiix , but will also incorporate the
improvements found in ref.\rvii , as well a number of our own. In
particular, we will use the analytic regulating function described in
\S 1.1, which will prove useful for the proof of the perturbative
unitarity and causality of the conventional theory in
\S 2.4, and vital when considering these aspects of the
effective theory in \S 5.3. We
will also adopt very general renormalization conditions, which are imposed
at $\Lambda =0$, i.e. on the Green's functions themselves; if the
proof were to be carried out for renormalization
conditions set at some scale $\Lambda_R\sim m$, and for zero external
momenta, it could be considerably simplified.\foot{While not vital for
proving the convergence of $\phi^4$ theory, this feature of our proof will be
essential when considering more complicated situations, such as
massless theories and theories with local symmetries, where it is necessary
for the low energy renormalization conditions to be directly related
to physical parameters.} We present the proof in some detail in order
to facilitate its extension to the effective theory in the rest of the
article.

The proof will be separated into four parts. It is first necessary to
give a detailed definition of the theory with which we will be
working, and in particular of the renormalization conditions.
We then prove that the magnitude of the vertices in the
effective Lagrangian at any $\Lambda$ are bounded to all orders in
perturbation theory. After this we prove, again to all orders in
perturbation theory, that the vertices, and thus Green's functions,
converge towards a finite limit as we take the cut--off to infinity.
Finally, we present a brief discussion of the perturbative analyticity,
unitarity and causality of the perturbative S--matrix of the conventional
theory.

\subsec{Defining the Theory.}

Using the concept of effective Lagrangian flow described in \S 1.2
above it is possible to see
that any effective quantum field theory may be completely defined by its field
content, its symmetries, the form of the regulator, and a full set of
boundary conditions on the renormalization group flow. The
regulator (which, as we saw in \S 1.1, amounts to a prescription
for the regulating function $K$)
dictates the precise form of this renormalization group flow;
different choices of $K$ yield different `renormalization
schemes'. The renormalization conditions are in effect
boundary conditions on the renormalization group flow of the
interaction part of the Lagrangian.

Euclidean $\phi^4$ scalar field theory is thus defined to be a theory with
one bosonic scalar
field $\phi(x)$, whose Lagrangian is invariant under translations and Euclidean
rotations and under the $Z_2$ field transformation \hbox{$\phi(x) \rightarrow
-\phi(x)$}. We only consider the unbroken symmetry in this article.
Such a theory has only three relevant operators;
$(\partial_{\mu}\phi)^2$, $\phi^2$ and $\phi^4$. The first two of
these, without the third, would describe a free theory. In the
conventional way of looking at quantum field theory, the $\phi^4$ term would
be the only interaction term in the bare Lagrangian; hence the name of
`$\phi^4$-theory'. From the point of view of the exact
renormalization group we call the theory $\phi^4$-theory because, as we
will prove, once we have specified the value of the mass of the
particle and the coupling constant associated with the $\phi^4$ term,
the perturbative S--matrix is determined completely, up to corrections
which vanish as $\Lambda_0/m\to\infty$.

The propagator is defined in accordance with \exxix\ and the succeeding
conditions, while
the interaction part of the action is then assumed to be a formal power
series in the fields; at the scale $0<\Lambda
<\Lambda_0$ all possible Lorentz invariant terms consistent with the
$Z_2$ global symmetry will appear in general, so we may write
the free and interacting parts of the action \classact\ as
\eqn\actgen{\eqalign
{\half\big(\phi,P_{\Lambda}\inv\phi\big)
&\equiv\sum^{\infty}_{j=0}\frac{\Lambda^{-2j}c_j}{j!}\int\!d^4x\phi(x)
(-\partial^2+m^2)^{j+1}\phi(x)\cr
S_{\rm int}[\phi;\Lambda]
&\equiv\sum^{\infty}_{m=1}\sum^{\infty}_{j=0}
\sum_{\{s_i:\sum_{i=1}^{2m-1}s_i=2j\}}\frac{\Lambda^{4-2m-2j}}{2m!}
c^{(m,j,s_i)}_{\mu_1\cdots\mu_{2j}}\cr
&\qquad\int\!d^4x\phi(x)
\big(\partial_{\mu_1}\cdots\partial_{\mu_{s_1}}\phi(x)\big)
\big(\partial_{\mu_{s_1+1}}\cdots\partial_{\mu_{s_1+s_2}}\phi(x)\big)\cdots
\big(\partial_{\mu_{2j-s_{2m}}}\cdots\partial_{\mu_{2j}}\phi(x)\big).\cr}}
For definiteness we take the order one regulating
function \exxx; this means that $c_n=1$.\foot{We will show how
this restriction may be lifted in \S 4.3 below.}

The dimensionless couplings $c^{(m,j,s_i)}_{\mu_1\cdots\mu_{2n}}$
are assumed to be expandable in terms of some `small' expansion
parameter $g$, beginning at first order.
It is conventional to let $g$ be equal to some
renormalized coupling constant, for example in \rix\
it is defined to be the coupling constant corresponding to the
four-point vertex at $\Lambda=\Lambda_R$ for zero momentum. Here we leave the
definition of $g$ open, simply treating it as a technical device with
which to order the perturbation series\foot{In practice a suitable
choice would be the value of the four point vertex defined at
$\Lambda=0$ for external momenta with
magnitude of order $\Lambda_R$}.

The expression \actgen\ is rather cumbersome, and for many purposes it
is much more convenient to write the interaction $S_{\rm int}$
in momentum space;
\eqn\exxxi{S_{\rm int}[\phi;\Lambda] \equiv \sum^{\infty}_{m=1} \sum
^{\infty}_{r=1} {g^r\over (2m)!}
\int {d^4p_1\cdots d^4p_{2m} \over (2\pi)^{4(2m - 1)}}
V^r_{2m}(p_1,\ldots ,p_{2m};\Lambda)
\delta^4\big(\hbox{$\sum^{2m}_{i=1}p_i$}\big)
\phi_{p_1}\cdots\phi_{p_{2m}},}
where $V^r_{2m}(p_1,\ldots ,p_{2m};\Lambda)\equiv
V^r_{2m}(p_i;\Lambda)$ is the value,
at $r$-th order in g, of the vertex in the effective action defined
at $\Lambda$ which has $2m$-legs. The canonical mass dimension of
the vertex function $V^r_{2m}$ is $4-2m$.

We take the vertex functions
$V^r_{2m}(p_1,\ldots ,p_{2m};\Lambda)$ to be infinitely
differentiable functions of the Euclidean momenta $p_i$.
In fact we shall see in
\S 2.4 below that when continued into the complex plane the
$V^r_{2m}(p_1,\ldots ,p_{2m};\Lambda)$ are regular functions for all
positive $\Lambda$, so that the quasi--local expansion \actgen\
converges whenever the field $\phi(x)$ is reasonably smooth. By
contrast at $\Lambda=0$,
$V^r_{2m}(p_1,\ldots ,p_{2m};0)$ are analytic everywhere except
on the boundary of the physical region; the representation
\exxxi\ is then only equivalent to the quasi--local expansion
\actgen\ within the radius of convergence of the Taylor expansions of
the vertex functions, which in practice means for \hbox{$(p_i+p_j)^2<m^2$}
for all $i$ and $j$.

The global symmetries of $S_{\rm int}$ ensure the following properties:
\item{(1)} Only even powers of $\phi$ appear (due to the $Z_2$
symmetry);
\item{(2)} Only the totally symmetric part of $V^r_{2m}(\Lambda)$
contributes; $V^r_{2m}(p_1,\ldots ,p_{2m};\Lambda)$ is invariant
under permutations of the $p_i$;
\item{(3)} $V^r_{2m}(\Lambda)$ is invariant under Euclidean
rotations. In particular $V^r_{2}(p,-p;\Lambda)$ depends on momentum only
through $p^2$. Therefore, denoting $(\partial/\partial p_{\mu})$ by
$\partial_{p_{\mu}}$,
\eqn\exxxii{ \partial_{p_{\mu}} V^r_{2m}(p ,-p;\Lambda)\vert_{p=0} = 0,}
and we can write
\eqn\exxxl{
\partial_{p_{\mu}}\partial_{p_{\nu}}V^r_{2}(p,-p;\Lambda) =
\delta_{\mu\nu} v^r_1(p^2;\Lambda) +
p_{\mu}p_{\nu}v^r_2(p^2;\Lambda).}

To define the theory completely we must impose a full set of
renormalization conditions on the vertex functions. These are chosen
to be of mixed type; we set the irrelevant couplings at the
regularization scale
$\Lambda_0$, which we assume to be much greater than the particle mass
$m$, and the relevant couplings at $\Lambda = 0$.
This means (see \exxiix) that we set the renormalization
conditions for the relevant
couplings on the Green's functions themselves; if we were to
go further and choose on--shell conditions (see \S 2.4 below),
these couplings would then be directly measurable.
Using the notation $\partial^j_p$ for any $j_{th}$ order
momentum derivative, where each derivative is with respect to any
particular component of any particular momentum, we
choose the following boundary conditions at $\Lambda_0$;
\eqn\exxxv{\partial^j_p V^r_{2m}(\Lambda_0) = 0 \qquad 2m + j > 4.}
As a consequence the only non-zero vertices at $\Lambda_0$ are
\eqn\exxxiii{V^r_2(p,-p;\Lambda_0) =
\Lambda_0^2 \lambda^r_1(\Lambda_0) + p^2\lambda^r_2(\Lambda_0),
\qquad V^r_4(p_1,p_2,p_3,p_4;\Lambda_0) = \lambda^r_3(\Lambda_0).}
This corresponds to the conventional way of choosing the bare
Lagrangian for $\phi^4$ theory;
only relevant operators appear in the local bare interaction
Lagrangian. In the conventional way of writing the theory
we would write these bare couplings as
\eqn\eccii{\Lambda_0^2 \lambda_1(\Lambda_0) =
m_0^2-m^2,\qquad\lambda_2(\Lambda_0)= Z-1,\qquad
\lambda_3(\Lambda_0)=\lambda_0,}
where $m_0^2$ is the bare mass, $Z$ is
the wavefunction renormalization and $\lambda_0$
is the bare coupling constant.\foot{Remember that the interaction has
been split from the propagator, hence the subtraction of $m^2$ in the
expression for $\lambda_1(\Lambda_0)$ and the subtraction of unity in
the expression for $\lambda_2(\Lambda_0)$.} It
should be emphasised that the $\lambda^r_i(\Lambda_0)$ are to be kept
completely arbitrary at the moment. They only become determined when
the renormalization conditions at $\Lambda = 0$ are specified.

We now have the freedom to choose the relevant coupling
constants. Defining a scale \hbox{$\Lambda_R\sim m$} such that
$0 < \Lambda_R \ll\Lambda_0$,\foot{As in \S 1,
$\Lambda_R$ is some scale similar to the energy at which we probe
the physics.} we take
\eqn\exxxvi
{\eqalign{\lim_{\Lambda\to 0}V^r_2(P_0,-P_0;\Lambda) &=
\Lambda_R^2\lambda^r_1,\cr
\lim_{\Lambda\to0}\bigl[
\partial_{p_{\mu}}\partial_{p_{\nu}}V^r_2(p,-p;\Lambda)
\vert_{p=P_0}\bigr]_{\delta_{\mu\nu}} &=\lambda^r_2,\cr
\lim_{\Lambda\to0}V^r_4(P_1,P_2,P_3,P_4);\Lambda) &=
\lambda^r_3,\cr}}
where $P_i^\mu$ are the external momenta at which the renormalization
conditions are set (with \hbox{$\sum_1^4 P_i^\mu=0$}), with magnitudes similar
to or less than $\Lambda_R$, and $\lambda^r_i$ are
some renormalization constants chosen independently of $\Lambda_0$.
$[\partial_{p_{\mu}}
\partial_{p_{\nu}}V^r_2(p,-p;\Lambda)]_{\delta_{\mu\nu}}$ is the part
of $\partial_{p_{\nu}} \partial_{p_{\nu}}V^r_2(p,-p;\Lambda)$
proportional to $\delta_{\mu\nu}$. A conventional, off-shell,  choice for these
renormalization conditions would be $P_i=0$,
$\lambda^r_1=0$, $\lambda_2^r = 0$ and $\lambda_3^r=\delta_{r,1}$.
However, here we will always consider the boundary conditions in their most
general form \exxxvi .

It is easy to see that within perturbation theory, and for finite
$\Lambda_0$, the two sets of boundary conditions \exxxiii\
and \exxxvi\ are consistent. At any order $r$ in the
expansion coefficient $g$, $V^r_2(p,-p;\Lambda)$ and
$V^r_4(p_i;\Lambda)$ can be expressed in
terms of diagrams containing the propagators $P_{\Lambda_0}(p)$ and
the coupling constants $\{\lambda^{s}_i(\Lambda_0);
\; i=1,2,3,\;s\leq r\}$. Now assume that the set
\hbox{$\{\lambda^{s}_i(\Lambda_0);\; i=1,2,3,\;s\leq r-1\}$} may
be expressed in terms of the set
$\{\lambda^t_i;\; i=1,2,3,\;t\leq s\}$. Writing the left
hand sides of \exxxvi\ in terms of the diagrams mentioned above, we
can then express the $\lambda^r_i(\Lambda_0)$ in terms of
$\{\lambda^s_i;\; i=1,2,3,\;s\leq r\}$.\footsy{Note that the diagrams
of order $r$ must be linear in $\lambda^r_i(\Lambda_0)$.}
Thus, if our assumption is true for $r-1$, it is also true for $r$.
The assumption is obviously
true for $r=1$, since all vertices
vanish for $s=0$. It follows that within
perturbation theory we can always determine
$\lambda_i^r(\Lambda_0)$ as an invertible function of $\Lambda_0$ and
$\{\lambda^{s}_i;\; i=1,2,3,\;s\leq r\}$

It should be stressed that it is not consistent to specify all
the relevant vertices; there are also
$\partial_{p_{\mu}}V^r_2(p,-p;\Lambda)$ and
$[\partial_{p_{\mu}}\partial_{p_{\nu}}V^r_2(p,-p;\Lambda)]_{p_{\mu}p_{\nu}}$
(the part of
$\partial_{p_{\mu}}\partial_{p_{\nu}}V^r_2(p,-p;\Lambda)$ proportional
to $p_{\mu}p_{\nu}$), but neither of these is an independent
variable. They both vanish
for $p=0$ and thus $\partial_{p_{\mu}} V^r_2(p,-p;\Lambda)$ can be
constructed from
$\partial^2_pV^r_2(p,-p;\Lambda)$, using Taylor's
theorem about $p=0$, and similarly
$[\partial_{p_{\mu}}\partial_{p_{\nu}}V^r_2(p,-p;\Lambda)]_{p_{\mu}p_{\nu}}$
can be reconstructed from
$\partial^3_p V^r_2(p,-p;\Lambda)$.

Since within perturbation theory we can express the
$\lambda_i(\Lambda_0)$ in terms of $\Lambda_0$ and the $\lambda_i$, we
see that the boundary conditions \exxxv\ and \exxxvi\ constitute a
complete set, and the theory is now completely defined.
We may therefore write the interaction
part of the action as $S_{\rm int}[\phi;\Lambda,\Lambda_0,\lambda_i]$.
(The dependence on the form of the
propagator and thus on $m$ will be left unstated from
now on; the dependence on $\Lambda_0$ and
$\lambda_i$ will also be frequently suppressed, as it was in \exxxi, to
simplify the notation.) Of course, since $\lambda_i(\Lambda_0)$
is an invertible function of $\Lambda_0$ and the
${\lambda_i}$, we may also write the interaction part of the action
as $S_{\rm int}[\phi;\Lambda,\Lambda_0,\lambda_i(\Lambda_0)]$ (as is
done in \rix ), but it is important to remember
that the independent variables in the theory are $\Lambda$,
$\Lambda_0$ and the ${\lambda_i}$. In an effective theory the action will also
depend on the irrelevant vertices at $\Lambda_0$, here we assume that
all of these are set to zero (\exxxv ). This restriction on the irrelevant
vertices will be eventually lifted in \S 3.1.

Note that if the theory is to be useful for constructing the S--matrix
of a single scalar particle over a reasonable range of
scales, it will be necessary to take $m\ll\Lambda_0$.
To achieve this requires a fine
tuning of $\lambda_1^r(\Lambda_0)$ to order $m^2/\Lambda_0^2$ at each
order in perturbation theory --- this is the well--known naturalness
problem for light scalars\nref\rvx{
K.~Wilson, \PR\vyp{D3}{1971}{1818}
\semi S.~Weinberg, \PR\vyp{D13}{1975}{974}; \vyp{D19}{1979}{1277}
\semi L.~Susskind, \PR\vyp{D20}{1979}{2619}.}\refs{\rvx,\rxx}. We will have
nothing new to say about
this problem here, save to say that it causes no obstruction to the bounding
arguments to follow.
\nref\rvi{
L.~D.~Landau in ``Niels Bohr and the Development of Physics'',
ed.~W.~Pauli (Pergamon, London, 1955)\nobreak
\semi K.~G.~Wilson, \PR\vyp{D6}{1972}{419} and ref.\rii
\semi M.~Aizenman, \PRL\vyp{47}{1981}{1}; \CMP\vyp{86}{1982}{1}
\semi J.~Fr\"ohlich, \NP\vyp{B200}{1982}{281}
\semi M.~L\"uscher and P.~Weisz, \NP\vyp{B290}{1987}{25}.}
Furthermore, if we go beyond perturbation theory it is no
longer obvious that boundary conditions of the type \exxxv\ and
\exxxvi\ are always consistent. Indeed, it is generally believed
that there is a
nonperturbative upper bound on $\lambda_3(\Lambda_0)$, which means
that as $\Lambda_0\rightarrow \infty$, $\lambda_3$ necessarily tends to
zero; conventional $\phi^4$ theory is `trivial'\rvi. The possible
implications of triviality for the effective theory will be discussed in
\S 5.3.

\subsec{Boundedness.}

Substituting the expansion \exxxi\ of $S_{\rm int}[\phi;\Lambda]$
into Wilson's flow equation \eiix\ leads to the following evolution
equations for the vertex functions $V^r_{2m}(\Lambda)$;
\eqn\exxxxi
{\eqalign{\Lambda {\partial \over \partial \Lambda} &\biggl(
 V^r_{2m}(p_1\ldots,p_{2m};\Lambda)\biggr)
=-{\Lambda^2 \over m(2m -1)}
\int {d^4p \over (2\pi\Lambda)^4}
K'_{\Lambda}(p) V^r_{2m+2}(p,-p,p_1\ldots,p_{2m};\Lambda) \cr
&\!\!\!+\sum^{m}_{l=1} \sum^{r-1}_{s=1}
{ \Lambda^{-2}2(2m)! \over (2m + 1 -2l)!(2l-1)!}
K'_{\Lambda}(P)
V^s_{2l}(p_1\ldots,p_{2l-1},P;\Lambda)V^{r-s}_{2m+2 -2l}
(-P,p_{2l}\ldots,p_{2m + 2};\Lambda),\cr}}
where $P \equiv \sum^{2l-1}_{i=1} p_{i}$, and (a
slight abuse of notation) $K'_{\Lambda}(p)\equiv
K'\big((p^2+m^2)/\Lambda^2\big)$, where
$K'(x)$ is the first derivative of the regulating function $K(x)$; with
the choice \exxx\ $K'_{\Lambda}(p)=-K_{\Lambda}(p)$.

We now construct a norm $\Vert V \VertL$  of a vertex function
$V^r_{2m}(p_i;\Lambda)$ by attaching
$[K_\Lambda(p_i)]^{1/4}$ to each leg of the
vertex and finding the maximum with respect to all the momenta;
\eqn\eNi{\Vert V\VertL \equiv
\max_{\{p_1,\ldots ,p_n\}}\biggl[\prod_{i=1}^{n}
[K_\Lambda(p_i)]^{1/4}
\vert V(p_1,...,p_{n};\Lambda)\vert\biggr].}
This is the appropriate analog for our analytic regulating function
of the norm defined in \rix . It is
shown in appendix A that it has the usual properties, and in particular
satisfies the triangle inequality and the Cauchy--Schwarz inequality. It is a
useful definition to take because, if the vertices
in a Feynman diagram absorb only a part of the damping factors $K$ on the
propagators, this is sufficient to allow the norm of the vertices to
be taken, while each internal propagator in that diagram will
still be left with sufficient damping that it too can be
bounded.\foot{We could have defined the norm such that the vertices
absorb factors of $K_\Lambda^{\alpha}$ where $\alpha$ is any real number
such that $0<\alpha<\half$; the vertices would then still have some
damping while sufficient damping is still associated with
the propagators. The value of a quarter is chosen simply because it
gives a sensible balance between the damping of the vertices and of
the propagator. With the choice \exxx\ $K_\Lambda^\alpha
=K_{\Lambda'}$, where $\Lambda'=\Lambda/\sqrt\alpha$.}
Thus, the integrals over internal momenta can be bounded at the same
time as the vertices, and the value of the whole diagram bounded.

It is not difficult to see that
\eqn\exxxxiii{\eqalign{\int {d^4p \over (2\pi)^4}
\vert [K^{1/2}_{\Lambda}(p)] \vert &< C
\Lambda^4 \vert K^{1/2}(m^2/\Lambda^2)\vert,\cr
\bigl\vert K_{\Lambda}^{-1/2}(p)\partial^k_p K_{\Lambda}(p)
\bigr\vert &< D_k \Lambda^{-k}\vert K^{1/2}(m^2/\Lambda^2)\vert,\cr}}
where $C$ and $D_k, k=1,2,3,\ldots$ are $\Lambda$-independent constants.
In all subsequent bounds we will ignore all such $\Lambda$-independent
constants, since they are unimportant for the argument, and only serve
to complicate our expressions. Considering all $\Lambda
\in[\Lambda_R,\Lambda_0]$, and forgetting the factor of $\vert
K^{1/2}(m^2/\Lambda^2)\vert$, which is approximately unity for all $\Lambda$
in this range, we can use \exxxxiii\ to bound the left hand side of \exxxxi ,
obtaining
\eqn\exxxxvi{\Biggl\Vert {\partial \over \partial \Lambda}
\biggl( V^r_{2m}(\Lambda)\biggr)\Biggr\Vert_{\Lambda}
\leq \Lambda \Vert V^r_{2m +2}(\Lambda) \VertL +\Lambda^{-3} \sum^{m}_{l=1}
\sum^{r-1}_{s=1} \Vert V^s_{2l}(\Lambda) \VertL \cdot \Vert
V^{r-s}_{2m +2 -2l}(\Lambda) \VertL.}
Similarly, by first acting with $j$ momentum derivatives on the flow
equation \exxxxi, then taking the norms, and using
\exxxxiii, we obtain
\eqn\exxxxvii{\eqalign
{\Biggl\Vert{\partial \over \partial\Lambda} \biggl( \partial^j_p
V^r_{2m}(\Lambda) \biggr) \Biggr\Vert_{\Lambda}
&\leq \Lambda \Vert \partial^j_p V^r_{2m}(\Lambda)
\VertL \cr
&\quad +\Lambda^{-3}\sum^m_{l=1} \sum^{r-1}_{s=1} \sum_{j_i;j_1 + j_2 + j_3 =
j} \Lambda^{-j_1} \Vert \partial^{j_2}_p V^s_{2l}(\Lambda) \VertL \cdot \Vert
\partial^{j_3}_{p} V^{r-s}_{2m + 2 -2l}(\Lambda) \VertL.\cr}}

When considering more
complicated situations such as massless particles and theories with a
number of particles with significantly different mass scales, direct bounds for
the vertices with $\Lambda$ below
$\Lambda_R$ will prove to be invaluable. Although such bounds
may be obtained indirectly from the vertices at $\Lambda_R$, we
believe it is more aesthetically pleasing and physically intuitive to
obtain them directly, an approach
made possible by the fact that our regulating function is
regular.\nobreak\foot{The
non--analytic regulating functions used in \rix\ lead to
ambiguities as $\Lambda \rightarrow 0$. For a method to overcome this
problem (though only for a finite number of momentum derivatives
and at the cost of further
complications) see \nobreak\ref\rxxv{G. Keller and C. Kopper,
\CMP\vyp{148}{1992}{445}.}.}

In order to obtain these bounds we derive an alternative inequality
to \exxxxvi. Considering $\Lambda\in[0,\Lambda_R]$, and this time
retaining the factors of $\vert
K^{1/2}(m^2/\Lambda^2)\vert$, we can use \exxxxiii\ to bound the left hand side
of \exxxxi, obtaining
\eqnn\esb
$$\eqalignno{\Biggl\Vert  {\partial \over \partial \Lambda}
\biggl( V^r_{2m}(\Lambda)\biggr)\Biggr\Vert_{\Lambda_R}
\leq \Lambda &\vert K^{1/2}(m^2/\Lambda^2)\vert\cdot\Vert V^r_{2m
+2}(\Lambda) \VertR \cr
&+\Lambda^{-3}\vert K^{1/2}(m^2/\Lambda^2)\vert \sum^{m}_{l=1}
\sum^{r-1}_{s=1} \Vert V^s_{2l}(\Lambda) \VertR \cdot \Vert
V^{r-s}_{2m +2 -2l}(\Lambda) \Vert_{\Lambda_R},&\esb \cr}
$$
where this time we define the norm with respect to $\Lambda_R$ rather than
$\Lambda$.\foot{It is easy to see that if in the integrand of the first
of the bounds \exxxxiii\ and the argument of the second
$K_{\Lambda_R}(p)$ were to be substituted for
$K_{\Lambda}(p)$, the bounds would remain true for all
$\Lambda < \Lambda_R$.} Remembering that $\Lambda_R \sim m$, and that $K(x)$
falls more quickly than any polynomial for large $x$, we can see that
for $\Lambda\in [0,\Lambda_R]$,
\eqn\es{\Lambda^{-n} K^{1/2}(m^2/\Lambda^2) \leq \Lambda_R^{-n}, }
where $n$ is any integer and we ignore a $\Lambda$- and
$\Lambda_R$-independent constant. Substituting \es\ into \esb\ we obtain
\eqn\esxxxxvi{\Biggl\Vert{\partial \over \partial \Lambda}
\biggl( V^r_{2m}(\Lambda)\biggr)\Biggr\Vert_{\Lambda_R}
\leq \Lambda_R \Vert V^r_{2m +2}(\Lambda) \VertR + \Lambda_R^{-3}\sum^{m}_{l=1}
\sum^{r-1}_{s=1} \Vert V^s_{2l}(\Lambda) \VertR \cdot \Vert
V^{r-s}_{2m +2 -2l}(\Lambda) \VertR.}
Similarly, by first acting with $j$ momentum derivatives on the flow
equation \exxxxi, then taking the norms, we obtain
\eqn\esxxxxvii{\eqalign
{\Biggl \Vert{\partial \over \partial\Lambda} \biggl( \partial^j_p
V^r_{2m}(\Lambda) \biggr) \Biggr\Vert_{\Lambda_R}
&\leq \Lambda_R \Vert \partial^j_p V^r_{2m}(\Lambda)
\VertR \cr
&\quad + \sum^m_{l=1} \sum^{r-1}_{s=1} \sum_{j_i;j_1 + j_2 + j_3 =
j} \Lambda_R^{-3-j_1} \Vert \partial^{j_2}_p V^s_{2l}(\Lambda) \VertR \cdot
\Vert
\partial^{j_3}_{p} V^{r-s}_{2m + 2 -2l}(\Lambda) \VertR .\cr}}

These equations coupled with the equations \exxxxvii\ and the boundary
conditions  \exxxv\ and \exxxvi\
are all we need to
prove the boundedness of the norms $\Vert V^r_{2m}(\Lambda)\VertL$.
These bounds are expressed in the following lemma;

\medskip
\vbox{\noindent{\it Lemma 1:}

\noindent (i) For all $\Lambda \in [\Lambda_R,\Lambda_0]$,
\eqn\exxxxiix{ \bigl\Vert \partial^j_p V^r_{2m}(\Lambda) \bigr\VertL \leq
\Lambda^{4-2m-j} \Biggl( P\log \biggl( {\Lambda \over \Lambda_R} \biggr)
+ {\Lambda \over\Lambda_0} P\log \biggl({\Lambda_0 \over
\Lambda_R}\biggr)\Biggr).}
\noindent (ii) For all $\Lambda \in [0,\Lambda_R]$,
\eqn\esxxxxiix{ \bigl\Vert\partial^j_p V^r_{2m}(\Lambda)\bigr\VertR \leq
\Lambda_R^{4-2m-j}. }}
Here $P\log(z)$ is a finite order polynomial in $\log(z)$,
whose coefficients depend neither on $\Lambda$ nor on $\Lambda_0$, but
may depend on the ratio $(\Lambda_R/m) $.\foot{The
order of the polynomials is $(r +1 - m)$ for $r\geq m-1$ and zero for
$r<m-1$; details can be found in
\rix\ and \rvii. In this article we will not keep track of the order of the
polynomials in order to make the proof as easy as possible to follow.
This involves no difference in technique, but a simplification of the
algebra.}
\medskip

The proof of this lemma will constitute most of the remainder of this
subsection, and will proceed by induction. The lemma is trivially true for
$r=0$, i.e. at zeroth order in the expansion coefficient $g$, because the
vertices are defined to be zero at this order. Furthermore, the boundary
condition at $\Lambda_0$ is such that only $V_2(\Lambda_0)$ and
$V_4(\Lambda_0)$ are non-zero; at $\Lambda_0$ there are only vertices
with two or four legs. From \eix\ and \exi, we see
that the vertices at all other
scales can be constructed from connected diagrams containing the vertices
defined at $\Lambda_0$ and the difference between propagators defined at
$\Lambda$ and $\Lambda_0$. It is
therefore a simple topological exercise to see that at any scale
$\Lambda$, a vertex with $2m$ legs must be constructed from a diagram
with at least $m-1$ $V_4(\Lambda_0)$ vertices, and must
therefore be at least this order in $g$. So at order $r$ in the
expansion coefficient $g$, $V^r_{2m}(\Lambda)$ vanishes if $m>r+1$.
Thus, at order $r$ the lemma is trivially satisfied for vertices
with $m>r+1$.

       The process of induction to be used to prove the lemma for all $r$
and $m$ should now be clear. The lemma is true trivially
at zeroth order in the expansion coefficient $g$. We then suppose
that the lemma is true up to order
$r-1$ in $g$. It is obviously true at order $r$ for $m>r+1$, since in this case
$V^r_{2m}(\Lambda)$ vanishes. We suppose that at order $r$ it is also
true down to $m +1$ for some
$m$ less than or equal to $r+1$. Then if $\Lambda\in
[\Lambda_R,\Lambda_0]$, all the vertices on the
right hand side of
\exxxxvii\ satisfy \exxxxiix, and \exxxxvii\ may be rewritten as
\eqn\exxxxix{\Biggl\Vert {\partial \over \partial \Lambda} \biggl(
\partial^j_p V^r_{2m}(\Lambda) \biggr) \Biggr\Vert_{\Lambda}
\leq \Lambda^{3-2m-j}\Biggl( P\log \biggl({\Lambda \over \Lambda_R}\biggr) +
{\Lambda
\over \Lambda_0} P \log \biggl( {\Lambda_0 \over \Lambda_R} \biggr)
\Biggr).}
Similarly, for $\Lambda\in [0,\Lambda_R]$, all the vertices
on the right hand side of
\esxxxxvii\ already satisfy \esxxxxiix, so \esxxxxvii\ may be rewritten as
\eqn\esxxxxix{\Biggl\Vert {\partial \over \partial \Lambda} \biggl(
\partial^j_p V^r_{2m}(\Lambda) \biggr) \Biggr\Vert_{\Lambda_R}
\leq \Lambda_R^{3-2m-j}.}
Since we intend now to proceed downwards in number of
legs, we must obviously deal in general with the irrelevant vertices before the
relevant vertices. In fact the proof will proceed in four distinct
steps: a) first we prove (i) for the irrelevant vertices; b) we prove
(ii) for the irrelevant vertices; c) we prove (ii) for the relevant
vertices; and finally d) we prove (i) for the relevant vertices.

\medskip

a) We thus begin with the irrelevant vertices, i.e. those
where the number of legs plus the number of momentum derivatives is
greater than four. For these we use the trivial boundary conditions
\exxxv\ at  $\Lambda_0$ to write
\eqnn\el
$$\eqalignno{\Vert \partial^j_p V^r_{2m}(\Lambda) \Vert_{\Lambda}
&= \Vert \partial^j_p V^r_{2m}(\Lambda) -
\partial^j_p V^r_{2m}(\Lambda_0) \Vert_{\Lambda}\cr
& = \Biggl\Vert \int^{\Lambda_0}_{\Lambda} d\Lambda' {\partial \over
\partial \Lambda'}\biggl(  \partial^j_p
V^r_{2m}(\Lambda') \biggr) \Biggr\Vert_{\Lambda}\cr
& \leq \int^{\Lambda_0}_{\Lambda} d \Lambda' \Biggl\Vert {\partial \over
\partial \Lambda'} \biggl(  \partial^j_p
V^r_{2m}(\Lambda')\biggr) \Biggr\Vert_{\Lambda}\cr
& \leq \int^{\Lambda_0}_{\Lambda} d \Lambda' \Biggl\Vert {\partial
\over \partial \Lambda'} \biggl(  \partial^j_p
V^r_{2m}(\Lambda') \biggr) \Biggr\Vert_{\Lambda'}, &\el\cr}
$$
where in the last line the norms go from being weighted by $K_\Lambda$
to being weighted by $K_{\Lambda'}$, which is made possible by the
fact that $K(x)$ is a monotonically decreasing function.
So using \exxxxix\ and \el,
\eqn\eli{\eqalign{ \Vert  \partial^j_p V^r_{2m}(\Lambda) &\Vert_{\Lambda}
\leq \int^{\Lambda_0}_{\Lambda} d \Lambda' (\Lambda')^{3-2m-j} \Biggl(
P\log \biggl({\Lambda' \over \Lambda_R} \biggr) + {\Lambda' \over
\Lambda_0} P \log \biggl( {\Lambda_0 \over \Lambda_R} \biggr) \Biggr).\cr
&=\Lambda^{4-2m-j} P \log \biggl( {\Lambda \over
\Lambda_R}\biggr)+\Biggl[\Lambda^{4-2m-j}\biggl(\sum^{2m + j-4}_{r=1}
\biggl({ \Lambda \over\Lambda_0}\biggr)^r
 + {\Lambda \over \Lambda_0} \biggr)- \Lambda_0^{4-2m -j}\Biggr] P \log
\biggl({\Lambda_0 \over \Lambda_R} \biggr).\cr}}

For all $\Lambda \in [\Lambda_R, \Lambda_0]$,
$\Lambda^{4-2m-j}(\Lambda/\Lambda_0)^r$ and $\Lambda_0^{4-2m-j}$ are
less than or equal to $\Lambda^{4-2m-j}(\Lambda/\Lambda_0)$. So we simply
combine the
terms in the square brackets and the bound \eli\ becomes precisely \exxxxiix\
for all the
irrelevant vertices at order $r$.

\medskip

b) We can now use our induction argument to prove (ii) for the
irrelevant vertices. The evaluation of \exxxxiix\
at $\Lambda=\Lambda_R$ gives us the bound
$\Vert \partial^j_p V^r_{2m}(\Lambda_R)\Vert_{\Lambda_R}
\leq \Lambda_R^{4-2m-j}$ for the
irrelevant vertices. (The second term may always be absorbed into the
first for any $\Lambda_R\leq \Lambda_0$, since it is always bounded by
some constant.) Furthermore for $\Lambda \in [0,\Lambda_R]$
$$
\partial^j_p V^r_{2m}(\Lambda)
= \partial^j_p V^r_{2m}(\Lambda_R)
- \int^{\Lambda_R}_{\Lambda} d\Lambda' {\partial \over
\partial \Lambda'}\biggl( \partial^j_p
V^r_{2m}(\Lambda') \biggr),
$$
so taking norms and using the bounds at $\Lambda_R$ we can write
\eqnn\esl
$$\eqalignno{\Vert  \partial^j_p V^r_{2m}(\Lambda) \VertR
& \leq \Lambda_R^{4-2m-j} + \Biggl\Vert \int^{\Lambda_R}_{\Lambda} d\Lambda'
{\partial \over
\partial \Lambda'}\biggl(  \partial^j_p
V^r_{2m}(\Lambda') \biggr) \Biggr\Vert_{\Lambda_R},\cr
& \leq \Lambda_R^{4-2m-j} + \int^{\Lambda_R}_{\Lambda} d \Lambda' \Biggl\Vert
{\partial \over
\partial \Lambda'} \biggl( \partial^j_p
V^r_{2m}(\Lambda')\biggr) \Biggr\Vert_{\Lambda_R}. &\esl\cr}
$$\nobreak
Combining \esxxxxix\ with \esl\ trivially gives \esxxxxix\ for
all the irrelevant vertices at order $r$.

\medskip

c) We are left with the problem of verifying the lemma for the relevant
vertices, i.e. those with $2m +j \leq 4$. This situation is
significantly different to that for the irrelevant vertices. In that
case the bound in \exxxxiix\ was found by integrating down from
$\Lambda_0$, and held for any `natural' bare vertices.
Following the same procedure for the relevant vertices, assuming that the
$\lambda_i(\Lambda_0)$ are `natural' would simply give $\Vert
V_4(\Lambda)\Vert_{\Lambda_R}
\leq P\log(\Lambda_0/\Lambda_R)$, $\Vert V_2(\Lambda) \Vert_{\Lambda_R} \leq
\Lambda_0^2P\log(\Lambda_0/\Lambda_R)$, etc., i.e. large
$\Lambda_0$-dependence
which would then feed into the bounds on the irrelevant
vertices via the flow equations.  In order to
verify \exxxxiix\ it is necessary to use the fact that the low energy
relevant coupling constants have been ``fine-tuned'' to unnaturally
low values by the renormalization conditions, and then fixed at these
values independently of $\Lambda_0$. Thus, we must find a way
of using the renormalization conditions on the relevant couplings in
order to obtain our bound.

We first do this for the four--point vertex, $V^r_4(\Lambda)$.
We have a bound on the four-point vertex defined at $\Lambda =0$ at the
particular momenta $P_i$ at which the renormalization condition
\exxxvi\ is set. However, in order to
satisfy the lemma we need to know that the norm of the
vertex satisfies this same bound; a much stronger condition. To do
this we use Taylor's formula
\eqn\elxi{ f(p_1.....p_n) = f(q_1.....q_n) + \sum^{n}_{i=1}
(p-q)_i^{\mu} \int^{1}_{0} d \rho
\;\partial_{k_{i}^{\mu}} f(k_1.....k_n),}
$$
k_i = q_i + \rho (p_i -q_i),\qquad 1 \leq i \leq n,
$$
to reconstruct $V^r_4(p_1,p_2,p_3,p_4;0)$ from
$V^r_4(P_1,P_2,P_3,P_4;0)$. $\Vert V^r_4(0)\Vert_{\Lambda_R}$ will
then satisfy the lemma if both terms on the right hand side of the Taylor
formula do. That the first does follows directly from the
renormalization condition. We also know (from
the induction argument for the irrelevant vertices) that
$\Vert\partial^j_k V^r_4(0) \Vert_{\Lambda_R} \leq
\Lambda_R^{-j}$ for $j \geq 1$. It
turns out that this is sufficient to show that the second term in the
Taylor formula also satisfies the required bound, and thus that
$V^r_4(0)$ satisfies (ii); the full argument is given in appendix B.

Now that we have the bound $\Vert V^r_4(0)\VertR \leq c$, where $c$ is
a constant, we can
verify (ii) for the four-point vertex for all $\Lambda \in
[0,\Lambda_R]$. Again since
\eqn\eslix{ V^r_4(\Lambda)=V^r_4(0)+\int^{\Lambda}_{0} d\Lambda'
{\partial \over \partial\Lambda'}V^r_4(\Lambda'),}
taking the norm of both sides we have
\eqn\eslixa{ \Vert V^r_4(\Lambda) \VertR
\leq c + \int^{\Lambda}_{0} d\Lambda'
\Vert {\partial \over \partial\Lambda'}V^r_4(\Lambda')\VertR.}
Using \esxxxxix\ with $m=2$, $j=0$
we thus find
\eqn\ess{\Vert V^r_4(\Lambda)\Vert_{\Lambda_R} \leq c.}

The remaining relevant vertices all have m=1, i.e two legs. The first is
$\partial^2_pV^r_2(\Lambda)$, and the method for proving that this
satisfies (ii) is a little more complicated than that for
$V^r_4(\Lambda)$. We have the renormalization condition \exxxvi\ for
$[\partial_{p_{\mu}} \partial_{p_{\nu}}
V^r_2(p,-p;0)\vert_{p=P_0}]_{\delta_{\mu\nu}}$, but it is necessary to
have some bound for the whole of
$\partial_{p_{\mu}}\partial_{p_{\nu}}V^r_2(p,-p;0)\vert_{p=P_0}$.
Again using Taylor's formula \elxi, but this time to construct
$\partial^2_pV^r_2(0)$, we have
\eqn\elxii{\eqalign{
\partial_{p_{\mu}}\partial_{p_{\nu}}
V^r_2(p,-p;0)\vert_{p=P_0} &= \partial_{p_{\mu}}
\partial_{p_{\nu}} V^r_2(p,-p;0)\vert_{p=0}\cr
&\qquad + P_{0}^{\lambda} \int^{1}_{0} {d\rho \over \rho^3} \>
{\partial \over \partial
p^{\lambda}}{\partial \over \partial p^{\mu}}{\partial \over \partial p^{\nu}}
V^r_2(\rho p,-\rho p;0)\vert_{p=P_0}.\cr}}
{}From \exxxii, the first term on the right only has a part proportional
to $\delta_{\mu\nu}$. Therefore, the full contribution to
$[\partial_{p_{\mu}}\partial_{p_{\nu}}V^r_2(p,-p;0)
\vert_{p=P_0}]_{p_{\mu}p_{\nu}}$
will come from the second term on the right, and using the result in
appendix B and the bounds already obtained on higher momentum
derivatives of $V^r_2(0)$, we see that
$[\partial_{p_{\mu}}\partial_{p_{\nu}}V^r_2(p,-p;0)
\vert_{p=P_0}]_{p_{\mu}p_{\nu}}$
satisfies the same type of bound as
$[\partial_{p_{\mu}}\partial_{p_{\nu}}V^r_2(p,-p;0)
\vert_{p=P_0}]_{\delta_{\mu\nu}}$.
Thus,
\eqn\elxii{\Bigl\vert\partial^2_p V^r_2(p,-p;0)
\vert_{p=P_0} \Bigr\vert \leq c.}
It is now straightforward to verify (ii) for $\partial^2_pV^r_2(\Lambda)$
using exactly the same procedure as that used to verify the lemma for
$V^r_4(\Lambda)$.

Because of \exxxii,
$\partial_{p_{\mu}}V^r_2(\Lambda)$ can be constructed
entirely from $\partial^2_p V^r_2(\Lambda)$ using Taylor's formula. So
using the bounds on $\partial^j_p V^r_2(\Lambda)$, $j \geq 2$, the
result in appendix B, it is easy to see that $\partial_{p_{\mu}}
V^r_2(\Lambda)$ also satisfies (ii).

Finally, the case $V^r_2(\Lambda)$ can be treated in just the
same way as $V^r_4(\Lambda)$. It should be mentioned that the fine
tuning is particularly severe in this case. If we were simply to
consider ``natural'' bare vertices, we would find that
$V^r_2(P_0,-P_0;\Lambda)$ was of order $\Lambda_0^2$, for
$\Lambda \in [0,\Lambda_R]$, and thus that the mass of the
particle would be of order $\Lambda_0$. Thus, by imposing \exxxvi\ we
have necessarily to tune $\lambda^r_1(\Lambda_0)$ to an
unnaturally precise value\refs{\rvx,\rxx}.

\medskip

d) The proof of (i) for the relevant vertices is now relatively
straightforward. We know that (ii) is true for the vertices at
$\Lambda_R$, and thus that for momenta with magnitudes $\sim
\Lambda_R$, e.g. the momenta at which the renormalization condition
were set, that the modulus of a relevant vertex is of the same
order of magnitude as its renormalization condition. Thus, we have in effect
derived new renormalization conditions on the coupling constants defined
at $\Lambda_R$. These can now be used to verify (i).

Starting with $V^r_4(\Lambda)$, we can use \exxxxix\ to write
\eqn\eliix{\Biggl\vert {\partial \over \partial \Lambda}
V^r_4(P_i;\Lambda) \Biggr\vert \leq  \Lambda\inv \Biggl(
P \log \biggl({\Lambda \over \Lambda_R}\biggr) + {\Lambda \over
\Lambda_0} P \log \biggl({\Lambda_0 \over\Lambda_R}\biggr)\Biggr),}
for all $\Lambda \in [\Lambda_R,\Lambda_0]$. Thus
\eqnn\esf
$$\eqalignno{
 \vert V^r_4(P_i;\Lambda)\vert
 &\leq \vert V^r_4(P_i;\Lambda_R) \vert + \int^{\Lambda}_{\Lambda_R} d
\Lambda' \Biggl\vert {\partial \over \partial \Lambda'}
V^r_4(P_i;\Lambda')\Biggr\vert \cr
&\leq c+ \int^{\Lambda}_{\Lambda_R} d
\Lambda' (\Lambda')\inv \Biggl( P
\log \biggl({\Lambda' \over \Lambda_R}\biggr) + {\Lambda \over
\Lambda_0}P\log\biggl({\Lambda_0 \over \Lambda_R}\biggr)\Biggr)\cr
&\leq P\log \biggl( {\Lambda \over
\Lambda_R} \biggr) + {\Lambda \over
\Lambda_0} P \log \biggl( {\Lambda_0 \over \Lambda_R} \biggr) - {\Lambda_R
\over \Lambda_0} P \log \biggl( {\Lambda_0 \over \Lambda_R} \biggr)\cr
&\leq P \log \biggl( {\Lambda \over
\Lambda_R} \biggr) + {\Lambda
\over \Lambda_0} P \log \biggl( {\Lambda_0 \over \Lambda_R} \biggr).&\esf\cr}$$
In order to complete the verification of (i) for the four--point
vertex we simply have to extend this result to a similar bound on
$\Vert V_4^r\Vert_\Lambda$ using Taylor's formula
and the methods of appendix B.

The verification of (i) for the remaining relevant vertices
proceeds in the
same manner, remembering to work downwards in the number of momentum
derivatives. We simply use (ii) to provide an effective renormalization
condition at $\Lambda_R$, integrate up from $\Lambda_R$ to $\Lambda \in
[\Lambda_R,\Lambda_0]$ and then use the Taylor formula
at $\Lambda$.

\medskip

Assuming the lemma to be true up to order
$r-1$ in the expansion coefficient $g$, and at order $r$ true down to
order $m + 1$ for some $m$, ($2m$ is the number of legs a vertex has),
it has now been proven that
it is true at order $r$ down to order $m$, for any $m$. Since it is true
for large enough $m$ for any $r$, at order $r$ in $g$ it is true for
all $m$, by induction in $m$. Thus the lemma is
true to order $r$ in $g$ and for large enough $m$ at order $r+1$ in
$g$, and the induction downwards in $m$ goes through in exactly the
same way as at order $r$. So since lemma~1 is trivially satisfied at
zeroth order in $g$, by induction it is satisfied at all orders in
$g$, and lemma~1 is proven. $\blackbox$

\medskip
In particular, at $\Lambda = 0$ lemma~1 tells us that,
\eqn\esc{\Vert
\partial^j_p V^r_{2m}(0) \VertR
\leq  \Lambda_R^{4-2m-j}.}
Remembering the relationship \exxiix\ between the dimensionful vertices at
$\Lambda=0$ and the connected amputated Green's functions, this can
equally be written as
\eqn\elxxv{\Vert
\partial^j_p \tilde G^{c,r}_{2m}(\Lambda_0,\lambda_i) \VertR
\leq  \Lambda_R^{4-2m-j},}
We therefore have a very
direct proof of the boundedness of the Green's functions and their
momentum derivatives. As long as the external momenta are all
$\sim\Lambda_R$, the damping factors on the external legs are all
$\sim 1$, and the bound on the norm of the Green's functions becomes a
bound on the Green's functions themselves. So, for external momenta
$\sim \Lambda_R$, the Green's functions, and their momentum
derivatives, are less than or equal to a $\Lambda_0$-independent
dimensionless constant multiplied by the appropriate power of
$\Lambda_R$.

This is not yet a proof of
renormalizability, since the values of the Green's functions
could be consistent with this bound while still fluctuating
significantly as $\Lambda_0$ is varied. Therefore, we still have
to examine the behaviour of the vertices, and hence the Green's
functions, as we vary $\Lambda_0$. In
particular, we need to show that they converge to definite limits as
$\Lambda_0 \rightarrow\infty$.

\subsec{Convergence.}

We wish to show that the Green's functions converge towards a finite
limit as $\Lambda_0 \rightarrow  \infty$. Once this has been achieved,
we will have proved perturbative
renormalizability in the conventional sense. In the more modern way of
thinking, we will show that with the specified renormalization conditions
held fixed the Green's functions are independent of $\Lambda_0$ up to
terms which vanish as powers of $\Lambda_R/\Lambda_0$.\foot{This was
the final result of Polchinski in \rix ; the method used here will be rather
different to his, being closer in spirit to that in \rvii.}

The quantity we need to bound is the rate of change of
$S_{\rm int}[\phi;\Lambda,\Lambda_0,\lambda_i]$ with respect to
$\Lambda_0$ at fixed $\lambda_i,\;i=1,2,3$; i.e. $\left(
\Lambda_0 {\partial \over\partial \Lambda_0} S_{\rm int}
[\phi; \Lambda,\Lambda_0,\lambda_i] \right)_{\!{\Lambda,\lambda_i}}$,
where the dependence on $\Lambda_0$ through the scale at which the
renormalization
conditions on the irrelevant vertices are set is displayed explicitly.

In order to do this, it is again easier to consider irrelevant
vertices first. Applying $j$ momentum derivatives to both sides of the
vertex evolution equations
\exxxxi, setting $\Lambda = \Lambda'$, integrating with respect to
$\Lambda'$ from $\Lambda \in [\Lambda_R,\Lambda_0]$ up to
$\Lambda_0$, differentiating with respect to $\Lambda_0$ at fixed
$\Lambda$ and ${\lambda_i}$, and taking bounds, we obtain
\eqn\elxv{\eqalign
{ &\Biggl\Vert \biggl( {\partial \over \partial \Lambda_0}
\partial^j_p V^r_{2m}(\Lambda,\Lambda_0,\lambda_i) \biggr)_
{{\!\Lambda,\lambda_i}} \biggr\Vert_{\Lambda}
\leq \Biggl[\Lambda_0 \Vert \partial^j_p
V^r_{2m+2}(\Lambda_0,\Lambda_0,\lambda_i) \Vert_{\Lambda_0} \cr
&\qquad+\sum^{m}_{l=1} \sum^{r-1}_{s=1} \sum_{j_i;j_1 + j_2 + j_3 = j}
\Lambda_0^{-3-j_1} \Vert \partial^{j_2}_p V^s_{2l}
(\Lambda_0,\Lambda_0,\lambda_i) \Vert_{\Lambda_0} \cdot
\Vert \partial^{j_3}_p V^{r-s}_{2m + 2 -2l}
(\Lambda_0,\Lambda_0,\lambda_i) \Vert_{\Lambda_0}\Biggr] \cr
&+ \int^{\Lambda_0}_{\Lambda} d \Lambda' \Biggl[\Lambda'
\Biggl \Vert \biggl( {\partial \over \partial \Lambda_0} \partial^j_p
V^r_{2m+2} (\Lambda',\Lambda_0,\lambda_i) \biggr)_{{\!\Lambda',\lambda_i}}
\Biggr\Vert_{\Lambda'}\cr
&+\sum^{m}_{l=1} \sum^{r-1}_{s=1} \sum_{j_i;j_1 + j_2
+ j_3 = j} (\Lambda')^{-3-j_1} \Biggl \Vert \biggl( {\partial \over
\partial \Lambda_0} \partial^{j_2}_p V^t_{2l}(\Lambda',\Lambda_0,\lambda_i)
\biggr)_{{\!\Lambda',\lambda_i}}
\Biggr\Vert_{\Lambda'} \cdot \Biggl \Vert \partial^{j_3}_p
V^{r-s}_{2m + 2-2l}(\Lambda',\Lambda_0,\lambda_i)
\Biggr\Vert_{\Lambda'}\Biggr] ,\cr}}
where $2m + j > 4$.
In order to investigate the $\Lambda_0$-dependence of the irrelevant
vertices for $\Lambda \in[0,\Lambda_R]$ we apply $j$ momentum derivatives to
both sides of the
vertex evolution equations
\exxxxi, set $\Lambda = \Lambda'$, integrate with respect to
$\Lambda'$ from $\Lambda \in [0,\Lambda_R]$ up to
$\Lambda_R$, differentiating with respect to $\Lambda_0$ at fixed
$\Lambda$ and ${\lambda_i}$, and take bounds to obtain
\eqn\eslxv{\eqalign
{ &\Biggl\Vert \biggl( {\partial \over \partial \Lambda_0}
\partial^j_p V^r_{2m}(\Lambda,\Lambda_0,\lambda_i) \biggr)_
{{\!\Lambda,\lambda_i}} \Biggr\Vert_{\Lambda_R}
\leq \quad  \Biggl\Vert \biggl( {\partial \over
\partial \Lambda_0} \partial^j_p
V^r_{2m}(\Lambda_R,\Lambda_0,\lambda_i)
\biggr)_{{\!\Lambda_R,\lambda_i}} \Biggr \Vert_{\Lambda_R}\cr
&+ \int^{\Lambda_R}_{\Lambda} d \Lambda'  \Biggl[
\Lambda_R\Biggl \Vert \biggl( {\partial \over \partial \Lambda_0} \partial^j_p
V^r_{2m+2} (\Lambda',\Lambda_0,\lambda_i) \biggr)_{{\!\Lambda',\lambda_i}}
\Biggr \Vert_{\Lambda_R}\cr
&+\sum^{m}_{l=1} \sum^{r-1}_{s=1} \sum_{j_i;j_1 + j_2
+ j_3 = j} \Lambda_R^{-3-j_1} \Biggl \Vert \biggl( {\partial \over
\partial \Lambda_0} \partial^{j_2}_p V^t_{2l}(\Lambda',\Lambda_0,\lambda_i)
\biggr)_{{\!\Lambda',\lambda_i}}
\Biggr \Vert_{\Lambda_R} \cdot \Biggl \Vert \partial^{j_3}_p
V^{r-s}_{2m + 2-2l}(\Lambda',\Lambda_0,\lambda_i) \Biggr \Vert_{\Lambda_R}
\Biggr],\cr}}
where as in \S 2.2 we bound
with respect to $\Lambda_R$ when $\Lambda\in [0,\Lambda_R]$.

When considering the relevant vertices, i.e. those with $2m + j \leq
4$, for $\Lambda \in [\Lambda_R,\Lambda_0]$, we apply momentum derivatives to
both sides of \exxxxi, put
$\Lambda$ equal to $\Lambda'$, integrate
with respect to $\Lambda'$ from $\Lambda$ down to $\Lambda_R$,
 differentiate with respect to $\Lambda_0$ for
fixed $\Lambda$ and ${\lambda_i}$, and take bounds to obtain
\eqn\elxvi{\eqalign
{\Biggr\vert \biggl(& {\partial \over \partial \Lambda_0}
\partial^j_p V^r_{2m}(p_i;\Lambda,\Lambda_0,\lambda_i)\vert_{p_i=P_i}
 \biggr)_{{\!\Lambda,\lambda_i}} \Biggr\vert
\leq \quad \Biggl \vert  \biggl( {\partial \over
\partial \Lambda_0} \partial^j_p
V^r_{2m}(p_i;\Lambda_R,\Lambda_0,\lambda_i)\vert_{p_i=P_i}
\biggr)_{{\!\Lambda_R,\lambda_i}} \Biggr \vert\cr
&\hskip -0.1in + \int^{\Lambda}_{\Lambda_R} d \Lambda'  \Bigg[\Lambda'
\Biggr \Vert \biggl( {\partial \over \partial \Lambda_0} \partial^j_p
V^r_{2m+2}(\Lambda',\Lambda_0,\lambda_i)
\biggr)_{{\!\Lambda',\lambda_i}} \Biggr\Vert_{\Lambda'}\cr
& \!\! +\sum^{m}_{l=1} \sum^{r-1}_{s=1} \sum_{j_i; j_1 +
j_2 + j_3 =j} \hskip -0.05in(\Lambda')^{-3-j_1} \Biggl \Vert \Biggl( {\partial
\over
\partial \Lambda_0} \partial^{j_2}_p V^s_{2l}(\Lambda',\Lambda_0,\lambda_i)
\biggr)_{{\!\Lambda',\lambda_i}}
\Biggr\Vert_{\Lambda'} \cdot \Biggl \Vert \partial^{j_3}_p
V^{r-s}_{2m+ 2 -2l}(\Lambda',\Lambda_0,\lambda_i)
\Biggr\Vert_{\Lambda'}\Bigg].\cr}}
For $\Lambda \in [0,\Lambda_R]$, we apply momentum derivatives to both sides of
\exxxxi, put
$\Lambda$ equal to $\Lambda'$, integrate
with respect to $\Lambda'$ from $\Lambda$ down to $0$, differentiate with
respect to $\Lambda_0$ for
fixed $\Lambda$ and ${\lambda_i}$, and take bounds with respect to
$\Lambda_R$ to obtain
\eqn\eslxvi{\eqalign
{\Biggr\Vert \biggl( {\partial \over \partial \Lambda_0}
&\partial^j_p V^r_{2m}(\Lambda,\Lambda_0,\lambda_i)
 \biggr)_{{\!\Lambda,\lambda_i}} \Biggr\Vert_{\Lambda_R}
\leq \quad \Biggl \Vert \biggl( {\partial \over
\partial \Lambda_0} \partial^j_p
V^r_{2m}(0,\Lambda_0,\lambda_i)
\biggr)_{{\lambda_i}} \Biggr \Vert_{\Lambda_R}\cr
&\hskip -0.3in + \int^{\Lambda}_{0} d \Lambda'  \Bigg[\Lambda_R
\Biggr \Vert \biggl( {\partial \over \partial \Lambda_0} \partial^j_p
V^r_{2m+2}(\Lambda',\Lambda_0,\lambda_i)
\biggr)_{{\!\Lambda',\lambda_i}} \Biggr \Vert_{\Lambda_R}\cr
& \hskip -0.28in +\sum^{m}_{l=1} \sum^{r-1}_{s=1} \sum_{j_i; j_1 +
j_2 + j_3 =j} \hskip -0.05in \Lambda_R^{-3-j_1} \Biggl \Vert \Biggl( {\partial
\over
\partial \Lambda_0} \partial^{j_2}_p V^s_{2l}(\Lambda',\Lambda_0,\lambda_i)
\biggr)_{{\!\Lambda',\lambda_i}}
\Biggr \Vert_{\Lambda_R} \cdot \!\Biggl \Vert \partial^{j_3}_p
V^{r-s}_{2m+ 2 -2l}(\Lambda',\Lambda_0,\lambda_i) \Biggr
\Vert_{\Lambda_R}\Biggl].\cr}}

These equations, along with
the boundary conditions \exxxvi\ and \exxxv\ on the flow equations, then give
the following bounds;
\medskip
\vbox{\noindent{\it Lemma 2}

\noindent (i) For all $\Lambda \in [\Lambda_R, \Lambda_0]$,
\eqn\elxvii{\Biggl\Vert \biggl( {\partial \over \partial \Lambda_0}
\partial^j_p
V^r_{2m}(\Lambda,\Lambda_0,\lambda_i) \biggr)_{{\!\Lambda,\lambda_i}}
\Biggr\Vert_{\Lambda} \leq
\Lambda_0^{-2} \Lambda ^{5-2m-j} P \log \biggl( {\Lambda_0 \over
\Lambda_R} \biggr).}

\noindent (ii) For all $\Lambda \in [0, \Lambda_R]$,
\eqn\eslxvii{ \Biggl\Vert \biggl( {\partial \over \partial \Lambda_0}
\partial^j_p
V^r_{2m}(\Lambda,\Lambda_0,\lambda_i) \biggr)_{{\!\Lambda,\lambda_i}}
\Biggr\Vert_{\Lambda_R} \leq
\Lambda_0^{-2} \Lambda_R^{5-2m-j} P \log \biggl( {\Lambda_0 \over
\Lambda_R} \biggr).}}
\medskip

The proof again proceeds by induction, the induction scheme being
exactly the same as the proof of lemma~1. Again we assume that the
lemma is true up to order $r-1$ in the expansion coefficient $g$, and
that at order $r$ in $g$ it is true down to $m+1$.

\medskip

a) We first consider the irrelevant vertices for $\Lambda\in
[\Lambda_R,\Lambda_0]$. Using $\Vert
\partial^j_p V^r_{2m}(\Lambda) \VertL \leq \Lambda^{-j} P \log
(\Lambda_0/\Lambda_R)$ (which is a weaker version of \exxxxiix), the
boundary conditions \exxxv, and the equations
\elxv, we immediately find
\eqn\elxiix{ \Biggl\Vert \biggl( {\partial \over
\partial \Lambda_0} \partial^j_p V^r_{2m}(\Lambda,\Lambda_0,\lambda_i)
\biggr)_{{\!\Lambda,\lambda_i}} \Biggr\Vert_{\Lambda}
\leq\cases{[\Lambda_0^{3-2m-j} + \Lambda_0^{-2} \Lambda^{5-2m-j}]
P\log\left({\Lambda_0 \over \Lambda_R}\right)
& if $2m + j> 5$, \cr
\Lambda_0^{-2}\log (\Lambda_0/\Lambda_R)
P\log \left( {\Lambda_0 \over \Lambda_R} \right)
& if $2m +j =5$.\cr}}
It is easy to see that both of these cases are consistent with \elxvii.

b) In order to verify lemma~2 for the irrelevant vertices we use \elxiix\
at $\Lambda=\Lambda_R$ to write
\eqn\esd{ \Biggl\Vert \biggl( {\partial \over \partial \Lambda_0}
\partial^j_p
V^r_{2m}(\Lambda_R,\Lambda_0,\lambda_i) \biggr)_{{\!\Lambda_R,\lambda_i}}
\Biggr\Vert_{\Lambda_R} \leq
\Lambda_0^{-2} \Lambda_R^{(4-2m-j+1)} P \log \biggl( {\Lambda_0 \over
\Lambda_R} \biggr), }
and use this bound along with lemma~2 and \eslxv\ in order to obtain
our desired result.

c) The proof for the relevant vertices is also very similar to that
for lemma~1.
Since the renormalization conditions on the relevant couplings
\exxxvi, are $\Lambda_0$-independent we know that
$( {\partial \over \partial \Lambda_0}
V^r_4(P_i;0,\Lambda_0,\lambda_i))_{{\lambda_i}}
=0$. Using Taylor's formula and the bounds already verified for the
$\Lambda_0$ derivatives of the irrelevant vertices we can then verify
(ii) for $\Vert( {\partial \over \partial \Lambda_0}
V^r_4(0,\Lambda_0,\lambda_i))_{{\lambda_i}}\VertR
$. Using \eslxvi\ we then verify (ii) for all $\Lambda
\in[0,\Lambda_R]$. The verification of (ii) for the
$\Lambda_0$--derivatives of the other relevant
vertices proceeds in the same manner (remembering the complication for the
part of $\partial^2_p V^r_2(0)$ proportional to $p_{\mu}p_{\nu}$ which is
solved using Taylor's formula about zero momentum).  Once
again we simply have to remember to work downwards in the number of
momentum derivatives, as for the verification of lemma~1.

d) Using (ii) at $\Lambda= \Lambda_R$ we can write
\eqn\esc{\Biggl\vert \biggl( {\partial \over \partial \Lambda_0}
\partial^j_p
V^r_{4}(P_i;\Lambda_R,\Lambda_0,\lambda_i)
\biggr)_{{\!\Lambda_R,\lambda_i}}
\Biggr\vert \leq
\Lambda_0^{-2} \Lambda_R P \log \biggl( {\Lambda_0 \over
\Lambda_R} \biggr),}
and thus obtain a boundary condition for the $\Lambda_0$ derivative of
the dimensionless vertex at $\Lambda_R$.
We then use \elxvi\ in order to obtain a bound on
$\vert( {\partial \over \partial\Lambda_0}
V^r_4(P_i;\Lambda,\Lambda_0,\lambda_i))_{{\!\Lambda,\lambda_i}}\vert$.
Using the bounds already obtained on the norms of the first momentum
derivatives
of $( {\partial \over \partial \Lambda_0}
V^r_4(\Lambda,\Lambda_0,\lambda_i))_{{\!\Lambda,\lambda_i}}$ and the Taylor
formula we prove that $\Vert ( {\partial \over \partial \Lambda_0}
\partial^j_p
V^r_{4}(\Lambda,\Lambda_0,\lambda_i) )_{{\!\Lambda,\lambda_i}}
\VertL$
is consistent with lemma (ii). As in part d) in the proof of lemma~1,
the procedure for the
$\Lambda_0$--derivatives of the other relevant vertices is the same. In
fact we see that the structure of the argument for the proof of
convergence is identical to that for boundedness. This same
type of argument will be used a number of further times in this article.

\medskip

We have therefore shown that the
lemma is true for all $m$ at order $r$ in the expansion coefficient.
At next order in $g$ it is true for large enough $m$, and again the
proof by induction goes through and the lemma is true at order $r +1$
in $g$. But the lemma is trivially satisfied at zeroth order in $g$,
and thus by induction, lemma~2 is true for all $m$ and $r$.\blackbox
\medskip
In particular, at $\Lambda = 0$,
\eqn\elxxiii{\Biggl\Vert \biggl( {\partial \over \partial \Lambda_0}
\partial^j_p V^r_{2m}(0,\Lambda_0,\lambda_i)
\biggr)_{{\lambda_i}} \Biggr\Vert_{\Lambda_R}
\leq \biggl({\Lambda_R \over \Lambda_0}\biggl)^2 \Lambda_R^{3-2m-j} P \log
\biggl( {\Lambda_0 \over \Lambda_R} \biggr).}
So if $\Lambda_R \ll\Lambda_0$, once we fix the relevant couplings at
$\Lambda = 0$, the dependence of the Green's functions at scale $\Lambda_R$ on
the high energy cut-off $\Lambda_0$ is very small, i.e. the Green's
functions are
determined almost completely by the low energy renormalization
conditions and only change very slowly as we change $\Lambda_0$.

Remembering the conventional meaning of renormalizability,
i.e. convergence, we wish to
show that the Green's
functions, have a finite limit as $\Lambda_0 \rightarrow \infty$.
To do this we use \elxxiii\ to show that
\eqnn\elxxiv
$$
\eqalignno{ \Vert \partial^j_p V^r_{2m}(0,\Lambda_0,\lambda_i) -
\partial^j_p V^r_{2m}(0,\Lambda_0',\lambda_i) \VertR
& = \Biggl\Vert \int^{\Lambda_0'}_{\Lambda_0} d\Lambda_0'' {\partial
\over \partial \Lambda_0''} \partial^j_p
V^r_{2m}(0,\Lambda_0'',\lambda_i) \Biggr\Vert_{\Lambda_R} \cr
&\leq  \Lambda_R^{4-2m-j}\Biggl[ \biggl( {\Lambda_R \over \Lambda_0} \biggr) P
\log
\biggl( {\Lambda_0 \over \Lambda_R}\biggr) - \biggl( {\Lambda_R \over
\Lambda'_0} \biggr) P \log \biggl( { \Lambda'_0 \over \Lambda_R} \biggr)
\Biggr] \cr
& \leq  \Lambda_R^{4-2m-j}\biggl( {\Lambda_R \over \Lambda_0}\biggr) P \log
\biggl(
{\Lambda_0 \over \Lambda_R} \biggr), &\elxxiv \cr}
$$
for $\Lambda'_0 > \Lambda_0$, and in particular for $\Lambda'_0
\rightarrow \infty$.
Obviously, stripping the external lines of their damping factors of
$[K_{\Lambda_R}(p_i)]^{1/4}$ will not introduce any $\Lambda_0$
dependence for external momenta $p^2_i\ll\Lambda_0^2$.
So remembering the relationship between the dimensionful vertices
defined at $\Lambda=0$ and the amputated connected Green's functions we
see that
\eqn\elxxvder{\Vert\partial_p^j
\tilde{G}_{2m}^{c,r}(p_1,\ldots,p_{2m};\Lambda_0,\lambda_i)
-\partial_p^j\tilde{G}_{2m}^{c,r}(p_1,\ldots,p_{2m};\Lambda_0',\lambda_i)\VertR
\leq
\Lambda_R^{4-2m-j} \biggl({\Lambda_R \over \Lambda_0}\biggr)
P \log \biggl( {\Lambda_0 \over \Lambda_R} \biggr),}
and at any finite order in perturbation theory any amputated Green's
function with  finite external
momenta tens to a finite limit as $\Lambda_0 \rightarrow \infty$. Therefore the
Euclidean theory is perturbatively renormalizable in the conventional sense.

\subsec{The Analytic S--Matrix.}

So far we have only been considering the Euclidean formulation of the
theory, obtaining bounds on Euclidean Green's functions (sometimes
referred to as Schwinger functions) $G_n^c(p_i)$ which are functions
of the real Euclidean momenta, $p_i$ such that $0<p_i^2<\infty$ and
$\sum_{i=1}^n p_i =0$. Because they are rotationally invariant, the
Green's functions are actually functions of the invariants $p_i\cdot p_j$;
these in turn are not all independent
since $\sum_ip_i=0$, and since any five or more four vectors are always
linearly related a Green's function with $n\geq 4$ external lines
depends in general on $10+4(n-4)-n=4n-10$ independent invariants.
Before Euclidean Green's functions can be
used to construct S--matrix elements, they must be analytically
continued to the boundary of the physical (timelike) region for which
$-\infty<p_i^2<0$ (where they are sometimes called Wightman
functions). We may then put
the external momenta on--shell, setting (taking $m$ as the
physical mass of the particle) $p_i^2 = -m^2 + i\epsilon$.\footsy{We use the
notation $p^2 = \vec p^2 - p_0^2$, where $p_0=ip_4$ is the energy in Minkowski
space and $\vec p$ is the spatial momentum three-vector.} The
on--shell Green's functions then depend only on the invariants
$p_i\cdot p_j$, $i\neq j$ (of which $3n-10$ are truly independent).
S--matrix elements constructed \rlsz\ from these on--shell Green's functions
will then be manifestly Lorentz invariant.

We thus consider in this section the Green's function
$\tilde{G}_n^c(z_{ij})$ where the complex numbers $z_{ij}$ are the set
of invariant momenta $\{z_{ij}\equiv (p_i+p_j)^2;
i=1,\ldots,n, j=1,\ldots,i\}$; $z_{ii}$ parameterize the
off--shellness of the external lines, while $z_{ij}$ with $i\neq j$
may be thought of as generalized Mandelstam invariants. In the
Euclidean formulation $z_{ij}$ all lie on the positive real axis; we
wish to analytically continue $\tilde{G}^c_n$ into a bounded
region of the complex
$z_{ij}$ planes, and in particular to the the poles and cuts (which
conventionally lie on the negative real axis when $i=j$, and
on both positive and negative real axes when $i\neq j$).\foot{Of
course the analytic continuation of a function of several complex
variables is a difficult subject, which we have no wish to go into
here. A useful pedagogical account of the famous `edge
of the wedge' theorem and other useful results may be found
in \ref\rsw{R.~F.~ Streater and A.~S.~ Wightman,
``PCT, Spin and Statistics, and All That'', W. A. Benjamin, Inc, New
York, 1964}.} Since
the Euclidean Green's functions are real,
$\tilde{G}_n^c(z_{ij})^{*}=\tilde{G}_n^c(z_{ij}^{*})$.

It is important to realize that it is only necessary to continue
the external momentum invariants $z_{ij}$, and that these
need only take values in some bounded region of the complex plane. The
loop momenta which are integrated over in each Feynman diagram
contributing to $G$ remain in Euclidean space. This is not merely a
matter of convenience; for the effective theory with finite cut--off the
Wick rotation of loop momenta is not well defined. This is because the analytic
cut--off function $K(z)$ necessarily has an essential singularity at
infinity, which means in turn that even though it vanishes
exponentially as $z$ goes to infinity along the positive real axis,
there must (because of the Casorati--Weierstrass theorem) be other
directions in the complex plane in which it grows arbitrarily large
as infinity is approached.\foot{Indeed the simple order one function \exxx\
$K$ {\it grows} exponentially on the negative real axis.} The
contribution to the loop integral from the circle at infinity can then
no longer be ignored; if it is the resulting Green's function is no longer
Lorentz invariant.

The position of the singularities of $G$ in the physical region is
given by the Landau rules\rlandau; the conventional derivation, which relies on
the continuation of the contribution of a typical Feynman diagram,
goes through without modification since both the regulating function
$K_{\Lambda_0}$ and the bare vertex functions $\{V_{2m}^r(\Lambda_0)\}$
(with the renormalization conditions \exxxv) are by
construction regular functions of the momentum invariants $z_{ij}$,
and thus introduce no new singularities. The position of all
singularities may thus be determined in terms of the single parameter
$m$. The Green's function ${G}_n^c(z_{ij})$ may then be
continued into the
complex plane by means of a Taylor expansion about some point
$z_{ij}^0$ on the positive real axis; the radius of convergence of
this series will be the distance to the nearest singularity. As this
singularity is a simple pole (in general in several or even all of the
$z_{ij}$) we may continue beyond it by considering instead the Taylor
expansion of the amputated Green's functions
$\prod_{i\neq j}(z_{ij}+m^2-i\epsilon)\tilde{G}_n^c(z_{ij})$.
In this way we also determine the residues at the poles. To continue
around the branch cuts, we may use several Taylor series which
converge in overlapping regions.

Now we may extend the results in \S 2.2 and \S 2.3 to a
consideration of the $\Lambda_0$-dependence of the on-shell
Green's functions, and thus of S--matrix elements.
Consider for example the bound \elxxvder; this may be rewritten to give
\eqn\elxxvderiv{\Vert\partial_z^n
\tilde{G}_{2m}^{c,r}(z_{ij};\Lambda_0,\lambda_i)
-\partial_z^n\tilde{G}_{2m}^{c,r}(z_{ij};\Lambda_0',\lambda_i)\VertL \leq
\Lambda_R^{4-2m-2n} \biggl({\Lambda_R \over \Lambda_0}\biggr)
P \log \biggl( {\Lambda_0 \over \Lambda_R} \biggr).}
When we Taylor expand the Green's function $\tilde{G}_n^c(z_{ij})$
about the Euclidean point $z_{ij}^0$ each of the coefficients in the
expansion is independent of $\Lambda_0$ up to terms of order
$(\Lambda_R/\Lambda_0)$. It follows that
$\tilde{G}_n^c(z_{ij})$ is independent
of $\Lambda_0$ up to order $(\Lambda_R/\Lambda_0)$ everywhere within
the circle of convergence of it's Taylor expansion, and as explained
above, this tells us that the residues of the poles are
independent of $\Lambda_0$.

It is even possible to set on-shell renormalization conditions, by
taking $P_i$ in \exxxvi\ on the boundary of the physical region. The
conventional choice is the symmetrical point $P_i^2=-m^2$,
$(P_i+P_j)^2=-\smallfrac{4}{3}m^2$, which we denote
by $Z_{ij}$. The vertex $V_4(Z_{ij})$ may be expressed as a
Taylor expansion about the Euclidean point $Z_{ij}^0$ where the
momenta satisfy the same conditions as those at which the
original renormalization condition \exxxvi\ was set. Now if we assume that
either lemmas 1 or 2 are true up to order $r-1$ in
$g$  then they will be true for all the irrelevant vertices at
order $r$ in $g$. Thus, all the terms beyond
the first in the Taylor expansion will be bounded. If we then
set the on--shell renormalization condition (which determines
$V_4^r(Z_{ij})$) consistently
with the lemma, then the first term in the Taylor series (which
defines an effective off--shell renormalization condition through
$V_4^r(Z_{ij}^0)$) is determined and may be expressed as a
convergent sum of an infinite number of terms all of which satisfy the lemma.
We can then use this effective off--shell renormalization condition to
prove the lemmas for the four point vertex in exactly the same way as we did
when
we had set the renormalization condition directly at the Euclidean point
$Z_{ij}^0$. Using a similar procedure for the
two--point vertex and its first two momentum derivatives, remembering
to work downwards in number of derivatives, then completes all the
remaining steps
in the proofs. Thus, we are able to prove lemmas 1 and 2
in their entirity while using on--shell
renormalization conditions.

Now that we have shown that the perturbative renormalizability of the
Green's functions on the boundary of the physical region (and thus of
the S--matrix elements themselves) follows directly from that of the
Euclidean Green's functions, we may consider the unitarity and
causality of the perturbative S--matrix. To do this we must first derive
the Cutkovsky rule \rui\ for determining the discontinuity of
the Green's functions across the cuts. This is nontrivial since, due to
the essential singularity in the propagator at infinity, we can
no longer perform a Wick rotation, so the usual position space
techniques which rely on the existence of a Lehmann representation for
the propagator (see for example \refs{\rvel,\rdiagrammar,\rxx}) are no
longer available.
However, since the Cutkovsky rule involves only the behaviour of the
propagator near the pole, stating in fact that each cut propagator is
replaced by its on--shell imaginary part $2\pi
i\Theta(p_0)\delta(p^2+m^2)$, it should be possible to prove it by
considering the behaviour of the singularities in a bounded region of
the complex plane; such a proof for normal singularities may be found in
\ref\refimov{G.~V.~Efimov, {\it Sov.~Jour.~Nucl.~Phys.~}\vyp{2}{1967}{309}.}.
A more general (and in our opinion intuitive) argument will be
presented in \S 5.3 below.

Once we have the Cutkovsky rule, Veltman's cutting formula \rvel\
follows immediately; applying it to $iT\equiv S-1$ shows that
$T-T^\dagger=iTT^\dagger$, and thus that $S$ is perturbatively
unitary. Indeed any theory whose propagator has no singularities in
any bounded region of the complex plane, save a single simple pole
with unit residue, and whose bare vertices are regular functions of momenta,
and are real when the momenta are Euclidean, can be shown to be
perturbatively unitary in this way.

A formal criterion for a causal S-matrix was established in \rbs, and is
known as the Bogoliubov condition. It is expressed by the equation
\eqn\ebc{{\delta \over \delta \phi_{\rm in}(x_1)} \biggl( {\delta
S^{\dag} \over \delta \phi_{\rm in}(x_2)}S \biggr) = 0 \qquad x_1 \leq
x_2,}
where $\phi_{\rm in}(x)$ represents an in--field satisfying
the linearized classical field equation, and $x_1 \leq x_2$ means that
$x_1$ and $x_2$ are spacelike separated or that $x_1^0 < x_2^0$.
A similar condition holds with $\phi_{\rm out}(x)$ substituted for
$\phi_{\rm in}(x)$. A much stronger condition\foot{Which may be shown
\rbs\ to be equivalent to the assumption of local commutativity in the
conventional operator formulation of quantum field theory (see, for
example, \rsw).} in terms of the interpolating
fields or localized sources may be found in \rbs, but will not be used
here since it is nonphysical, except in the sense in which it reduces
asymptotically to \ebc; strictly speaking the only observable
quantities in a relativistic quantum theory are the momenta and
internal quantum numbers of free particle states \ref\rpl{L.~Landau and
R.~Peierls,
\ZP\vyp{69}{1931}{56}\semi L.~D.~Landau in ``Theoretical Physics in
the Twentieth Century, a Memorial Volume to Wolfgang Pauli'', ed.
M.~Fierz and V.~F.~Weisskopf (Interscience, New York 1960).}.
Since the space--time positions $x_i$ of the incoming particles are
not themselves physical observables \ref\rdirac{P.~A.~M.~Dirac,
``Principles of Quantum Mechanics'' (O.U.P., 1930).}, in \ebc\ $x_i$ must be
understood to mean the positions of their wave packets; the condition
\ebc\ is thus only precise up to the minimal uncertainty $\Delta x_i
>E_i\inv$, where $E_i$ is the particle's energy \rpl.

It was shown in \rvel\ that \ebc\ is a simply a special case of the cutting
formula, from which it follows that we are
guaranteed to have causal perturbative S-matrix elements.
Therefore we can define a renormalizable
(in the conventional sense), unitarity and causal perturbative S-matrix. This
is
essentially because our regularization maintains the analytic
structure of the classical theory as we evolve. Because of the form of the
regularized propagator, the bare action of the classical theory
\classact\ is necessarily quasi--local when the cut--off $\Lambda_0$
is finite. Furthermore, although the interaction $S_{\rm int}[\phi;\Lambda]$ is
local at the regularization scale $\Lambda_0$ (because of the renormalization
conditions \exxxv), at scales $0<\Lambda<\Lambda_0$ it too becomes
quasi--local. It is interesting to note that $S_{\rm int}[\phi;\Lambda]$
becomes truly non--local (by which we mean that the vertex
functions $V_{2m}^r(\Lambda)$ are no longer regular) only at the point
$\Lambda=0$. This is because the $\Lambda$--derivative of the regulated
propagator, $-\Lambda^{-3}K'_\Lambda(p)$, is regular (see
\exxix,\kdef), so when the vertices are also regular
the right hand side of the evolution equation \exxxxi\ consists of
sums of products of regular functions or convergent integrals of such
(the first term), and is thus itself regular. Indeed the formal
solution \eix\ also shows that when $\Lambda$ is positive the
evolution preserves the regularity of the vertex functions;
$P_\Lambda-P_{\Lambda_0}$ is regular for $\Lambda,\Lambda_0>0$.
By contrast the solution \ewpert\ for the generating functional for
amputated connected Green's functions shows that at $\Lambda=0$
the Green's functions (and thus the vertex functions $V_{2m}^r(0)$,
because of the identification \exxiix) are no longer regular, due to
the pole in the propagator $P_\Lambda$.

This has several interesting consequences. The regularity of the vertex
functions for all positive $\Lambda$ means that their Taylor
expansions have an infinite radius of convergence; the expansion
\actgen\ will then always converge provided that the field $\phi(x)$
is the Fourier transform of a regular function. Furthermore, it
suggests that the local renormalization conditions we imposed at
$\Lambda_0$ were unnecessarily restrictive; we could instead consider
the renormalization conditions as being fixed at a nearby scale
$\alpha\Lambda_0$, where $\alpha$ is a number close to one, and the
interaction therefore being only quasi--local, though still natural. Indeed we
will
prove in \S 3.1 below that the effective theory constructed from
a general quasi--local bare action \classact\ is indeed bounded and
convergent in just the same sense as the conventional one considered
above.

This sudden change in the analytic structure of the vertex functions
at vanishing $\Lambda$ is due to the essential singularity in the regulating
function \kdef; the essential singularity in $K_\Lambda(p)$ necessary
for the convergence of the Feynman integrals is necessarily mirrored by
one of the same type at $\Lambda=0$. The value of the regulating
function in the limit depends in fact on the order $\sigma$ of the essential
singularity\foot{The simplest example of a regulating function with
essential singularity of order $\sigma$ is $K(x)=e^{-x^\sigma}$;
\exxx\ is then the special case $\sigma=1$.}:
\eqn\ereglim{\lim_{\Lambda\rightarrow 0} K_\Lambda(p)=
\cases{0,&if $\Real\, p^2>-m^2$;\cr
       1,&if $p^2=-m^2$;\cr
       0,&if $\Real\, p^2<-m^2$ and $\sigma$ is even;\cr
       \infty,&if $\Real\, p^2<-m^2$ and $\sigma$ is odd.\cr}}
When $\Real\, p^2=-m^2$ but $\Imag\, p^2\neq 0$ the limit is not well defined.
The only regions that matter are the Euclidean axis and the position
of the pole; $K_0(p)=0$ on the Euclidean axis guarantees that the
Euclidean vertex functions tend uniformly to the amputated connected
Green's functions (as expressed by \exxiix), while $K_0(-m^2)=1$
guarantees that the position of the emerging singularities, and the
discontinuities across the cuts, are those given by the Landau and
Cutkovsky rules. Indeed the Cutkovsky rule and
Veltman's cutting formula may be derived using the effective action at
any positive scale $\Lambda$, not just the bare action at $\Lambda_0$,
since all that is necessary is that the vertex functions are regular
and, on the boundary of the physical region, real. If we take
$\Lambda$ infinitesimal, so that the Euclidean momenta in the loops
are negligible compared with the momenta in the external lines, we see
that the singularities arise at just those kinematical points where
internal lines have gone on shell in just such a way that the event
represented by the diagram could occur classically\ref\rcn{
S.~Coleman and R.~E.~Norton, \NC\vyp{38}{1965}{438}.} in a
theory with the classical (and regular) action $S[\phi,0^+]$. The
causality of the perturbative S--matrix is therefore a direct consequence of
the
causality of the classical theory with propagator $P_\infty(p)$.

However, the classical field equations derived from the quasi--local classical
action $S[\phi,\Lambda]$ will involve terms with more than two time
derivatives, which means that in general they will admit additional
solutions which lead to instabilities\rpu.
If the asymptotic states are unstable, then the Fock spaces of in--
and out--particle states no longer
exist (though a perturbative S--matrix may be still be constructed about the
conventional solution). In the conventional formulation of the theory
this problem is usually ignored; after all, the extra solutions
disappear in the strict infinite cut--off limit. However in an effective
theory we wish to keep the regularization scale finite, and to choose
(with as much freedom as possible) non-zero irrelevant coupling
constants in the bare action. In the following section we will discuss
the consequences of the presence of such couplings as far a
renormalizability is concerned, and in \S 5 below we will show how we
can use some of this freedom to eliminate unstable solutions in
order to construct an effective theory with a unitary causal
S--matrix.

\newsec{Perturbative Renormalizability of Effective Scalar Theory.}

The proof of conventional renormalizability in the previous section is
quite sufficient if one is only interested in field theories in the
infinite cut--off limit. However, for an effective theory we wish to keep the
regularization scale finite, and thus consider an action of the general form
\actgen\ at the regularization scale. This means in particular that we must
consider more general
renormalization conditions on the irrelevant vertices than those
adopted for the conventional theory (namely \exxxv ). We will show
below that we can indeed
add irrelevant operators to the bare Lagrangian without substantially
changing the Green's functions at scales far below the cut--off
provided that these operators have `natural' dimensionless
coupling constants $c^{(m,j,s_i)}_{\mu_1\cdots\mu_{2j}}$
of order unity or less. Then we discuss how the accuracy of the
theory for processes of a given energy may be systematically
improved by setting
low energy renormalization conditions on some of the canonically
irrelevant couplings (which thereby become `physically relevant'), and
similarly what
happens to the theory when we consider energies approaching
$\Lambda_0$.

\subsec{Universality.}

According to lemma~2 if we define the irrelevant couplings at
$\Lambda_0$ to be zero, then once we fix the renormalization
conditions on the relevant coupling constants $\lambda_i$ at $\Lambda=0$,
all the other couplings at $\Lambda_R$ depend only weakly on $\Lambda_0$, and
converge to a limit as $\Lambda_0 \rightarrow \infty$. This is
depicted schematically in \fig\flowcon{The flow diagram for a
conventional theory, the notation being the same as for
\flow. Each trajectory now corresponds to a different value of
$\Lambda_0$, and crosses the axis $\eta =0$ when $\lambda =
\lambda(\Lambda_0)$, the bare coupling. The uppermost trajectory is
the limiting one as $\Lambda_0\to\infty$.}. However, as explained in the
introduction, in order to have a full proof of
renormalizability for an effective theory it is also necessary to
show that once we have fixed the
low energy relevant couplings we can vary both the value of
$\Lambda_0$, and the boundary conditions on the irrelevant couplings
at $\Lambda_0$ still producing only changes in the values of the couplings
at $\Lambda_R$ which are suppressed by powers of
$(\Lambda_R/\Lambda_0$).

The style of our proof is similar to that in \rvii, but avoids
some unnecessary assumptions made there. We will begin by
defining two different field theories, both with the same field
content, global symmetries and
propagators, but with different interactions. The first theory is
the one already considered in \S 2; the bare interaction is local,
$S_{\rm int}[\phi,\Lambda_0]$ being defined by the boundary conditions
\exxxv\ and \exxxvi. The second theory has a quasi--local interaction which we
call $\bar S_{\rm int}[\phi,\Lambda]$. This has exactly the same boundary
conditions \exxxvi\ on the relevant couplings as $S_{\rm int}[\phi,\Lambda]$,
but has different boundary conditions on the irrelevant couplings. The
bare interaction $\bar S_{\rm int}[\phi,\Lambda_0]$ of the second theory
is thus defined by an expansion of the form \exxxi\ but with
\eqn\elxxvi{\eqalign{\bar V^r_2(p,-p;\Lambda_0) &=
\bar\lambda^r_1(\Lambda_0) \Lambda_0^2 +
p^2 \bar\lambda^r_2(\Lambda_0) + \Lambda_0^2 \eta^r_2(p,-p;\Lambda_0),\cr
\bar V^r_4(p_1,p_2,p_3,p_4;\Lambda_0) &=
\bar\lambda^r_3(\Lambda_0) + \eta^r_4(p_1,p_2,p_3,p_4;\Lambda_0),\cr
\bar V^r_{2m}(p_1.....p_{2m};\Lambda_0) &=
\Lambda_0^{4-2m}\eta^r_{2m}(p_1.....p_{2m};\Lambda_0),\qquad m>2.\cr}}
The bare relevant couplings of the second theory,
$\bar\lambda^r_i(\Lambda_0), i=1,2,3$ are not specified. The
irrelevant bare couplings $\eta^r_{2m}$ are arbitrary,
but specific, real functions of Euclidean momenta, which
have to satisfy a number of conditions
(which correspond to the conditions (1)--(3) on the
regulating function $K(z)$ described in \S 1.1):
\item{(1)} $\eta^r_{2m}(p_1.....p_{2m};\Lambda_0)$ do not grow more
quickly than $[K_{\Lambda_0}(p_i)]^{-1/4}$ for large Euclidean momenta
$p_i^2$; this ensures that all Feynman diagrams are finite;\foot{A
weaker condition than this will be given at the end of \S 4.3 below.}
\item{(2)} $\eta^r_{2m}(z_{ij})$ are regular
functions of the momentum invariants $z_{ij}$; this guarantees that
no new singularities are introduced;\foot{It also means that since their
Taylor expansions in powers of $z_{ij}$ have an arbitrarily large radius of
convergence, $S_{\rm int}[\phi,\Lambda_0]$ may always be expanded as an
infinite sum of local operators of the form \actgen. This will prove
invaluable when we consider the removal of redundant operators in
\S 4.3 below.} according to the Schwarz reflection principle
$\eta^r_{2m}(z_{ij}^*)=\eta^r_{2m}(z_{ij})^*$;
\item{(3)}
\eqn\elxxix{ \Vert \partial^j_p \eta^r_{2m}(\Lambda_0) \VertL \leq
\Lambda_0^{-j} P \log \biggl( {\Lambda_0 \over \Lambda_R} \biggr)}
for $2m + j > 4$, which is simply the requirement that the irrelevant
couplings at $\Lambda_0$ are natural \rxx, or more precisely that
none of them are unnaturally large.

\noindent We also insist that the vertices are ordered such that
$\eta^r_{2m}(\Lambda_0)=0$ for $m>r+1$ (and remember that
$\eta_2^0(\Lambda_0)$ is defined to be zero),
so that a similar perturbative induction procedure to that used to prove lemmas
1 and 2 can be used.\footsy{We choose this ordering
requirement simply to agree with the relationship between $r$ and $m$
used in \S 2; it is only really necessary to specify
that $\eta^r_{2m}(\Lambda_0)$
vanishes at each $r$ for large enough $m$.} If we were to make a
trivial field redefinition $\phi\rightarrow g^{-\half}\phi$ at
$\Lambda_0$ and extract an overall factor of $g\inv$ from the
action, $\eta^r_{2m}(\Lambda_0)$ would be nonvanishing for all $r\geq
0$; the perturbation series would then be just
the loop expansion.

Now that both theories are completely defined we can consider their
difference. We introduce the quantity
\eqn\elxxxs{U^r_{2m}(\Lambda) = V^r_{2m}(\Lambda)
-\bar V^r_{2m}(\Lambda).}
Working in the range $\Lambda \in [\Lambda_R,\Lambda_0]$, and subtracting the
flow equation for the $\bar V^r_{2m}(\Lambda)$ away
from that for the $V^r_{2m}(\Lambda)$ and taking norms, we
easily obtain
\eqn\elxxx{\eqalign{
\Biggl\Vert {\partial \over \partial \Lambda} &\Bigl(
\partial^j_p U^r_{2m}(\Lambda)\Bigr) \Biggr\Vert_{\Lambda} \leq
\Biggl[\Lambda \Vert \partial^j_p U^r_{2m+2}(\Lambda)\VertL\cr
&\hskip -0.2in +  \sum^{m}_{l=1} \sum^{r-1}_{s=1} \sum_{\{j_i;j_1 +
j_2 + j_3 = j\}} \hskip -0.05in \Lambda^{-3-j_1}\Vert \partial^{j_2}_{p}
U^s_{2l}(\Lambda)\VertL \! \cdot \! \biggl( \Vert \partial^{j_3}_p
U^{r-s}_{2m + 2 -2l}(\Lambda)\VertL + \Vert
\partial^{j_3}_p V^{r-s}_{2m+2 -2l}(\Lambda)\VertL\biggr)\Biggr].\cr}}
Similarly, working in the range $\Lambda \in [0,\Lambda_R]$, subtracting the
flow equation for the $\bar V^r_{2m}(\Lambda)$ away
from that for the $V^r_{2m}(\Lambda)$ and taking norms, we
also have
\eqn\eslxxx{\eqalign{
\Biggl\Vert {\partial \over \partial \Lambda} \Bigl(
&\partial^j_p U^{r}_{2m}(\Lambda)\Bigr) \Biggr\Vert_{\Lambda_R} \leq
\Biggl[\Lambda_R \Vert \partial^j_p U^{r}_{2m+2}(\Lambda)\VertR\cr
& \hskip -0.26in +\sum^{m}_{l=1} \sum^{r-1}_{s=1} \sum_{\{j_i;j_1 +
j_2 + j_3 = j\}}\hskip -0.05in \Lambda_R^{-3-j_1}\Vert \partial^{j_2}_{p}
U^{s}_{2l}(\Lambda)\VertR \! \cdot \! \biggl( \Vert \partial^{j_3}_p
U^{r-s}_{2m + 2 -2l}(\Lambda)\VertR + \Vert
\partial^{j_3}_p V^{r-s}_{2m+2 -2l}(\Lambda)\VertR\biggr)\Biggr].\cr}}

{}From the equality
\eqn\elxxxi{  \partial^j_p
U^r_{2m}(\Lambda) = \partial_p^j
U^r_{2m}(\Lambda_0)
+\int^{\Lambda_0}_{\Lambda} d \Lambda' {\partial \over \partial
\Lambda'} \biggl( \partial^j_p
U^r_{2m}(\Lambda')\biggr),}
where $\Lambda \in [\Lambda_R, \Lambda_0]$, we easily derive the inequality
\eqn\elxxxii{\big\Vert \partial^j_p U^r_{2m}(\Lambda)
\big\Vert_\Lambda
\leq \Vert \partial^j_p U^r_{2m}(\Lambda_0)\Vert_{\Lambda_0} +
\int^{\Lambda_0}_{\Lambda} d \Lambda' \Biggl\Vert {\partial \over
\partial\Lambda'} \biggl(  \partial^j_p
U^r_{2m}(\Lambda')\biggr)\Biggr\Vert_{\Lambda'}.}
For $\Lambda \in [0,\Lambda_R]$, we can show that similarly
\eqn\eslxxxii{\big\Vert  \partial^j_p U^{r}_{2m}(\Lambda)
\big\Vert_{\Lambda_R}
\leq \Vert \partial^j_p U^{r}_{2m}(\Lambda_R)\VertR +
\int^{\Lambda_R}_{\Lambda} d \Lambda' \Biggl\Vert {\partial \over
\partial\Lambda'} \biggl(  \partial^j_p
U^{r}_{2m}(\Lambda')\biggr)\Biggr\Vert_{\Lambda_R}.}

 In order to
obtain an equation useful for finding bounds on the difference between
the relevant coupling constants in the range $\Lambda \in
[\Lambda_R,\Lambda_0]$ it is necessary to integrate with
respect to $\Lambda'$ from $\Lambda$ down to $\Lambda_R$, put the momenta
equal to those at which the renormalization conditions on the relevant coupling
constants are set, and take bounds to obtain
\eqn\elxxxiii{\big\vert \partial^j_p
U^r_{2m}(\Lambda)\vert_{p_i=P_i}\big\vert
\leq \big\vert \partial^j_p
U^r_{2m}(\Lambda_R)\vert_{p_i=P_i}\big\vert
+ \int^{\Lambda}_{\Lambda_R} d\Lambda'
\biggl\Vert {\partial \over \partial \Lambda'} \partial^j_p
U^r_{2m}(\Lambda') \biggr\Vert_{\Lambda'}.}
For an equation useful for finding bounds on the difference between
the relevant vertices in the range $\Lambda \in [0,\Lambda_R]$ we integrate
with respect to
$\Lambda'$ from $\Lambda$ down to $0$ and take bounds with respect to
$\Lambda_R$ to obtain
\eqn\eslxxxiii{\Vert \partial^j_p
U^{r}_{2m}(\Lambda)\VertR
\leq \Vert  \partial^j_p
U^{r}_{2m}(0)\VertR
+ \int^{\Lambda}_{0} d\Lambda'
\biggl\Vert {\partial \over \partial \Lambda'} \partial^j_p
U^{r}_{2m}(\Lambda') \biggr\Vert_{\Lambda_R}.}

These four inequalities, together with the boundary
conditions on the flow equation, will now be shown to lead
to the following bounds on the difference
between the vertices in the two theories:
\medskip
\vbox{\noindent {\it Lemma 3}

\noindent (i) For all $\Lambda \in [\Lambda_R, \Lambda_0]$,
\eqn\elxxxiv{\Vert \partial^j_p U^r_{2m}(\Lambda)\VertL \leq {\Lambda
\over \Lambda_0} \Lambda^{4-2m-j} P \log \biggl( {\Lambda_0 \over
\Lambda_R} \biggr).}

\noindent (ii) For all $\Lambda \in [0, \Lambda_R]$,
\eqn\eslxxxiv{\Vert \partial^j_p U^{r}_{2m}(\Lambda)\VertR \leq {\Lambda_R
\over \Lambda_0} \Lambda_R^{4-2m -j} P \log \biggl( {\Lambda_0 \over
\Lambda_R} \biggr).}}
\medskip
Once again, the method of proof is the same induction scheme as that
used to prove lemma~1. We assume that the lemma is true to order $r-1$
in $g$, and down to vertices with $2m+2$ legs at order $r$, and then
proceed downwards in the number of legs, following the sequence of
steps a)--d).

a) For the irrelevant vertices we
use \elxxxii, \elxxx\ and the boundary conditions \elxxvi\ on the vertices at
$\Lambda_0$ to obtain
\eqn\eua{\big\Vert \partial^j_p U^r_{2m}(\Lambda)
\big\Vert_\Lambda
\leq  \Lambda_0^{4-2m-j} P \log \biggl(
{\Lambda_R \over \Lambda_0}\biggr) +
\int^{\Lambda_0}_{\Lambda} d \Lambda' {(\Lambda')^{4-2m-j}
\over \Lambda_0} P \log \biggl( {\Lambda_0 \over
\Lambda_R} \biggr),}
and remembering that $2m+j-4 >0$ for the irrelevant vertices we see
that (i) is true at order $r$
for vertices with $2m$ legs.

b) Using the bound on the dimensionful irrelevant
vertices at $\Lambda_R$ obtained from (i) along with \eslxxxii\
and \elxxx, we easily see that (ii) is also true at  order $r$.

c) When considering the relevant vertices, we have to find bounds on the
difference between the relevant vertices at $\Lambda=0$, before
\elxxxiii\ and \eslxxxiii\ are any use. Using the fact that both theories have
the same
renormalization conditions for the relevant couplings, and thus
$\partial^j_p U^r_{2m}(p_i\ldots;0)\vert_{p_i=P_i}=0$, we can then use the
bounds
already obtained on the irrelevant vertices along with Taylor's
formula to prove (ii) for $U^r_4(0)$. Using this bound and
equations \eslxxxiii\ and \eslxxx\ we easily prove that
\eqn\eub{\Vert U^r_4(\Lambda)\VertR \leq
\biggl({\Lambda_R\over\Lambda_0}\biggr)
P\log\biggl({\Lambda_0\over\Lambda_R}\biggr)}
for all $\Lambda \in [0,\Lambda_R]$. Similar considerations hold for
the other differences of relevant vertices.

d) Using these results we can now derive a bound on the difference
between relevant coupling constants at $\Lambda_R$:
\eqn\elxxxv{\big\vert U^r_{4}(P_i;\Lambda_R)
\big\vert \leq {\Lambda_R\over \Lambda_0}
P \log \biggl( {\Lambda_0 \over\Lambda_R}\biggr) \biggr.}
Feeding this into \elxxxiii\ we get
\eqn\eucss{\big\vert
U^r_4(P_i;\Lambda)\big\vert
\leq \biggl({\Lambda_R\over \Lambda_0}\biggr)
 P\log\biggl({\Lambda_R\over\Lambda_0}\biggr) +
\int^{\Lambda}_{\Lambda_R} d\Lambda'
\Lambda_0^{-1} P\log\biggl({\Lambda_R\over\Lambda_0}\biggr),}
Remembering that in this particular case $4-2m-j =0$ (and that $4-2m -j\geq0$
for
all the relevant couplings), we see that
\eqn\euc{\vert
U^r_{4}(P_i;\Lambda)\vert
\leq \biggl({\Lambda\over \Lambda_0}\biggl)
P\log\biggl({\Lambda_R\over\Lambda_0}\biggr),}
for all $\Lambda \in[\Lambda_R,\Lambda_0]$. Thus, we now prove
(i) for the relevant vertex corresponding to this coupling
constant using Taylor's formula and the
bounds already obtained on the the irrelevant vertices.
The verification of (ii) for the remaining relevant vertices
proceeds in an exactly analogous manner, where, once again, we
remember to work downwards in number of momentum derivatives.

Once this is done, then as explained in these previous sections, the
proof by induction is complete, and lemma~3 is true for all $r$ and
$m$.\blackbox
\medskip
Setting $\Lambda = 0$, lemma~3 yields
\eqn\elxxxvi{\big\Vert \partial^j_p ( V^r_{2m}(0) -
\bar{V}^r_{2m}(0))\big\VertR \leq {\Lambda_R \over \Lambda_0}
\Lambda_R^{4 -2m-j} P \log \biggl({\Lambda_0 \over \Lambda_R}\biggr).}
So for given $\Lambda_0$, if we introduce any irrelevant bare
couplings, satisfying the conditions outlined following \elxxix, while
keeping the renormalization conditions on the relevant couplings at
$\Lambda =0$ fixed, we only change the norm of the vertices at $\Lambda=0$ (and
thus the Green's functions of the theory for energies of order $\Lambda_R$) by
terms of order $(\Lambda_R/\Lambda_0)$. We should note that the
irrelevant couplings include the higher derivative pieces of
$V^r_2(p,-p;\Lambda_0)$, and we have the
freedom to choose the form of $V^r_2(p,-p;\Lambda_0)$ at order greater
than one in $p^2$ and order greater than or equal to one in $r$, as
long as our choice is consistent with the conditions described
immediately following \elxxvi.
Therefore, if we change the high $p^2$
behaviour of the bare two--point function the low energy Green's
functions of the theory only change by
terms of order $(\Lambda_R/\Lambda_0)$, provided that the change is defined
to be at first or greater order in $r$.\foot{This means that at low
energies the Green's functions are approximately independent of
small perturbations to
the regulating function. In fact as we will show in
\S 4.3 the S--matrix is completely independent of the regulating
function to any order in its Taylor expansion.}

Combining the above results with those determined from equation
\elxxiv\ we see that the low energy theory is extremely insensitive
to both the value of the high energy cut-off and to the irrelevant bare
couplings, provided that the cut-off is very large compared to the
scale at which we probe the physics (as depicted in \flow).
The renormalization conditions \elxxvi\ for the effective theory show
that it actually contains an infinite number
of parameters. However, the results described above mean that
at energies $\sim \Lambda_R$ the Green's functions and hence the S-matrix
elements are actually determined by a finite number
of these parameters (the relevant coupling constants $\lambda_i$)
up to corrections
vanishing as $(\Lambda_R/\Lambda_0)$ becomes very small; the
irrelevant bare couplings $\eta^r_{2m}$ really are irrelevant
as far as the low energy physics is concerned.
Furthermore, as $(\Lambda_R/\Lambda_0) \rightarrow 0$ the theory
converges towards a unique limit which is determined entirely by the
renormalization conditions on the relevant coupling constants. This
convergence is generally known as universality.

So, whatever the form of the Lagrangian at $\Lambda_0$ actually is, we
can describe the low energy physics to very good accuracy by setting
all the bare irrelevant couplings equal to zero and letting $\Lambda_0/m
\rightarrow \infty$. If we do this then the bare couplings are simply
functions of the renormalization conditions on the relevant
couplings, and there are the same number of each. Thus, if we now view
the bare parameters as the independent parameters of the theory we are
back to the conventional view of quantum field theory.

This explains why the conventional way of looking at
quantum field theory works. However, it now
seems as if this may not be the best way to view a quantum field
theory. Since we always in practice deal with theories designed to work
over a limited range of scales, without regard for the
physics at much higher scales (which may in any case be unknown), and
since in general the effective Lagrangian at just below the scale
of the new physics will always contain irrelevant couplings (it
would require a remarkable degree of fine tuning to ensure
that all such couplings vanished simultaneously at the same scale), we will
always expect deviations from the predictions of the conventional quantum field
theory when we can make measurements with sufficient accuracy at low
scales, or alternatively as we approach the scale of the new physics.

\subsec{Systematic Improvement.}

The previous two sections have shown that if we look at physics at
scales $\sim \Lambda_R$, then once we set renormalization conditions on the
relevant couplings which are independent of the naturalness scale $\Lambda_0$,
then there is only very weak dependence on
$\Lambda_0$ as long as $\Lambda_0\gg\Lambda_R$.
The predictions made by the theory, using just
the renormalization conditions on the relevant couplings as input, have
accuracy of order ${\Lambda_R\over\Lambda_0}P\log({\Lambda_0\over
\Lambda_R})$, and may therefore start to show deviations from experiment
when measurements are made with this type of precision. However, this does not
mean that the
theory is no longer any good. It simply needs more input, in the form
of the experimental determination of couplings which, though canonically
irrelevant, have now become relevant physically, in the sense that
they are measurable. By measuring such couplings the theory may be
systematically improved.

We therefore consider specifying further renormalization conditions,
at energies comparable with the scale at which the physics
is being probed, for all the coupling constants corresponding to
operators with canonical dimension $\leq D$, where $D >4$.\foot{For our
scalar theory there are (as we show in
\S 4.2 below) only $N_D=2,4,6,10,\ldots$ physically relevant nonredundant
couplings for $D=6,8,10,12,\ldots$.
The values of these couplings are further constrained by the
requirement that the theory be stable (\S 5.2).}
This in turn automatically determines the value of all bare coupling
constants corresponding to operators with canonical dimension
$\leq D$. We want this to be done in such a way that the theory remains
natural, i.e. so that all the bare couplings are of order unity at the
naturalness scale and lemma~1 is obeyed. In practice we would expect
to be able to specify these renormalization
conditions by matching them to experimentally determined quantities.
In this case the renormalization conditions are automatically
consistent with naturalness. Indeed, they tell us what the
naturalness scale is (see \S 3.3 below).

Once we have specified these renormalization conditions
it is entirely straightforward to repeat the induction procedure used
in \S 2.3 and \S 3.1 to prove the appropriate generalizations of
lemmas 2 and 3, namely
\eqn\esiapre{\Biggl\Vert \biggl({\partial\over\partial\Lambda_0}\partial^j_p
V^r_{2m}(\Lambda,\Lambda_0,\lambda_i)\biggr)_{{\lambda_i}}
\Biggr\Vert_{\Lambda_R} \leq\Lambda_0^{2-D}\Lambda_R^{D+1-2m-j}
P\log\biggl({\Lambda_0\over\Lambda_R}\biggr),}
and
\eqn\esiaxpre{\big\Vert \partial^j_p
(V^r_{2m}(\Lambda,\Lambda_0',\lambda_i) -
\bar V^r_{2m}(\Lambda,\Lambda_0,\lambda_i))\big\Vert_{\Lambda_R}
\leq \Lambda_0^{3-D}
\Lambda_R^{D+1-2m-j} P \log \biggl({\Lambda_0 \over \Lambda_R}\biggr),}
for all $\Lambda\in [0,\Lambda_R]$, with similar bounds for
$\Lambda\in [\Lambda_R,\Lambda_0]$. So the physically
relevant coupling constants at $\Lambda_R$ depend on the naturalness scale and
the bare coupling constants with canonical dimension greater than $D$ (the
`physically irrelevant' couplings) only at a level of
$({\Lambda_R\over\Lambda_0})^{D-3}P\log ({\Lambda_0\over\Lambda_R})$.

Thus, we have the
intuitively obvious result that the precision of an effective
theory is systematically improved if more
renormalization conditions on physically relevant couplings are
determined experimentally; to a given order in
$({\Lambda_R \over \Lambda_0})$ we need determine only a finite
number of new couplings. For the case of $Z_2$ symmetric scalar field
theory we actually increase our predictive power in steps of
$({\Lambda_R\over\Lambda_0})^2$ because all operators are of
even dimension. From the Lorentz structure of the
theory all vertices which are differentiated an odd number of times
vanish at zero momentum (as we saw for $\partial_{p_{\mu}}V_2(\Lambda)$ in \S
2.1). Thus, if we have set renormalization conditions for couplings
of dimension $4-D$ corresponding to operators up to dimension $D$,
where $D$ is even, we
have in effect set them up to dimension $D+1$, since we know that
the renormalization conditions on couplings of dimension $3-D$ are
constrained to vanish for all momenta zero. In
particular, we see that we can improve the bounds in lemma~2 and
lemma~3 by a factor of $({\Lambda_R/\Lambda_0})$ without having to
change the renormalization conditions at all. Combining this with the
results \esiapre\ and \esiaxpre, we have the rather stronger

\medskip
\vbox{\noindent{\it Lemma 4:}\hfill\nobreak

\noindent For all $\Lambda \in [0,\Lambda_R]$, and $D\geq 4$,
\eqn\esia{\Biggl\Vert \biggl({\partial\over\partial\Lambda_0}\partial^j_p
V^r_{2m}(\Lambda,\Lambda_0,\lambda_i)\biggr)_{{\lambda_i}}
\Biggr\Vert_{\Lambda_R} \leq\Lambda_0^{1-D}\Lambda_R^{D+2-2m-j}
P\log\biggl({\Lambda_0\over\Lambda_R}\biggr),}
and
\eqn\esiax{\big\Vert \partial^j_p
(V^r_{2m}(\Lambda,\Lambda_0',\lambda_i) -
\bar V^r_{2m}(\Lambda,\Lambda_0,\lambda_i))\big\Vert_{\Lambda_R}
\leq \Lambda_0^{2-D}
\Lambda_R^{D+2-2m-j} P \log \biggl({\Lambda_0 \over \Lambda_R}\biggr),}
with obvious extension to $\Lambda\in [\Lambda_R,\Lambda_0]$.}
\medskip

Since we only set renormalization conditions on a finite number of
couplings corresponding to
operators of up to a given canonical dimension the renormalized
effective theory is, in a sense, always local, even though the
bare theory was only quasi--local. The renormalized effective theory
would only be quasi--local if we required it to be arbitrarily precise.

This systematic improvement of an effective field theory is rather
closely related to
Symanzik's proposal to speed up the convergence to the
continuum limit of a lattice field theory by adding to the lattice
action a finite number of irrelevant terms, thus constructing an
`improved action'\ref\rx
{K. Symanzik: ``Some Topics in Quantum Field Theory'', in:
Mathematical Problems in Theoretical Physics, p.47; eds. R. Schrader
et al.; Springer, Berlin, 1982 (Lecture Notes in Physics 153) \semi K.
Symanzik, \NP\vyp{B226}{1983}{187,~205}.}.
The bounding arguments leading to lemma~4 constitute a
proof (within perturbation theory, of course) of a continuum
version of Symanzik's hypothesis for lattice field theories.

Instead of systematically improving in the accuracy of the theory for
processes at a given scale, we could consider instead attempting
similarly to
maintain the accuracy of the theory when we consider processes of
energy $E\sim |z_{ij}|$ much greater than $\Lambda_R\sim m$. It is
tempting to suppose that this may be done simply by finding bounds
on Green's functions with norms taken
with respect to some new scale, $\Lambda_H>\Lambda_R$. Naively we
can repeat the same
arguments as in \S 2.2, \S 2.3 and \S 3.1 to find the same bounds, simply with
$\Lambda_R$ replaced with $\Lambda_H$. However, the coefficients in
the polynomials in these bounds will depend
on the ratio $(\Lambda_H/\Lambda_R)$, and if this ratio is very large
then the effect of these factors may compensate for, or even
overwhelm, the powers of $(\Lambda_H/\Lambda_0)$. In other words, the
bounding arguments may be spoiled if there are severe (by which we
mean stronger than logarithmic) infrared divergences. In order to address
this problem we have to look at the details of the bounding argument
far more carefully than we have done so far, and we
postpone this investigation to a future publication\rbtir.
Here we simply note the result: lemma~4
remains true if $\Lambda_R$ is replaced by a new scale $\Lambda_H\sim
E$, $\Lambda_H\in [\Lambda_R,\Lambda_0]$, up to logarithmic factors
$P\log({\Lambda_H\over\Lambda_R})$, and away from momenta and partial
sums of momenta with magnitudes much less than $\Lambda_H$.
So we may maintain the predictive power at high
energies in exactly the same way as we improve it while staying at the
same energy scale.\foot{It is also possible to find bounds on the Green's
functions which have some momenta with magnitude $\Lambda_H$ and some
with magnitude $\Lambda_R\sim m$, i.e. with a given set of (almost)
exceptional momenta\rbtir.}

As the energy of the physics we probe approaches $\Lambda_0$ we need
more and more renormalization conditions to maintain the same
level of accuracy, and if $\Lambda_H\sim\Lambda_0$ we need
in principle an infinite number (namely the complete vertex functions
$V_{2m}^r(p_1,\ldots,p_{2m};\Lambda_0)$). In the absence of some deep
underlying principle which could determine these vertex functions, our
quasi--local field theory is now too general to be very useful,
although as we will show in \S 5.3 it is still completely
consistent; in fact it is equivalent, as conjectured by Weinberg
\riii, to an analytic S--matrix theory.

\subsec{Unnaturalness and Nonrenormalizability}

It is important to note that for a particular value of $\Lambda_0$ we did not
have an entirely free choice in
specifying the renormalization conditions at $\Lambda=0$. As stated, the low
energy
renormalization conditions on the coupling constants must be
consistent with the naturalness \elxxix\ of the bare couplings.
This does not mean that it is sufficient that the dimensionless
vertices at $\Lambda_R$ are constants of order unity, satisfying
lemma~1. Rather, it is also a necessary consequence of naturalness that
they must satisfy lemma~2, or more accurately, the improved form of
lemma~2, \esia\ above, i.e. if we have already set the renormalization
conditions on couplings corresponding to operators with canonical
dimension $\leq D$, then we only have freedom of order
$({\Lambda_R\over\Lambda_0})^{D-3} P\log ({\Lambda_0\over\Lambda_R})$
when setting the renormalization conditions on the dimensionless
couplings corresponding to operators of dimension $D+2$. Perhaps the
easiest way to see this is to use the improved form of lemma~3,
\esiax\ above, which expresses our limited freedom to change the
theory through natural changes in the couplings corresponding to
operators of dimension greater than $D$.

If we were to ignore this increasingly tight restriction on the low
energy renormalization conditions we would violate naturalness. For
example, specifying the dimensionless coupling constant associated with the
$\phi^6$ term to be an arbitrary
$\Lambda_0$-independent constant of order $\Lambda_R^{-2}$ would lead to
the corresponding bare coupling being unnatural. We can see this
explicitly by following through the same argument \eliix\esf\
for $V^r_6(\Lambda)$, which then yields
\eqn\eunii{ \vert V^r_6(P_i;\Lambda)\vert \leq
\Lambda_R^{-2}
\Bigl(P\log\biggl({\Lambda\over\Lambda_R}\biggr) +
{\Lambda\over\Lambda_0}P\log\biggl({\Lambda_0\over\Lambda_R}\biggr
)\Bigr).}
This bound on
$V^r_6(\Lambda_0)$ is clearly not `natural' when $\Lambda\gg\Lambda_R$.
It is only by specifying $V^r_6(\Lambda_R)$ in a manner consistent
with the naturalness of the theory at $\Lambda_0$, which means in
practice within the tight limits already set
by the renormalization conditions on the physically relevant coupling
constants of lower dimension, that we can maintain the natural bound on
$V^r_6(\Lambda)$ all the way up to $\Lambda_0$.

We can think of this necessary fine tuning of the low energy
renormalization conditions as in some sense the inverse of the
usual fine tuning problem\rxx. There we have to
fine tune the bare mass to an unnaturally precise value in order
to obtain a predictive theory of a particle with
mass $m\ll\Lambda_0$. Now we see that conversely we have to
fine tune the low energy renormalization conditions on irrelevant
vertices in order to get natural (i.e. untuned) bare irrelevant couplings.

If $V^r_6(\Lambda)$ were in reality such as to saturate the bound \eunii,
this would feed into the flow equations, causing the low energy theory
to depend strongly on $\Lambda_0$. Then no matter how many low
energy renormalization conditions we were to set, all unspecified low
energy physics would still depend on the unknown couplings times
positive powers of $\Lambda_0/\Lambda_R$. The
only way to remove this dependence would be to to specify all the low
energy couplings, and thus lose any predictive power. The easiest way to
see this is to make the theory natural again by choosing a new cut--off scale
$\Lambda_0'$ such that $V^r_6(\Lambda_0')$ is of order $(\Lambda_0')^{-2}$; we
can then use the bounding lemmas as before.
Unfortunately to do this would require $\Lambda_0'\sim\Lambda_R$, so
there would no
longer be energy region in which the theory depended, to a given level
of accuracy, only on a finite
number of parameters. This would then be what we commonly call a
nonrenormalizable theory. The distinction between a
`renormalizable' effective theory (which has
$\Lambda_0\gg\Lambda_R$) and a `nonrenormalizable'
effective theory (which has $\Lambda_0\gsim\Lambda_R$) is thus rather
vague, resting as it does entirely on the predictivity of the theory;
it could be made much more precise however by specifying the
accuracy to which we wish to work and the dimension D of the physically
relevant couplings we were then prepared to determine.\nobreak\footsy{Unless
of course we were to insist (rather puritanically) that a
theory should only be called `renormalizable' if we can work in
the limit $\Lambda_0/\Lambda_R\to\infty$; such a definition would not
be very useful here though as then all the scalar theories we consider
would probably be nonrenormalizable, with the sole exception of the
free one.}

Thus, once we have specified the low energy renormalization conditions on
the relevant couplings, we may maintain a renormalizable theory by
allowing only natural irrelevant vertices,
and by being careful to set low energy renormalization conditions on
any further physically relevant couplings in a manner
consistent with this requirement. Turning this around, we may
determine the naturalness scale $\Lambda_0$ experimentally by
measuring coupling constants corresponding to canonically irrelevant
operators. The first observation of such a dimensionful coupling
thus inevitably seals the ultimate fate of the theory,
leading us to expect (or at least hope for) new physics by
the time we reach the new scale $\Lambda_0$ as
determined (albeit very approximately) by the value of the new coupling.

\newsec{Redundancy.}

In \S 3.2 above we showed that the precision of a theory
can be improved systematically by specifying the values of more and more
irrelevant coupling constants as low energy renormalization conditions;
the values of these coupling constants could be determined in principle by
comparison with S--matrix elements. In fact, this last statement is not
strictly correct. Not all the coupling constants in the theory correspond
to physical observables; those that do not, and the operators
they correspond to, may be described as `redundant'.\foot{This is in
the same spirit as, but not precisely the same as the definition of a
redundant operator in statistical mechanics (\ref\rwegner{F.~J.~Wegner,
\JP\vyp{C7}{1974}{2098}.}; see \ref\rAD{J.~Alfaro and P.~H.~Damgaard,
\AP\vyp{220}{1992}{188}.} for a more modern discussion); here a
redundant operator is one which does not contribute to S--matrix
elements, rather than one which does not contribute to the partition
function.}

Even in conventional quantum field theory, where we only consider the
coupling constants corresponding to relevant operators, we notice a
special (and as we will see later, slightly unusual) example of a
redundant variable. The wavefunction renormalization, which we denote
\exxxvi\ by $\lambda_2$, has no physical meaning since it cancels out in the
construction of the S-matrix elements. Since the renormalized
wavefunction renormalization is a redundant variable, so too is the bare
wavefunction renormalization $\lambda_2(\Lambda_0)$ (\exxxiii),
because the latter is defined by the former, as we saw in \S 2.1.

In an effective field theory we have many redundant variables; new ones
appearing each time we consider higher dimension operators. In \riii\
the absence of a redundant term in the effective Lagrangian is
explained in a footnote by remarking that it
is possible to remove it by making a field redefinition. The
systematic elimination of redundant variables has recently been
addressed in \ref\rG{H. Georgi, \NP\vyp{B361}{1991}{339}.}. In this
section we will describe exactly which terms in our effective Lagrangian are
redundant, and then prove that they may all be removed from the effective
Lagrangian by making field redefinitions. To do this it will first be
necessary to formulate the necessary equivalence theorem which permits
us to make such redefinitions without changing S--matrix elements.

\subsec{Field Redefinitions and the Equivalence Theorem.}

The formal invariance of S--matrix elements under point transformations of
the interpolating fields is a fundamental property of quantum field
theory, and as such has a long history\ref
\reqt{R.~Haag, \PR\vyp{112}{1958}{669}\semi
H.~J.~Borchers, \NC\vyp{25}{1960}{270}\semi
J.~S.~R.~Chisholm, \NP\vyp{26}{1961}{469}\semi
S.~Kamefuchi, L.~O'Raifeartaigh and A.~Salam, \NP\vyp{28}{1961}{529}.}.
This `classical' equivalence theorem is not sufficient for our
purposes however; we also need to show that the S--matrix is invariant under
redefinitions of the quantum field --- a `quantum' equivalence
theorem\nref\rKT{R.~E.~Kallosh and I.~V.~Tyutin,
\SJNP\vyp{17}{1973}{98}\semi E.~S.~Abers and B.~W.~Lee,
\PRep\vyp{9C}{1973}{1}.}\refs{\rdiagrammar,\rKT}.

We will consider field redefinitions of the form
\eqn\eredef{\phi_p\rightarrow\phi'_p=\phi_p+{\cal F}[\phi_p;\Lambda],}
where the functional\foot{Actually ${\cal F}[\phi_p;\Lambda]$
is both a function
and a functional of $\phi_p$; we hope that the abuse of notation is
not confusing.} ${\cal F}[\phi_p;\Lambda]$ possesses the
same properties as the interaction Lagrangian, which means that we
may write it as
\eqn\efdef{{\cal F}[\phi_{p};\Lambda]\hskip -0.02in \equiv \hskip
-0.02in \sum^{\infty}_{m=0}\sum
^{\infty}_{r=m} \hskip -0.02in \frac{g^r}{(2m+1)!}
\hskip -0.02in \int \hskip -0.02in {d^4p_0\cdots d^4p_{2m} \over (2\pi)^{8m}}
f^r_{2m}(p_0,p_1,\ldots ,p_{2m};\Lambda)
\delta^4\hskip -0.01in \big(\hbox{$\sum^{2m}_{i=0}p_i\hskip -0.02in -
\hskip -0.02in p$}\big)
\phi_{p_0}\phi_{p_1}\hskip -0.01in \cdots\phi_{p_{2m}}.}
for all positive $\Lambda$.\foot{It makes no sense to make a field
redefinition when $\Lambda=0$, since all of the field has then been
integrated out.} So in particular ${\cal F}$ is analytic in $\phi$
(which means that it may be expanded in powers of $\phi$), it does not
spoil the $Z_2$ symmetry, so ${\cal F}[-\phi]=-{\cal F}[\phi]$,
and it may be expanded as a power series in the coupling $g$,
beginning at order $m$. This latter assumption ensures that for $m>0$
the redefinition is perturbatively invertible, and furthermore
that it leaves the free theory unaltered, while generating new
vertices which obey the
technical restriction $r\geq m-1$; for $m=0$ we require only that,
with $f_0(p^2;\Lambda)\equiv\sum_{r=0}^{\infty}f_0^r(p;\Lambda)$,
 $1+f_0(p^2;\Lambda)$ has no zeros in any bounded region of the
complex plane. The coefficient functions
$f^r_{2m}(p_0,p_1,\ldots,p_{2m};\Lambda)$ are assumed to have the
same properties as $\bar{V}^r_{2m}(p_1,\ldots,p_{2m};\Lambda)$; in
particular they are regular functions
of the momentum invariants $z_{ij}$ (so that no new singularities are
introduced by the field redefinition), and they are natural at the scale
$\Lambda_0$. Finally we impose the additional requirement that
$f^r_{2m}(p_0,p_1,\ldots,p_{2m};\Lambda)$ vanishes as
$p^2\rightarrow\infty$ faster than
any power of the `external' Euclidean momentum $p^2$, where
$p\equiv-\sum_{i=0}^{2m}p_i$; in practice we take functions containing the
regulating factor $K_{\Lambda}(p)$.

To prove the classical equivalence theorem, it is sufficient to show that
on--shell the amputated connected Green's functions (cf. \exvi,\exxiii)
\eqn\eagf{\tilde{G}_n(p_1,\ldots,p_n)\equiv P\inv_{\Lambda}(p_1)\cdots
P\inv_{\Lambda}(p_n)\big\langle\phi_{p_1}\cdots
\phi_{p_n}\big\rangle_{\!\Lambda}}
are equivalent to the transformed Green's functions
$$\tilde{G}_n'(p_1,\ldots,p_n)\equiv P\inv_{\Lambda}(p_1)\cdots
P\inv_{\Lambda}(p_n)\big\langle\phi'_{p_1}\cdots
\phi'_{p_n}\big\rangle_{\!\Lambda},$$
since for transformations of the form \eredef\ $\tilde{G}_n$ and
$\tilde{G}_n'$ differ only by terms of the form
\eqn\eterms{P\inv_{\Lambda}(p_1)\cdots P\inv_{\Lambda}(p_n)
\big\langle {\cal F}[\phi_{p_1},\Lambda]\cdots {\cal F}[\phi_{p_m},\Lambda]
\phi_{p_{m+1}}\cdots\phi_{p_n}\big\rangle_{\!\Lambda},}
with $m>0$.
It now suffices to show\rKT\ that on--shell the amputated terms \eterms\
always vanish.\foot{Although such terms contain composite operator insertions,
it is not necessary to show that such insertions are renormalizable;
it is sufficient for present purposes that they are finite for finite
$\Lambda$. The renormalization of Green's functions with composite
operator insertions will be considered elsewhere\ref\rbtope{R.~D.~Ball
and R.~S.~Thorne, CERN-TH.????/93, OUTP-93-??P.}.}

To do this we expand the operator
${\cal F}[\phi_{p},\Lambda]$ as in \efdef, and consider separately
those parts which are nonlinear in the field, and those that
are linear (so $m>0$ and $m=0$ respectively in \efdef).
For the nonlinear part, each insertion ${\cal F}[\phi,\Lambda]$
in \eterms\ creates several particles, and
$\langle F\cdots\rangle$ is thus lacking the one--particle pole
present in $\langle\phi\cdots\rangle$. In the amputated connected
Green's functions \eagf\ the amputating inverse propagators cancel the
one--particle poles on--shell, leaving a finite result; since in
\eterms\ some of these poles are missing, it must vanish on--shell.
The linear part of ${\cal F}[\phi_{p},\Lambda]$ will be of the form,
$f_0(p^2)\phi_p$. If, as suggested in \rG, we further
assume that when analytically continued into the
complex plane $f_0(z)$ has a zero at $z=-m^2$, \eterms\ will vanish
trivially on--shell. This extra assumption is not necessary however;
if $f_0(-m^2)$ is nonzero, on--shell
$\tilde{G}_n'=(1+f_0(-m^2))^n\tilde{G}_n$ and the extra factors of
$1+f(-m^2)$ may be absorbed by a finite renormalization of the
external wave functions when we construct the S--matrix element.
The classical equivalence theorem is then proven for any redefinition
of the interpolating fields of the form \eredef,\efdef.

To prove the quantum equivalence theorem, we
consider the changes in the functional integral representation \exiii\
of the generating functional $\tilde{W}[J]$ of the amputated connected
Green's functions due to the change of variables \eredef:
\item{(a)} The coupling of the source
to the field, $\big(\phi,P\inv_{\Lambda}J\big)\rightarrow
\big(\phi',P\inv_{\Lambda}J\big)$, which thus induces the
transformation $\tilde{G}_n\rightarrow\tilde{G}_n'$ described above.
\item{(b)} The action $S[\phi;\Lambda]\rightarrow
S[\phi';\Lambda]$; because \efdef\ begins at first order in the
coupling constant, the free action
remains unchanged, while $S_{\rm int}[\phi;\Lambda]\rightarrow
S_{\rm int}'[\phi;\Lambda]$. By construction $S_{\rm int}'[\phi;\Lambda]$
has the same symmetry, analyticity, convergence and naturalness
properties as the original effective action $S_{\rm int}[\phi;\Lambda]$.
The new vertices $(V_{2m}^r)'$ obey the same technical ordering
requirement (they vanish for $m>r+1$) as the old ones, by virtue of
the similar requirement on $f_{2m}^r$.
\item{(c)} The measure
${\cal D}\phi\rightarrow{\cal D}\phi{\cal J}[\phi,\Lambda]$,
where the Jacobian
\eqn\ejacob{{\cal J}[\phi,\Lambda]={\rm Det}
\Big[\delta_{pp'}+{\partial {\cal F}[\phi_p] \over
\partial\phi_p'}\Big]=\exp\;\Tr\ln\Big[\delta_{pp'}+
{\partial {\cal F}[\phi_p] \over\partial\phi_p'}\Big]\equiv
\exp\;-S_{\cal J}[\phi,\Lambda].}
Expanding the logarithm\footsy{It is not appropriate to use a ghost field
here, because the transformation \eredef, and thus the Jacobian, is
assumed to be regular; if the field transformation had introduced an
unphysical pole, writing the Jacobian in terms of a propagating ghost
might be have been useful when checking that all the unphysical singularities
cancelled out.}, it is not difficult to see that
\eqn\esj{
S_{\cal J}[\phi,\Lambda]=\sum_{n=1}^{\infty}\frac{(-)^{n+1}}{n}
\frac{g^{\bar{r}}}{\prod {(2m_\alpha)!}}
\int\!\prod_{\alpha =1}^{n}\prod_{i=1}^{2m_\alpha}
\Big[\frac{d^4p_i^\alpha}{(2\pi)^4}\phi_{p_i^\alpha}\Big]
\;(2\pi)^4\delta^4\Big(
\sum_{\alpha =1}^{n}\sum_{i=1}^{2m_\alpha}p_i^\alpha\Big)
\;\bar{V}_{2\bar{m}}^{\bar{r}}(\{p_i^\alpha\};\Lambda),}
where $\bar{m}\equiv\sum_{\alpha =1}^{n}m_\alpha$,
$\bar{r}\equiv\sum_{\alpha =1}^{n}r_\alpha$, and
\eqn\evbar{
\bar{V}_{2\bar{m}}^{\bar{r}}(\{p_i^\alpha\};\Lambda)\equiv
\int\!\frac{d^4q}{(2\pi)^4\Lambda^4}\prod_{\alpha =1}^{n}
\biggl(\Lambda^4 f_{2m_\alpha}^{r_\alpha}\Big(q+\sum_{\beta =1}^{\alpha -1}
\sum_{i=1}^{2m_\beta}p_i^\beta,p_1^\alpha,\ldots,p_{2m_\alpha}^\alpha;
\Lambda\Bigr)\biggr).}
The Jacobian is thus ultraviolet finite whenever
$f_{2m}^r(p_0,p_1,\ldots,p_{2m};\Lambda)$
vanishes faster than any power as $p^2\rightarrow\infty$; this
rather novel feature will be illustrated by some explicit examples
in \S 4.3 below. Note that if $f_{2m}^r(p_0,p_1,\ldots,p_{2m};\Lambda)$
were independent of $p$, and thus the regulating factor absent, as
it would be in dimensional regularization,
$S_{\cal J}[\phi,\Lambda]$ would be proportional to $\delta^d(0)$,
where d is the dimension of spacetime. All such terms are then set
formally to zero in dimensional regularization
(and indeed the Jacobian was ignored in \rG).
Using our form of regularization we are able to deal with the Jacobian
factor in a much less formal (and therefore, we feel, rather more
satisfactory) manner. Indeed, here it is necessary for the
invariance of the generating
functional; it may be seen from the explicit representation
\esj\evbar\ above that it has the same structure and properties
as the interaction $S_{\rm int}'[\phi,\Lambda]$, to which
it contributes further terms.

\nref\rzj{J.~Zinn-Justin, in ``Trends in Elementary
Particle Theory'', ed. H.~Rollnik and K.~Dietz (Springer, 1975).}
If we were to make all three of the above changes, it should
be clear that $\tilde{W}[J]$, and thus in particular the S--matrix, remain
unchanged (see \refs{\rzj,\rfs}
for formal algebraic arguments in the context of the
path integral, and \rdiagrammar\ for a purely diagrammatic argument);
with our regularization everything remains finite and well-defined
(at least in perturbation theory), and
no new singularities are introduced. The classical equivalence theorem
as proven above shows that if instead of the complete change of
variables we were to change only the
coupling of the source to the field, then since this amounts to a
transformation of the interpolating fields alone, the S--matrix is
still unchanged. We could thus consider making the last two changes only;
if we let $S_{\rm int}[\phi,\Lambda]\rightarrow
S_{\rm int}'[\phi,\Lambda]+S_{\cal J}[\phi,\Lambda]$, but keep
the coupling of the field to the source unchanged,
the S--matrix is still the same. This is then an equivalence theorem for
transformations of the quantum fields\refs{\rdiagrammar,\rKT}. As formulated
above, the two equivalence theorems are equivalent, at least in
perturbation theory.

We conclude by resolving a paradox. In general when one
makes a field redefinition in a quantum field theory it is not
sufficient to consider only the three changes discussed above;
sometimes there are extra terms generated
\ref\rGJ{S.~F.~Edwards and Y.~V.~Gulyaev,
{\it Proc.~Roy.~Soc.~}\vyp{A279}{1964}{229}
\semi D.~W.~McLaughlin and L.~S.~Sculman, \JMP\vyp{12}{1971}{2520}
\semi J.~M.~Charap, {\it J.~Phys.~}\vyp{6}{1973}{393}
\semi J.~L.~Gervais and A.~Jevicki, \NP\vyp{B110}{1976}{93}.}, at
precisely two loops \rAD. A particularly clear exposition may be found in
\ref\rS{P.~Salomonson, \NP\vyp{B121}{1977}{433}.}, where an explicit
computation of two-loop diagrams is considered, and it is found that
extra terms are generated by field redefinitions
when using a momentum cut-off (but not when using dimensional
regularization; no such terms were found in \rdiagrammar).
On the other hand, we also find no such terms -- hence the paradox. It is
resolved by remembering that in the conventional way of
setting up a quantum field theory it is the bare
Lagrangian which undergoes the field redefinition, and the resulting
Feynman diagrams that are then regularized. In \rS, two formally
equivalent two-loop vacuum diagrams are shown to be different when
regularized by a momentum cut-off; it is, however, very easy to check
that if the regularization is explicitly included in the bare
Lagrangian before making the field redefinition, the two diagrams
are indeed equal (since one of the vertices induced by
the field redefinition now has an extra damping factor), and no extra
terms arise.
It thus appears that for continuum quantum field theory, within
perturbation theory, regularization and field redefinition do not
commute. Obviously, in order to obtain
meaningful results without introducing extra terms it makes sense to
regularize the theory first.

\subsec{On--shell Effective Action}

If we could determine off--shell Green's functions experimentally then
we would be able to measure all the coupling constants corresponding to
operators up to canonical dimension $D$ by performing experiments to
an accuracy of order $(\Lambda_R/\Lambda_0)^{D-3}$. However, since
it is actually only possible to measure on--shell S-matrix
elements\foot{We discuss this further in \S 5.4 below.}, these coupling
constants are not all independent; when we put all external particles
on shell we set $p_i^2=-m^2+i\epsilon$, so the S--matrix elements
depend only on the momentum invariants $\{p_i\cdot p_j; i\neq j\}$ (or
equivalently the generalized Mandelstam invariants
$\{z_{ij}, i\neq j\}$ used in \S 2.4), continued to the
boundary of the physical region.
Thus terms in the Green's functions which depend on the magnitude of
individual momenta are redundant when we go on shell, since they may
be absorbed into similar terms which do not; similarly terms which
are related when we impose overall momentum conservation should not
be regarded as independent.
But (as shown at the end of \S 1.2 above), the connected amputated Green's
functions may be constructed from the vertices in the effective action
defined at $\Lambda =0$. Thus, if there is redundancy in the
former, there is also redundancy in the latter, and (using the
flow equations) the same degree of redundancy in the effective
action at any scale $\Lambda$ (and in particular at $\Lambda_0$,
i.e. in the bare action $S[\phi,\Lambda_0]$).

Since whenever $\Lambda$ is positive the vertex functions are regular
functions of the momentum invariants $z_{ij}$, we may expand them in
Taylor series which converge in any bounded region.
Any terms in these expansions which contain a $p_i^2$, or which are
related through $\sum_ip_i=0$, are then redundant. So if instead we
consider the expansion \actgen\ in terms of local
operators in coordinate space, any operators
of the form $\partial^2\phi f[\phi]$, where
$f[\phi]$ is a local functional of $\phi(x)$, or which
can be written in this form by integrating by parts, are redundant.
For example, at dimension six we have four independent operators
(ignoring those which may be obtained by integration by parts), namely
$\{\phi^6,\phi^2(\partial_{\mu}\phi)^2;
\phi^3\partial^2\phi,\phi\,\partial^4\phi\}$. Of these the latter two are
both redundant, and of these two the first is what we will call a
vertex redundancy, since it may be absorbed into the $\phi^4$ vertex,
while the second is a propagator redundancy since it is only quadratic
in $\phi$. Similarly, at dimension eight there are four nonredundant
operators,
$\{\phi^8,\phi^4(\partial_{\mu}\phi)^2,(\partial_{\mu}\phi)^4,
\phi^2(\partial_{\mu}\partial_{\nu}\phi)^2\}$
while there are three vertex redundancies,
$\{\phi^5\partial^2\phi,\phi^2(\partial^2\phi)^2,
\phi\,\partial^2\phi\,(\partial_{\mu}\phi)^2\}$
and one propagator redundancy $\phi\,\partial^6\phi$.

Before proving that all the redundant terms in the effective action
may be systematically eliminated by repeated field redefinitions, we
first pause to consider the implications of such a result.
{}From the form of the flow equations \exxxxi\ for the vertex functions
it is not difficult to see that, even if the redundant terms are eliminated
at the scale $\Lambda_0$, they will tend to reappear at
other scales $\Lambda$.
This does not mean that we increase the number of coupling constants
(to a given dimensionality) by changing scales, but rather that there
are complicated constraints on the coupling constants at the
scale $\Lambda$ such that the total number of independent coupling
constants is held fixed. Indeed, the redundant couplings at the scale
$\Lambda$ could also be removed by further field redefinitions.\foot{It
would be interesting, but perhaps difficult, to try to write down a
form of the evolution equation which incorporated these field redefinitions
infinitesimally, so that the nonredundant couplings evolved as a
closed set. This could be particularly useful for attempting
nonperturbative evolution using a truncated form of the equations.}
However when calculating perturbatively it is most useful to
eliminate the redundancy, up to a given dimension $D$
and to a given order in $g$, in the Lagrangian defined
at $\Lambda_0$ rather than at some other $\Lambda$, since the bare
Lagrangian can then be used to calculate amplitudes to this order in
$g$ to accuracy $(\Lambda_R/\Lambda_0)^{D-3}$ with the minimum of
complication.

Now if we wish to calculate S--matrix elements to an accuracy
of order $(\Lambda_R/\Lambda_0)^{D-3}$ we showed in \S 3.2 that
we would need to fix the coupling constants corresponding to
operators up to dimensionality $D$. We have to obtain the bare couplings
constants indirectly by fixing renormalization conditions on the
Green's functions and their momentum derivatives, which means in
practice on the vertices $(\partial_{p_{i_1}}\partial_{p_{i_2}})^j
V_{2m}(p_i;0)$ with $2m+2j\leq D$.\foot{This makes it rather easy to count
the number $N_D$ of physically relevant operators with a given dimension
$D$; it is just $N_D=\sum_{m=2}^{D/2}\langle
D/2-m,\min\{6m-10,D/2-m\}\rangle$ where
$\langle n,j\rangle$ is the number of different ways of decomposing a
non--negative integer $n$ into a sum of $j$ non--negative integers,
with the convention $\langle 0,0\rangle =1$ (so
for example $\langle n,2\rangle =1,2,2,3,3,4,\ldots$, while
$\langle n,j\rangle =1,2,3,5,7,11,\ldots$ for $n=j=1,2,3,\ldots$).}
If we use on--shell renormalization conditions,
it will not be possible to set redundant conditions directly since
$\partial_{p_i^{2}}^jV_{2m}(p_i;0)\vert_{p_i=P_i}$ is not accessible
experimentally. Setting off--shell renormalization conditions must be
done with more care, since we must ensure that the conditions are
truly independent; if we set as many
conditions as there are physically relevant operators, we must
recognize that some of these conditions will be redundant.
In principle we should be able to use the freedom
in choosing off--shell renormalization conditions to ensure that the
redundant terms of the same dimension in the bare Lagrangian are
absent, but in practice this will be very difficult to accomplish,
and it is easier to remove the redundancy in the bare Lagrangian
by using field redefinitions. This will then allow us to calculate
with the minimum number of physically relevant bare couplings, and thus
with the least complication. In particular, all bare two--point couplings
may be completely eliminated.

\subsec{Systematic Elimination of Redundant Terms.}

We first consider redundant vertices with $m>2$, and show that they may be
systematically eliminated by considering a nonlinear transformation
of the form \eredef\efdef\ with $m>0$. Two-point redundancies will be
considered later.

Consider first the redundant terms
$a_0(g)\Lambda^{2-2m}\phi^{2m-1}\partial^2\phi$,
where $a_0(g)$ is a power series in $g$ beginning at order $m-1$ in
$g$ and consistent with naturalness;
$a_0(g) = \sum_{r=m-1}^{\infty} a_0^r g^r $ , where the $a_0^r$
are of order unity, depending at most on logarithms of
$\Lambda/\Lambda_R$. We wish to make a field
redefinition such that the $r_{th}$ order piece of this term will be
eliminated by the change in the free part of the effective Lagrangian,
$\phi\,K\inv_{\Lambda}(-\partial^2)(-\partial^2 +
m^2)\phi$. To this end we make the field redefinition \eredef\ with
\eqn\eriii{{\cal F}[\phi(x);\Lambda] = \half a_0^r g^r
\Lambda^{-2m}K_{\Lambda}(-\partial^2)\phi^{2m-1}(x).}
Clearly this satisfies the conditions listed after \efdef; in
particular it necessarily contains the damping factor necessary for
the convergence of the Jacobian. The free part of the Lagrangian now becomes
$$
\eqalign{\phi\,K\inv_{\Lambda}(-\partial^2)(-\partial^2 +
m^2)\phi &+ a_0^r g^r
\Lambda^{-2(m-1)}\phi^{2m-1}(-\partial^2+m^2)\phi\cr
&+\quarter{a_0^{2r}g^{2r}\Lambda^{-4(m-1)}}
\phi^{2m-1}K_{\Lambda}(-\partial^2)(-\partial^2 +
m^2)\phi^{2m-1}.}
$$
Thus, we have a term $-a_0^rg^r\Lambda^{2-2m}\phi^{2m-1}
\partial^2\phi$ which cancels the term we wish to eliminate,
replacing it by $\Lambda^{2-2m}m^2a_0^rg^r\phi^{2m}$. We
also generate a natural vertex with $4m-2$ legs which begins at
$2r_{th}$ order in $g$. The interaction Lagrangian also changes
under the field redefinition. But since it begins at first order in
$g$, this change is at $(r+1)_{th}$ or larger order in $g$.
For example a term of the form
$g^{r'}\phi\partial^{2j}\phi$ develops a piece proportional to
$g^{r+r'}\phi^{2m-1}\,\partial^{2j+2}\phi$ due to the transformation.
However, we can simply absorb this into the $(r+r')_{th}$ order term in
terms already present in the Lagrangian and only concern ourselves about
this term when removing redundancies at this higher order.
So, as far as our effective Lagrangian is concerned there is no change
from zeroth to $(r-1)_{th}$ order in $g$; at $r_{th}$ order in $g$ we have
simply eliminated the term we wanted to by changing the coupling
constant of the $\phi^{2m}$ operator; at higher orders in $g$ we have
changed the interaction Lagrangian in such a way that
renormalizability is still manifest.

We next consider the Jacobian of the transformation.
As we have already seen, since we use a quasi--local
operator in the free part of our Lagrangian, we must make a quasi--local field
transformation in order to eliminate local redundant terms by using the
variation of this free part of the Lagrangian; this makes our Jacobian
a well defined object. Indeed the same reasoning that led to the
general expressions \esj\evbar\ now gives
\eqn\erxxx{S_{\cal J}[\phi,\Lambda]=\sum_{n=1}^{\infty}
{(-)^{n+1} \over n} g^{nr}
\Lambda^{4-8m'}\int \frac{d^4p_1 \ldots d^4p_{2m'}}{(2\pi)^{4(2m'-1)}}
\delta^4\Big(\sum_{i=1}^{2m'} p_i\Big)
\bar{V}^{nr}_{2m'}(p_1,\ldots,p_{2m'};\Lambda)
\phi_{p_1}\cdots\phi_{p_{2m'}},}
where $m' =(m-2)n$, and
\eqn\erxxxv{\bar{V}^{nr}_{2m'}(p_1,\ldots ,p_{2m'};\Lambda)
=[\half(2m-1)a_0^r]^n \!\int\! \frac{d^4q}{(2\pi)^4\Lambda^4}
K_{\Lambda}(q) K_{\Lambda}\Big(q+\sum_{i=1}^{2(m-1)}p_i\Big)
\cdots K_{\Lambda}\Big(q + \sum_{i=1}^{2(m'-m+1)}p_i\Big).}
These new vertices are manifestly finite, since the integral
over $q$ converges\foot{If the regulating factor
is set to one, then $\bar{V}^{nr}_{2m'}=
[\half(2m-1)a_0^r]^n (2\pi)^{-4}\Lambda^{-4}\delta^4(0)$.}, and indeed
satisfy all the conditions necessary for renormalizability. The new
vertex at order $r$ is not redundant (it is proportional to
$\phi^{2m-2}$); all the other vertices are higher order in $r$. It
follows that
the change in the Lagrangian brought about by the
Jacobian of the field transformation maintains the manifest
renormalizability of the theory, and does not introduce any new
redundancies; $S_{\cal J}$ may be absorbed into the transformed
interaction $S'_{\rm int}$.
Thus, by the equivalence theorem, our field redefinition \eriii\ has
removed the redundant term $\phi^{2m-1}\partial^2\phi$ at
$r_{th}$ order in $g$ without changing the S--matrix.

We now consider a more general type of redundancy. The most
general form for a redundant term with $2m$ $\phi$ fields is
$b_0(g)\Lambda^{-(d+1)}\partial^2\phi\,{\cal O}_d[\phi]$, where
$d$ is the canonical dimension of the local operator ${\cal O}_d[\phi]$,
$2m-1$ plus the number of derivatives in ${\cal O}_d[\phi]$. In order
to eliminate such a term at $r_{th}$ order in $g$ we must make a
field redefinition of the form
\eqn\erxxxx{{\cal F}[\phi(x),\Lambda] = \half b_0^r g^r\Lambda^{-(d+1)}
K_{\Lambda}(-\partial^2){\cal O}_d[\phi(x)].}
This type of field redefinition does not
upset the form of the Lagrangian any more than the
transformation \eriii\ did. Again, by expressing the Lagrangian in
terms of the new field, besides for removing the redundant
term, we only change the Lagrangian at orders greater than $r$ in
$g$, and in such a way that the renormalizability is still manifest.
The field redefinition gives a Jacobian of the same form as
\erxxx\ except that the expression for $\bar{V}^{nr}_{2m'}$
now contains polynomials in the momenta (including $q$) as well as
the factors of $K_{\Lambda}$ present
in \erxxxv. Since $K_{\Lambda}$ falls of more quickly for large
momenta than any polynomial, we can see that the integral
over $q$ is still convergent. Besides eliminating the redundant term
the field redefinition introduces new $r_{th}$ order
terms proportional to $\phi\,{\cal O}_d[\phi]$ (from the free Lagrangian)
and $\delta{\cal O}_d[\phi]/\delta\phi$, which, though possibly
still redundant, contain one fewer factors of
$\partial^2$ acting on single fields; they have a lower `order of
redundancy'. We can therefore absorb these redundant terms into other
terms of the same form which already exist, and eliminate them with a
further field redefinition.
Therefore, by working in steps down from a given order in redundancy,
it is clear that at given order in $g$ we can systematically eliminate
the redundancy from arbitrarily complicated vertices.
Since each redefinition satisfies the conditions for the equivalence
theorem, we can remove all the redundant terms with four
or more fields and up to any finite order in derivatives, and at
any finite order in $g$.

Finally, we consider interaction vertices quadratic in $\phi$.
All of these vertices, besides for the piece proportional to $\phi^2$, may be
written in the form
$c(g)\Lambda^{2p-2}\phi\,(\partial^2)^p\phi$; they are thus
all redundant. The field transformation which eliminates them is
linear in $\phi$;
\eqn\erxiix{{\cal F}[\phi(x);\Lambda] =
\half c_0^rg^r\Lambda^{2-2p}K_{\Lambda}(-\partial^2)
(\partial^2)^{p-1}\phi(x).}
This reduces the level of redundancy at order $r$ by two; the Jacobian
is a field--independent constant, and may thus be ignored.
As noted in \S 4.1,
the amputated Green's functions only remain unchanged on--shell under
a linear field redefinition if ${\cal F}$ vanishes on--shell;
\erxiix\ clearly does not. However, this is of no consequence for the
S--matrix, since the extra factors (here of
$1+\half c_0^rg^r(m^2/\Lambda^2)^{p-1}$) multiplying the on--shell
Green's functions may be absorbed by a finite renormalization of the
wave--function.

Instead of \erxiix\ we could, as in \rG,
consider redefinitions of the form
\eqn\erxx{{\cal F}[\phi(x),\Lambda] =-\half c_0^rg^r\Lambda^{2-2p}
K_{\Lambda}(-\partial^2)(-\partial^2+ m^2)(\partial^2)^{p-2}\phi(x).}
This now vanishes on--shell, so the redundant term is eliminated
only at the expense of changing terms with a
lower level of redundancy or at higher order in $g$. Clearly all
quadratic redundancies with $p\geq 2$  may be eliminated in this way.
This leaves only redundant terms of the form $c(g)\phi\,\partial^2\phi$.
However it is precisely such terms which, together with the free action,
specify the wavefunction renormalization; they may be readily
eliminated by including them in the free action and renormalizing the
field so that the residue of the pole in the propagator remains unity.
As mentioned at the beginning of this section, the
$\phi(x)\partial^2\phi(x)$ term in the interaction Lagrangian is a
rather special example of a redundant term.

To summarize, at given order in $g$, by
making repeated field redefinitions and working downwards in order of
redundancy we can remove all the redundant
terms in the Lagrangian, up to any order in the dimensionality of
the operators, without introducing any redundant terms at lower orders
in $g$ and without changing the S--matrix.
Therefore, once we have removed all the redundant terms up to
order $r$ in $g$, we can remove all the redundant terms up to order
$r+1$ in $g$. Since the interaction Lagrangian vanishes
at zeroth order in $g$, by induction we can remove all
redundant terms, to a given dimensionality, from the Lagrangian
to any order in $g$.

We close this section with the observation that, up to any finite
order in its Taylor expansion, the regulating function
$K_\Lambda(p^2)$ is itself redundant. To prove this we
first use the equivalence theorem to
rewrite the regularized action \classact\ in a new way by making
the simple field redefinition
$\phi_p\to[K_\Lambda(p^2)]^{1/2}\phi_p$. It is not
difficult to see that this satisfies all the conditions for the
equivalence theorem.\footsy{Indeed it was in order to ensure
that this redefinition is invertible that we chose to make $1/K(z)$
regular.} Moreover it has the useful property of putting
all the dependence on $K_\Lambda$ into the vertex functions; after the
transformation
\eqn\enewclassact{S[\phi;\Lambda]
= \half\big(\phi,P\inv_{\infty}\phi\big)
+ S_{\rm int}[K_{\Lambda}^{1/2}\phi;\Lambda].}
This representation of the action is peculiar in the sense that in it
the flow equation \eiix\ becomes trivial; $\frac{\partial S_{\rm
int}}{\partial\Lambda}=0$. It is thus useless for
proving renormalizability but, as we will see in \S 5.3, very
convenient for proving the unitarity of the effective theory.
The effective theory will then be simultaneously
renormalizable and unitary because the equivalence theorem tells us
that both representations \classact\ and \enewclassact\ yield the
same S--matrix. Indeed the
representation \enewclassact\ should perhaps be christened the `unitary
representation', in close analogy with the `unitary gauge' discussed in
\refs{\rdiagrammar,\rxx}.

Another useful consequence of this form of the action is that we
can use it to loosen
the condition on the rate of growth of the bare vertex functions given
in \S 3.1; all Feynman diagrams will be finite provided only that
$\prod_{i=1}^{2m}[K_\Lambda(p_i)]^{1/2}V^r_{2m}(p_1,\ldots,p_{2m};\Lambda)$
falls faster than any power of $p_i^2$ as $p_i^2\to\infty$, which
means when combined with its regularity that this quantity must have an
essential singularity at the point at infinity, and then be uniformly bounded.

Now consider a new regulating function
$\tilde{K}_\Lambda(p)=T_\Lambda(p)K_\Lambda(p)$. In perturbation theory
we may approximate $T_\Lambda(p)$ by its Taylor expansion; then
$S_{\rm int}[\tilde{K}_{\Lambda}^{1/2}\phi;\Lambda]$ differs from
$S_{\rm int}[K_{\Lambda}^{1/2}\phi;\Lambda]$ only in redundant terms
which may all be removed by field redefinitions as shown above.\foot{The
Jacobians of such redefinitions are still well defined, even though there
is now no factor of $K^{-1}_{\Lambda}$ in the inverse propagator, because
we need an extra factor of $K^{1/2}_{\Lambda}(p)$ in the field transformation
since there is now a factor of this type in the redundant vertex.} Thus the
only truly essential feature of the regulating function is the order
of the essential singularity at the point at infinity, and of course
this is entirely unobservable. Since different regulating functions
correspond to different `renormalization schemes', this amounts to a
proof that S--matrix elements are `scheme independent'.

\newsec{Consistency.}

In the conventional theory of \S 2 the only non--zero bare
interaction vertex was the marginal $\phi^4$ vertex. However, it is
possible, as we showed in \S 3, to set up many different
quasi--local bare interaction Lagrangians which contain irrelevant
vertices satisfying
the finiteness, analyticity and naturalness conditions stated at
the start of \S 3.1, and still obtain a renormalizable theory.
In \S 4.3 above we found that many of these new vertices are
redundant, since they may be eliminated by field redefinitions which
leave S--matrix elements unchanged. However there are still an
infinite set of vertices which are not redundant, and, as we showed in
\S 3.2, these may be used to systematically improve the accuracy
of the theory. In this section we consider the unitarity, analyticity,
and causality of perturbative S--matrix elements in such theories.
Since we assume that our bare vertices are analytic in the
momentum invariants, it may seem from the arguments in \S 2.4
that this is relatively straightforward. Before we can do this
however, we must first ask whether the S--matrix actually exists.

The bare classical action for our theory, \classact, is intrinsically
quasi--local on the naturalness scale $\Lambda_0$, and in
particular must involve terms where more than one time derivative acts on a
particular field, even after integrations by parts and the elimination
of redundant terms. It is well--known \refs{\rpu,\rbs}\ that such actions in
general have unstable solutions to their equations of motion.
Such solutions may be
interpreted as tachyons, or complex poles, or poles with negative
residue, but as emphasized in ref.\riv\ this only serves to obscure
the underlying problem. When the number of time derivatives becomes
infinite the initial value problem can break down altogether,
since an arbitrarily
large amount of initial data is then required to determine the subsequent
solution. Thus, for an arbitrarily selected classical action
containing our nonlocal propagator the `in' and `out' states necessary for
the construction of the S--matrix may no longer exist.

Of course one may still construct a well--behaved perturbation series
about some perturbatively stable solution of the field equations, and
attempt thereby to
formally construct a perturbative S--matrix.
Indeed the classical equations of motion for the free theory in the
two representations \classact\ and \enewclassact\ are simply
$$
{(p^2+m^2)[K_\Lambda(p^2)]\inv\phi_\Lambda(p)=0,\qquad\qquad
(p^2+m^2)\phi(p)=0.}
$$
The second of these equations is of course the usual Klein--Gordon
equation, with solutions $\phi^0=0$ (the vacuum), and
$\phi^0=e^{ip\cdot x}$, $p^2=-m^2$ (the one--particle states). The first
equation has an equivalent set of solutions $\phi^0_\Lambda$ since both
$K_\Lambda(p)$ and $1/K_\Lambda(p)$ are regular, so we
have a one--to--one correspondence $\phi^0_\Lambda=K_\Lambda\phi^0$.
The free inhomogeneous equations are even simpler, since if the source
is introduced as in \exiii, the condition \exiv\ means that any
factors of $K_\Lambda$ cancel; if we choose $Q(p)=P_\infty\inv(p)$,
the inhomogeneous equations of motion for the free theory are just
$(p^2+m^2)\phi_p=-J_p$ for all $\Lambda$.

This argument cannot be regarded as entirely satisfactory, however, since
the `free' external states should really be determined not
perturbatively but by solving the full effective
field equations (which include all interactions of the particle
with itself). In any attempt to go beyond
perturbation theory (for example by considering solutions to the
full classical equations of motion, or some approximation to the
effective field equations)
the unstable solutions will then generally reappear.
The situation is similar to conventional local scalar
field theory in six dimensions with only a $\phi^3$ interaction
(or as was shown in \riv, to open string field theory). Ideally we
would like to be able to construct a manifestly stable quasi--local bare
classical action which has a unique propagating solution to
its equations of motion. Perturbation theory about this solution
should then yield a theory of a single scalar particle in one--to--one
correspondence with the single scalar field.

Fortunately, as we will demonstrate in \S 5.2 below, there is at
least one way in which this can be achieved. First however we will
consider the evolution of solutions to the equations of motion as
dictated by the exact renormalization group equation.

\subsec{The Effective Field Equations.}

Here we consider the behaviour of generating functionals for
one--particle--irreducible, or proper, diagrams\foot{We refer to these as
`proper generating functionals' rather than `effective
actions'; for us the effective action is always $S[\phi,\Lambda]$.}
within the context of effective Lagrangian flows.
This is a worthwhile exercise in itself since it shows that, while the proper
generating functional $\Gamma[\phi,\Lambda]$ calculated using the
effective action defined at scale $\Lambda$ does not evolve
with $\Lambda$ in such a simple way (\exxii) as
the generating functional for connected Green's functions
$W[J,\Lambda]$, the solutions to the effective field equations
are preserved as we change $\Lambda$.

We define $\Gamma[\phi,\Lambda]$ by the usual Legendre transformation
\eqn\ecxxiix{\Gamma[\bar{\phi}_\Lambda;\Lambda]=W[J_\Lambda;\Lambda]
-(J_\Lambda,\bar\phi_\Lambda),
\qquad{\delta W[J_\Lambda;\Lambda]\over\delta J_\Lambda}
=\bar\phi_\Lambda,\qquad
{\delta\Gamma[\bar\phi_\Lambda;\Lambda]\over\delta\bar\phi_\Lambda}
=-J_\Lambda,}
where $\bar\phi_{\Lambda}=\big\langle\phi\rangle_{\!\Lambda}$ is
the expectation value of the field calculated at the scale $\Lambda$
in the presence of the source $J_\Lambda$. We may also define a
$\Lambda$-independent proper generating functional
$\tilde\Gamma[\phi]$ by Legendre
transformation of $\tilde{W}[J]$. Since $\tilde{W}[J]$ generates amputated
connected diagrams, we include a factor of the bare inverse propagator
$P\inv _\infty$ in the Legendre transform to ensure that
$\tilde\Gamma[\phi]$ still generates proper diagrams (though these
are rather unusual proper diagrams since they may be constructed from
the $\tilde G^c_n(p_1,\ldots,p_n)$, and (cf. \exxiv) $\tilde
G^c_2(p,-p)$ is not actually the inverse propagator);
\eqn\ecxxix{\tilde\Gamma[\bar{\phi}]=\tilde{W}[J]
-(J,P\inv _\infty \bar\phi),\qquad{\delta\tilde{W}[J]\over\delta J}
=P\inv _\infty \bar\phi,\qquad
{\delta\tilde\Gamma[\bar\phi]\over\delta\bar\phi}=-P\inv _\infty J.}
If we now take $J_\Lambda=P_\Lambda\inv J$, we may use \exxii\ and
\exvii\ to relate $\Gamma[\phi,\Lambda]$ to $\tilde\Gamma[\phi]$;
\eqn\ecxxx{\tilde\Gamma[\bar\phi]=\Gamma[\bar\phi_{\Lambda};\Lambda]
+\big(J,P\inv _\Lambda \bar\phi_\Lambda -P\inv _\infty \bar\phi\big)
+\half\big(J,P\inv _\Lambda J\big)+{\cal E}_\Gamma(\Lambda),}
where ${\cal E}_\Gamma(\Lambda)$ is just a field and source
independent function of $\Lambda$. The `classical' fields are related by
\eqn\eeff{\bar\phi_\Lambda=K_\Lambda \bar\phi - J,}
where we have used the definition \exxix\ of the regulating function.
Using eqn.\eeff\ in \ecxxx\ we then find
\eqn\ecxxxvi{\tilde\Gamma[\bar\phi]
=\Gamma[K_\Lambda \bar\phi - J;\Lambda] -\half\big(J,P_\Lambda\inv J\big)
+{\cal E}_\Gamma(\Lambda),}
where $J$ may be related to $\Gamma$ using the effective
field equations
\eqn\eeffe{K_\Lambda
{\delta\Gamma[\bar\phi_\Lambda;\Lambda]\over\delta\bar\phi_\Lambda}
={\delta\tilde\Gamma[\bar\phi]\over\delta\bar\phi}=-P_\infty\inv J.}
The relationship \ecxxxvi\ between
$\tilde\Gamma[\phi]$ and $\Gamma [\phi;\Lambda]$ is not particularly
useful (at least in comparison to the relationship \exxii\ between the
connected Green's functions), essentially because the classical fields
$\bar\phi$ and $\bar\phi_\Lambda$ are themselves implicit functions of
the source $J$.

However, we can now see that there is a simple relationship between
the solutions to the effective field equations, \eeffe\ with $J=0$;
\eqn\eefffeqn{
{\delta\Gamma[\varphi_\Lambda;\Lambda]\over\delta\varphi_\Lambda}
={\delta\tilde\Gamma[\varphi]\over\delta\varphi}=0.}
Denoting these solutions by
$\varphi_\Lambda^0$ and $\varphi^0$ respectively,  we see from \eeff\ that
\eqn\ekkd{\varphi_\Lambda^0=K_\Lambda\varphi^0.}
Just as in the free classical theory, the regularity of both
$K_\Lambda(p)$ and $1/K_\Lambda(p)$ ensures that we
have a one--to--one correspondence between the solutions of
the invariant effective field equation and
those for the effective field equation calculated using the
effective action at $\Lambda$.\foot{Also, at the solution we
see from \ecxxxvi\ that
$\tilde\Gamma[\varphi^0]=\Gamma[\varphi_\Lambda^0;\Lambda]$; the
value of the proper generating functional at the solution is
unchanged by the flow.} This means that the latter are stable
to variations in $\Lambda$; they evolve in a predictable well--defined
way, with no new solutions being generated by the evolution. On the
mass shell $K_\Lambda(-m^2)=1$ for all $\Lambda$, and the two
solutions are always equal. This shows that the external line wave
functions constructed from the on--shell solution are
$\Lambda$-independent, and in fact equal to those of the unregulated
theory (let $\Lambda\rightarrow\infty$).

At given $\Lambda$ we may calculate the
proper Green's functions merely by evaluating vacuum diagrams in the
presence of a background field $\varphi$, i.e we let $\phi \rightarrow
\phi +\varphi$. In particular
this means that if we expand $\Gamma[\varphi,\Lambda]$ in a loop
expansion (which is equivalent to an expansion in powers of $\hbar$
provided that we ignore the implicit $\hbar$-dependence in the
parameterization of the effective action),
\eqn\eloopexp{\Gamma[\varphi,\Lambda]=S[\varphi,\Lambda]+O(\hbar).}
Thus provided the quantum corrections are genuinely `small', the
solutions $\varphi_{\Lambda}^0$ to the effective field equations are also in
one--to--one correspondence with the solutions to the classical
equations of motion at the scale $\Lambda$;
\eqn\eclassem{\frac{\delta S[\varphi_\Lambda,\Lambda]}
{\delta\varphi_\Lambda}=0,}
just as is in a conventional quantum field theory.
\nref\rSim{J.~Z.~Simon, \PR\vyp{D43}{1990}{3308}.}\foot
{It has been shown in \rSim\ that when
considering a truncated loop expansion of the form \eloopexp\
there may be spurious solutions to the equations of motion which are
not expandable in powers of $\hbar$. However the specific examples
constructed in this paper suggest that, as one might expect, such
solutions have nothing to do with nonperturbative solutions to the
full effective field equations, and should thus be ignored.}

Of course, there may
be situations in which, for a particular range of scales (and in
particular at $\Lambda_0$), the quantum corrections cannot be ignored
because they destabilize the vacuum, or lead to the formation of
bound states. When $\Lambda$ is very small this is not possible,
however, since then the loops are suppressed by the cutoff:
when calculating loops all momenta are taken in the Euclidean region
so $K_0$, $P_0$, and thus all the loops, vanish identically. Of course
they then remain zero even when the external lines are continued to
the boundary of the physical region, so in the
limit $\Lambda\rightarrow 0$ the proper generating functional\footsy{
By which we mean $\Gamma$; the relationship between $\tilde\Gamma$
and $\Gamma$ is rather delicate in this limit, due to the singular
nature \ereglim\ of
the regulating function away from the Euclidean region.
In practice we only use $\tilde\Gamma$ to
establish the simple evolution \ekkd\ of $\bar \phi_{\Lambda}^0$.} and the
effective action actually coincide; combining with \exxiix\ we thus have
\eqn\ezerocutlim{
\lim_{\Lambda\to 0}[\Gamma[\varphi,\Lambda]
-\half(\varphi,P_\Lambda\inv\varphi)]
=S_{\rm int}[\varphi,0]=\tilde{W}[\varphi]-{\cal E}_W(0).}

Now from eqn.\ezerocutlim\ we can see
that in the limit $\Lambda\rightarrow 0$ the solutions of the effective field
equations \eefffeqn\ must coincide with those of the classical equations of
motion \eclassem. If the quantum corrections are indeed large enough to
destabilize the bare classical action or lead to bound states being formed,
there must thus be a critical scale
$0<\Lambda_{\rm crit}<\Lambda_0$ at which the solutions to \eclassem\ change
discontinuously. In this paper we tacitly assume that no such
transition takes place, since if it did then below $\Lambda_{\rm
crit}$ perturbation theory would have to be reformulated to take
account of the new solutions.

\subsec{Classical Stability.}

A particular example of a quasi--local bare interaction Lagrangian
which satisfies the necessary conditions for boundedness and
convergence, yet has only stable solutions to its equations of motion
may be obtained using the classical part of the exact renormalization
group equation \eiix. As we noted in \S 1.2 the first term in
this equation corresponds to a tree graph, the second to a single loop
(see \frge); the second is thus of order $\hbar$ with respect to the
first (as may be seen formally by the simple substitution
$P_\Lambda\to\hbar P_\Lambda$, $S_{\rm int}\to\hbar\inv S_{\rm int}$),
and should thus be dropped in the classical limit $\hbar\to 0$ to give
\eqn\eCA{{\partial S^{\rm cl}_{\rm int}\over\partial\Lambda}
={1\over 2}\int{d^4p\over(2\pi)^4}{\partial P_{\Lambda}\over\partial\Lambda}
{\delta S^{\rm cl}_{\rm int}\over\delta\phi_p}
{\delta S^{\rm cl}_{\rm int}\over\delta\phi_{-p}}.}
We use this equation to evolve a primordial classical action
\eqn\eSinfty{S^\infty[\phi]=\half\big(\phi,P_\infty\inv \phi\big)
+S^\infty_{\rm int}[\phi],}
from infinity down to $\Lambda_0$ to give a bare regularized action
$S[\phi;\Lambda_0]$ of the required form \classact\ with
$S_{\rm int}[\phi,\Lambda_0]
=S^{\rm cl}_{\rm int}[\phi,\Lambda_0]$.
Were we to use the full quantum renormalization group equation \eiix\
the loop integrals would diverge and $S[\phi;\Lambda_0]$ would be
infinite, but using the classical evolution equation \eCA\ means that
these loops do not form, and $S[\phi;\Lambda_0]$ is well-defined.
In fact, using the classical limit of the formal solution \eix\ we
see that we have a simple
algorithm for constructing $S[\phi;\Lambda_0]$ from an unregularized but
stable primordial action: we simply construct all tree diagrams using the
vertices from the primordial action and the `propagator'
\eqn\eldef{(P_{\infty}-P_{\Lambda_0})=P_\infty(1-K_{\Lambda_0})\equiv
L_{\Lambda_0}.}
{}From the properties of $K(x)$ defined in
\kdef, we see that $L_\Lambda(p)=\Lambda^{-2}
L\big((p^2+m^2)/\Lambda^2\big)$,
where $L(x)=(1-K(x))/x$. $L(x)$ is thus a function with the
same convergence properties as $K(x)$, while its analytic continuation
$L(z)$ is regular, with an essential singularity at the point at
infinity of the same order as that of $K(z)$.

We now consider a primordial action $S^\infty[\phi]$ which is Lorentz
and $Z_2$ invariant, either
local or quasi--local (which means that the interaction vertices are
regular), and which has vertices natural with respect to the
scale $\Lambda_0$, satisfying the bounding requirements \elxxix, and
the technical ordering requirement. Then it is not difficult to see
that the regularized action $S[\phi,\Lambda_0]$ will also be
quasi--local (because $L(z)$ is regular), natural, bounded and ordered.
Moreover, the solutions to the classical equations
of motion of the quasi--local bare action
$S[\phi;\Lambda_0]$ are in one--to--one correspondence with those of
the primordial action $S^\infty[\phi]$, so in particular if the latter
is stable, so is the former.

It is easy to prove this latter result by exploiting the similar result
obtained in the previous section for the solutions to the effective
field equations \eefffeqn. Taking the classical limit of the loop
expansion \eloopexp, to correspond to the classical limit \eCA\ of
the exact renormalization group equation, and noting that
$S^\infty[\phi]$ is independent of $\Lambda$, \eefffeqn\ becomes
\eqn\ekkc{{\delta S^\infty[\phi] \over \delta\phi}=
{\delta S^{\rm cl}[K_{\Lambda}\phi;\Lambda] \over \delta\phi_{\Lambda}}
={\delta S[K_{\Lambda_0}\phi;\Lambda_0] \over \delta\phi_{\Lambda_0}}=0.}
The solutions $\phi^0$, $\phi_\Lambda^0$ and $\phi_{\Lambda_0}^0$ to
the classical equations of motion of the primordial, running
classical, and bare actions respectively are thus related by (cf \ekkd)
\eqn\ekkf{ \phi^0=K_\Lambda\inv \phi^0_{\Lambda}
=K_{\Lambda_0}\inv \phi^0_{\Lambda_0},}
which establishes the correspondence. The theories obtained by
classical evolution of a stable primordial action
will therefore be free from the nonperturbative problems due to
unstable solutions; the field equations for the quasi--local action
$S[\phi;\Lambda_0]$ have no such solutions.

Another method for obtaining a regularized, quasi--local
classical action from the conventional local classical
action (for which the interaction Lagrangian contains only
a single operator, $\phi^4$), while maintaining the solutions to the classical
equations of motion in one--to--one correspondence with the
solutions for the local action, was developed recently by Woodard and
Kleppe \rv. Their `nonlocal regularization' is constructed by
introducing an auxiliary field with `propagator' \eldef; the trees
described in the algorithm following \eSinfty\ are then formed when
the auxiliary field is eliminated using its equation of motion. Our
procedure is thus exactly equivalent to theirs if we restrict the
primordial classical action \eSinfty\ to be local (this is proven
formally in appendix C).
However we can now see how to construct a much wider class of stable
quasi--local regularized actions by considering more general primordial
actions than just the conventional local one; we can choose any action
containing an arbitrary number of irrelevant terms satisfying the
conditions outlined above, which contains no terms with more than one
derivative acting on each field, and which has only stable solutions
to its equations of motion. In practice this is best achieved by
constructing a quasi--local Hamiltonian ${\cal H}[\phi,\pi_\mu]$
which is Lorentz
and $Z_2$ invariant (which means it actually depends on $\phi^2$ and
$\pi_\mu\pi_\mu$), regular in both $\phi$ and $\pi_\mu$, bounded below
with a single minimum when $\pi_\mu=0$ (which we take to be at
$\phi=0$, so that the $Z_2$ symmetry is unbroken); the Hilbert space
may then be constructed in the usual way, while the primordial
Lagrangian is obtained by Legendre transformation. Quasi--local
primordial Lagrangians
obtained in this way will in general contain all operators constructed
from $\phi^2$ and $(\partial_\mu\phi)^2$, with the couplings
unconstrained beyond the basic stability requirement.

Now the evolution procedure
which generates the regularized classical action does not introduce any new
parameters; it simply creates new interactions at higher orders
in $g$ with coupling constants expressed in terms of the original
coupling constants of the primordial action. Thus the resulting
regularized action has only the same number of independent parameters as
the primordial action.\foot{Discounting parameters in the regulating
function since to any finite order in its Taylor expansion these are
redundant.} So to a given order in the canonical
dimension of the operators, the number of independent couplings is
identical to the number
we could have in a consistent classical field theory.
Indeed, if we were to work entirely at tree level, suppressing all
quantum corrections so that $S_{\rm int}[\phi,\Lambda]
=S^{\rm cl}_{\rm int}[\phi,\Lambda]$ for $\Lambda <\Lambda_0$, then
since the exact renormalization group equation \eCA\ guarantees that
the connected amputated Green's functions are
$\Lambda$-independent (except for the trivial correction \exxiv\ to the two
point function which in any case vanishes on shell), the
tree level S--matrix elements calculated using the
running classical action for any $\Lambda$ must be identical to
those calculated directly from the primordial action $S^\infty[\phi]$.

Despite its generality this construction is still unnecessarily
restrictive; although some operators with more than one derivative
on each field (we call these `higher--derivative' operators) are generated
by the classical evolution, their couplings must at present be specified
entirely in terms of those of the operators appearing in the
primordial action, while others are simply not generated
at all. For example while at dimension six there are no higher--derivative
non--redundant operators, at dimension eight there is one,
$\phi^2(\partial_\mu\partial_\nu\phi)^2$,
while at dimension ten there are three,
$\{\phi^4(\partial_\mu\partial_\nu\phi)^2,
(\partial_\mu\phi)^2(\partial_\alpha\partial_\beta\phi)^2,
\phi^2(\partial_\alpha\partial_\beta\partial_\gamma\phi)^2\}$. It is
not difficult to see, by constructing trees, that the construction
as described above generates none of these operators except for
$\phi^4(\partial_\mu\partial_\nu\phi)^2$, essentially because each
vertex involves an even number of lines.

This rather undesirable restriction may be lifted by quantizing the
theory adiabatically over a range of scales of order $\Lambda_0$,
rather than suddenly at $\Lambda_0$. This may be formulated rather
naturally in terms of the generalized evolution equation
\eqn\eevolgen
{{\partial S_{\rm int} \over \partial \Lambda} = {1 \over
2} \int {d^4 p \over (2 \pi )^4}{\partial P_{\Lambda} \over
\partial\Lambda} \biggl[{\delta S_{\rm int} \over \delta \phi_p}
{\delta S_{\rm int}\over\delta \phi_{-p}} -
h(\Lambda/\Lambda_0){\delta^2 S_{\rm int}\over\delta\phi_p
\delta \phi_{-p}} \biggr].}
The discussion above may then be summarized by saying that we evolve
according to \eevolgen,\footsy{Strictly speaking we can then only use the
effective action with $\Lambda \leq \Lambda_0$ to calculate Green's
functions. Of course we can still take the regularization scale higher
than $\Lambda_0$ by evolving the action defined at $\Lambda_0$ upwards using
the full evolution equation with $h=1$; such an action would not in
practice be very useful however, since when calculating Green's
functions there would be large cancellations between contributions
from unnaturally large vertices.} with the boundary condition
$S_{\rm int}[\phi,\Lambda]\Lam2inf S^\infty[\phi]$ and
with the dimensionless function $h(x)=\Theta(-x)$. More generally
however we may take
\eqn\eadiabatic{h(x)=\cases{1,&if $x<1$;\cr
                            0,&if $x>1+\alpha$;\cr}}
while for $x\in[1,1+\alpha]$, $h(x)$ interpolates
smoothly between $1$ and $0$. The parameter $\alpha$ may be any
positive real number of order one. As long as $\alpha$ is not too
large the quantum corrections generated during the evolution from
$(1+\alpha)\Lambda_0$ down to $\Lambda_0$ will not corrupt the
naturalness of the bare action $S[\phi,\Lambda_0]$,
nor upset its stability. They will however make contributions to all
possible terms in the action (subject of course to the global
symmetries, which are still preserved by \eevolgen), and in particular
generate the terms which were not generated by the classical evolution
alone. The couplings in $S[\phi,\Lambda_0]$ will now depend not only
on those in the primordial action $S^\infty[\phi]$ but also (in a
rather complicated way) on the particular choice of function $h(x)$,
and it is not difficult to see that by varying $h(x)$ we can loosen
the tight constraints inherent in $S^{\rm cl}[\phi,\Lambda_0]$.\foot{
Although all the new couplings of the higher derivative operators will
necessarily have expansions beginning at first order in $\hbar$ and
order $m$ or more in $g$; the couplings generated by the classical
evolution begin at order $m-1$.}
The couplings can still by no
means be chosen entirely arbitrarily however; just as the coupling
in conventional $\phi^4$ theory must be positive, here we are restricted to a
highly non--trivial subspace of the (infinite dimensional) space of couplings.

Note that we have not been able to show that the stable effective field
theories constructed by evolution from stable primordial actions
are actually complete, in the sense that all
stable regularized bare actions can be constructed in this way.
To do this it would be necessary to
show that there always exists some choice of $h(x)$ such
that when the stable action $S[\phi,\Lambda_0]$ is evolved up to
infinity using \eevolgen, all terms with more than
two derivatives acting on a single
field will disappear. What we have instead is an existence proof for a rather
large subset of the total set of possible stable regularized
actions. In \S 5.4 below we will show that equivalence classes of
these actions are perturbatively in one--to--one correspondence
with all possible
S--matrices consistent with the basic postulates of S--matrix theory.

We conclude this section by summarizing the sequence of formal steps
(shown schematically in \fig\fstabact{Constructing a stable field
theory: the horizontal axis is the value of a typical vertex function,
the vertical axis the scale $\Lambda$; the open arrows show
the evolution of the effective vertex as $\Lambda$ is reduced from
infinity to zero, while the solid arrows show the evolution of the
corresponding amputated connected Euclidean Green's function with which it
coincides at $\Lambda=0$. The dotted lines show the nonadiabatic
construction in which the theory is quantized instantaneously at
the naturalness scale $\Lambda_0$. The
numbers on the right hand side refer to the different stages described
in the text.}) by which
we can construct a stable effective quantum field theory:
\item{1.} We construct a stable Hamiltonian, and thus a stable
quasi--local primordial action \eSinfty, which will in general
be an infinite sum of terms each of which is an invariant product of
fields and their first derivatives. The irrelevant coupling constants are
chosen to be natural with respect to the cut-off scale $\Lambda_0$,
while the relevant coupling constants (which may include couplings
which while canonically irrelevant are still physically relevant, as
explained in \S 3.2) are left undetermined; they will eventually become
determined by the physical renormalization conditions.
\item{2.} We use the flow equation \eevolgen\ to run down from
infinity to $(1+\alpha)\Lambda_0$; this gives us a stable regularized
quasi--local classical action $S[\phi,(1+\alpha)\Lambda_0]$. In
practice this is best done perturbatively for a truncated subset of
interactions; the coupling constants are written as natural power
series in $g$, beginning at order $m-1$ for a vertex with $2m$ legs,
and the evolution performed consistently to a given order.
\item{3.} We prepare this regularized classical action for
quantum evolution below $(1+\alpha)\Lambda_0$, in perturbation theory about the
stable solution $\phi^0$ of the classical field equation. In general
this will mean that we truncate the action by removing the entire
infinite series of irrelevant interactions beyond a given order in
$g$. This truncation will only change perturbative Green's
functions by an amount suppressed by powers of
$\Lambda_R^{D-2}/\Lambda_0^{D-2}$ (lemma~4). It may give rise to extra
unphysical
(and in general unstable) solutions to the equations of motion; but in
perturbation theory these spurious solutions may be consistently
ignored.
\item{4.} We use the flow equation \eevolgen\ to evolve perturbatively from
$(1+\alpha)\Lambda_0$ down to $\Lambda_0$.
For each choice of $h(x)$ this gives us a
regularized quasi--local classical action of the general form
\classact\ which is stable apart from the spurious solutions
introduced by the truncation described above.
\item{5.} We remove all redundant operators to some finite
order in irrelevancy by means of field redefinitions, as described
in \S 4.3. This will change the off--shell Green's functions but
not the S--matrix; it may introduce further spurious instabilities.
The theory is now totally defined in perturbation theory except for
the relevant coupling constants.
\item{6.} We use the full exact renormalization group equation \eiix\ to
run down to $\Lambda=0$. If the quantum corrections become large enough
to destabilize the classical action at some $\Lambda_{\rm crit}$, the
theory must be reformulated; we will assume here that the
couplings are chosen such that this does not happen.
\item{7.} At $\Lambda=0$
the vertices of the effective action are equal to the amputated
connected Euclidean Green's
functions \exvi. The physically relevant renormalization conditions may
thus be rewritten as renormalization conditions on these Green's functions.
To construct S--matrix elements we must analytically continue the
Euclidean Green's functions to the physical region, and
add external wavefunctions (given by the solution $\varphi^0$ of the effective
field equations which corresponds to the original solution $\phi^0$ of the
classical field equations about which we perturbed). The physically
relevant renormalization conditions may be rewritten as conditions on
the on--shell Green's functions, or S--matrix elements.
\medskip
Of course in a practical calculation one does not have to perform
all of these steps in the above sequence. In particular perturbative
calculations may be performed
in just the same way as in a conventional theory; indeed
in \ref\rKWII{G. Kleppe and R.P. Woodard, \AP\vyp{221}{1993}{106}.}
it has been demonstrated by explicit computation to two loops in
$\phi^4$-theory that it is hardly (if at all) more difficult to calculate
with the quasi--local effective theory than with a local theory
with some analytic regulator (a direct comparison being made with
dimensional regularization). The additional terms introduced by
the classical evolution from the primordial to the regularized action
(or in \rKWII\ by the auxiliary field) are taken into account simply by
including extra graphs with lines which never form closed loops, and
correspond to `propagators' $L_{\Lambda_0}$. Furthermore,
if we assume that
$K_{\Lambda_0}$ has an essential singularities of order
one, so that it takes the simple exponential form \exxx, we may
use the proper--time representations
\eqn\eproptim{
\eqalign{P_{\Lambda_0} &= \int_1^{\infty} {d \tau \over \Lambda_0^2}
\exp \biggl[ {-\tau (p^2 +m^2)\over\Lambda_0^2}\biggr],\cr
L_{\Lambda_0} &= \int_0^1 {d \tau\over\Lambda_0^2}
\exp \biggl[ {-\tau (p^2 +m^2)\over\Lambda_0^2}\biggr];\cr}}
both propagators have the same exponential form, but are integrated
over different proper--time regions. Performing the integrals over
internal momenta (which are now all Gaussian), the graph then reduces
to a series of proper--time integrations over some complicated region;
if these integrations had been regulated analytically (for example
using dimensional regularization), then in the limit
$\Lambda_0\to\infty$ they would be precisely those which are
obtained in the conventional treatment. Of course, in an effective
theory we prefer to keep
$\Lambda_0$ fixed, and evaluate the integrals to a given order in
$\Lambda_R^2/\Lambda_0^2$, where $\Lambda_R$ is of the same order as
the external momentum invariants.

\subsec{Unitarity, Causality and Triviality.}

Constructing our theory using the steps outlined above,
we guarantee that the quasi--local regularized action
at $\Lambda_0$ has a unique stable minimum (in practice this minimum
being at $\phi =0$ for the unbroken theory) about which
we can consider small stable fluctuations; the path integrals \eiv\
and \exiii\ are then well defined, at least perturbatively, and
the perturbative Euclidean Green's functions, and thus by analytic
continuation (as discussed in \S 2.4 above) the on--shell
Green's functions and S--matrix elements, will exist
for scattering processes with in--coming states of arbitrary energies E.
We can then see from the results in \S 2 and \S 3 that the perturbative
S--matrix will be bounded (and thus finite, even in the limit
$\Lambda_0/E\to\infty$), convergent (the dependence on $\Lambda_0$ being
suppressed for $E\ll\Lambda_0$ by some power of $E^2/\Lambda_0^2$, the power
being given by lemma~4) and universal (the dependence on the irrelevant
couplings being suppressed in the same way).

The Landau and Cutkovsky rules, and thus perturbative unitarity and causality
of S--matrix elements, follow for the effective theory in just the same way as
for the regularized conventional theory discussed in \S 2.4, basically
because the vertex functions are regular at all scales $\Lambda>0$, and
thus introduce no new singularities over and above those due to the
simple pole in the bare propagator.
A perhaps more intuitive proof of the Cutkovsky rule than that presented in
ref.\refimov\ may be constructed using the `unitary'
representation \enewclassact\ of
the regularized action in which the regulating factors are associated
with the vertices rather than the propagators. The latter thus retain their
unregulated form $P_\infty(p)$, and may be separated into positive and
negative energy pieces in the usual way. Replacing the cut propagators by
their on--shell imaginary parts $2\pi i\Theta (p_0)\delta(p^2+m^2)$, the
delta--functions kill the regulating factors on the adjoining vertices, but
this does not result in new divergences because the cut loops are always
finite (each loop contains two delta functions, one of which takes care of
the integral over $p_0$, while the other restricts the length of $\vec p$;
the remaining integrals are then all angular and thus finite).\foot
{The same is not true if only single propagators are cut, as, for
example, in the Feynman tree theorem; the phase space integrals over
$\vec p$ will then need regularizing, and there will be extra contributions
from the circle at infinity.} Since the vertex functions are regular, they
remain real when some of their arguments are analytically continued
from the Euclidean region to the boundary of the physical region;
the proof of the cutting formula can then go
through using the same arguments as those in \refs{\rvel,\rdiagrammar}. (If a
position space proof is desired the vertices may be Fourier transformed
keeping the all momenta Euclidean; since the order of the
regulating function is at least one and $\Lambda$ is positive the
Euclidean momentum integrals will remain convergent when
$x_4\to ix_0$, and the position--space vertices will be manifestly real.)

We thus have a Lorentz invariant analytic S--matrix which is perturbatively
unitary and causal for
all processes, even those with energy $E$ above $\Lambda_0$, essentially
because our regulating function has no singularities except at the point at
infinity. This is in stark contrast to theories with, for example,
higher derivative or Pauli--Villars regulators, for which the S--matrix
only exists in perturbation theory, and then unitarity and causality are
manifestly violated, not only at the regulator scale, but also at
scales $E\ll\Lambda_0$, albeit by amounts suppressed by powers of
$E/\Lambda_0$. Such theories make sense physically only in the limit
$\Lambda_0/E\to\infty$, and this limit is itself rather delicate because
strictly speaking the S--matrix only exists in the limit. By contrast
in the stable effective theory as
constructed in the previous section we
can keep $\Lambda_0$ finite, and consider what happens as the energy $E$
of the process approaches and then exceeds the naturalness scale $\Lambda_0$.

As we increase $E/m$ we must, as explained in \S 3.2, increase
systematically the number of couplings we regard as physically relevant
in order to maintain the precision of the theory, but in most other respects
we calculate just as we would for the conventional theory in which only the
canonically relevant operators are regarded as physically relevant. The
new couplings must be such that their bare values are natural at the
scale $\Lambda_0$, and furthermore such that there must exist natural
irrelevant couplings such that the theory remains stable; as we saw in
the previous section this latter restriction will not be very difficult to
satisfy (indeed the positivity of the bare coupling $\lambda_0$ of the
conventional theory may now be relaxed a little, depending on the couplings
of the $\phi^{2n}$ terms, $n\ge 3$). A modified form of Weinberg's theorem will
be obeyed up to the naturalness scale (as proven in \rbtir).
However, violations may begin to appear when we
exceed the scale $\Lambda_0$.
The `tree unitarity bounds' on tree--level S--matrix elements \ref\rclt{
J.~M.~Cornwall, D.~N.~Levin and G.~Tiktopoulos \hbox{\PRL\vyp{30}{1973}{1268}};
\hfill\break\hbox{\PR\vyp{D10}{1973}{1145}}\semi
C.~H.~Llewellyn Smith \PL\vyp{46B}{1973}{233}.} will also be violated,
though by only small amounts at low scales, in accordance with the
systematic increase in the number
of physically relevant couplings. However bounds which can be derived
directly from the analyticity and unitarity of the S--matrix, such
as the Froissart bound on the total cross-section, or the
partial--wave unitarity bounds,
will always be satisfied, no matter how high the energy, provided
we work to sufficiently high order in perturbation
theory.\foot{At tree level,
all nontrivial partial wave unitarity bounds are necessarily violated,
while to a given finite order in perturbation theory we must be prepared
for small violations of the order of the next order contribution.}
Ad hoc unitarization procedures are thus no more necessary in
effective theories than they are in conventional local field theory.

When we reach $\Lambda_0$, the number of physically relevant couplings will
become infinite; the theory then depends in general on the complete vertex
functions.\foot{Although the vertex functions at $\Lambda_0$ are regular, and
thus have convergent Taylor series, there is no reason to believe that these
Taylor series converge particularly rapidly.} This is not to say that the
theory is entirely lacking in predictivity however; it is actually equivalent
(as we show in the next section) to S--matrix theory. Of course, it would be
desirable to find extra conditions (for example symmetries, or underlying
more `fundamental' structures) on the vertex functions to increase
the predictivity of the theory, but there is no a priori guarantee
that such conditions actually exist.

As we take $E/\Lambda_0\to\infty$ the behaviour of the theory will depend
on the form of the vertex functions in the deep Euclidean region. The form of
vertices required
for finiteness, as described in \S 4.3, does not restrict the growth
of the vertex functions with $E$, since they may have essential
singularities at the point at infinity at arbitrarily high order
provided the root of the regulating function has a singularity of higher order.
If the vertex functions  rise (exponentially), perturbation theory
will eventually break down, (signaled perhaps by a severe violation of
the Froissart bound or some partial wave unitarity bound), and the
theory will at best require reformulating; at worst it may be inconsistent.
However if they fall (again exponentially), the theory is asymptotically
free in a rather general sense; it may then be perturbative over all
energy scales, and considered as a (very unattractive) candidate for a
`fundamental' theory.
The marginal case seems less interesting, as it can presumably only be
achieved by a lot of fine tuning.

It should be clear from the above discussion that in the effective
theory the `triviality' of the conventional theory \rvi\ is no
longer so significant, simply because the regularization scale
$\Lambda_0$ remains finite. The running four--point coupling
still grows logarithmically at small scales; to one loop
\eqn\elrun{
\lambda(\Lambda)=\lambda_R\big/\big(1-\smallfrac{3\lambda_R}{16\pi^2}
\ln(\Lambda/\Lambda_R)\big)}
for $\Lambda\ll\Lambda_0$, since there the running is determined almost
entirely determined by the canonically relevant couplings, and is
thus given by the usual Gell-Mann--Low equation. However, when
$\Lambda$ becomes a significant fraction of $\Lambda_0$ the running
of the couping will depend more and more on canonically irrelevant
couplings, but since the vertex functions at $\Lambda_0$ are by
construction regular there is no Landau pole; the only singularity is
the essential one at infinity. Above $\Lambda_0$ the four--point
coupling will in general rise or fall exponentially depending on the
form of its essential singularity. If it falls, it will exhibit
a maximum value at around
$\Lambda_0$. Furthermore there is no reason why this maximum value
should not be sufficiently small for perturbation theory to be valid
at all scales; insisting on this puts an upper limit on the
renormalized coupling, which we may (very) loosely estimate from the
one--loop expression \elrun\ as
$[1+\smallfrac{3}{16\pi^2}\ln(\Lambda_0/\Lambda_R)]\inv $.
Nonperturbative studies of triviality bounds for lattice regulated
scalar theory suggest that such a bound may actually remain even if
$\lambda(\Lambda_0)$ is allowed to become arbitrarily large, although
it is difficult to see how such a result could be proven in the present
context. In any case the bound is of no physical interest, since
it is only when $\Lambda_0/\Lambda_R\gsim 10^{23}$ that it acquires
real significance. What we can show however is that in the limit
$\Lambda_0/\Lambda_R\to\infty$ (which is now well defined) the
presence of the irrelevant operators cannot cure the triviality (at
least in perturbation theory). Consider the growth
of the coupling from $\Lambda_R$ up to an intermediate scale
$\Lambda_H$, which is a finite multiple of $\Lambda_R$; then using the
universality proved in
\S 3.1 extended to this higher scale $|\bar \lambda\big(\Lambda_H\big)
-\lambda\big(\Lambda_H\big)|
\leq O\big(\Lambda_H/\Lambda_0\big)$. For a scalar field
theory, the only alternative to the effective theory is a free one.

All the discussion in this section (and indeed so far in this paper)
was based on the phenomenologically neutral assumption that
we have only a single particle pole, with vertex functions which are
regular. Of course it is quite possible that in the real world this
assumption may have broken
down by the time we reach $\Lambda_0$, and new analytic structures,
and in particular new poles describing new particles may be
discovered. These may then be incorporated into a
new more `fundamental' effective theory, the vertices of the old
effective theory (still valid for scales below $\Lambda_0$ if the new
singularities decouple)
being now largely calculable by matching to the new theory at a matching
scale of order $\Lambda_0$. This situation is clearly a rather common
one, and in a future article \ref\rbtdec{R.~D.~Ball
and R.~S.~Thorne, CERN-TH.????/93, OUTP-93-??P.} we will discuss in
some detail the simple paradigm of
two scalar particles, one light and the other heavy.

\subsec{Weinberg's Conjecture.}

The effective field theories we have constructed are extremely
general, and it is tempting to suppose that they are the most
general theories possible consistent with all the various constraints
we have imposed: analyticity in the space of fields, and a small coupling
constant, so that we may construct Fock spaces of `in' and `out'
states; stability and boundedness, so that the S--matrix
actually exists; analyticity in momentum space, so that the S--matrix
is unitary and causal; and then the symmetry constraints, in this case
Lorentz invariance and the $Z_2$ internal symmetry. Indeed it was
conjectured by Weinberg some years ago \riii\ that there is in fact an
equivalence between such theories and all possible S-matrices
consistent with the basic postulates of S--matrix theory\ref
\rsmatrix{R.~J.~Eden, P.~V.~Landshoff, D.~I.~Olive and
J.~C.~Polkinghorne, ``The Analytic S--Matrix'' (C.U.P. 1966)\semi
G.~F.~Chew, ``The Analytic S Matrix'' (Benjamin, New York,
1966)\semi
D.~Iagolnitzer, ``The S Matrix'' (North Holland, 1978).};
cluster decomposition, unitarity,
analyticity\foot{Or more precisely that the transition amplitudes are
the real--boundary values of analytic functions; this is sufficient
for causality, but it is not known whether it is strictly necessary.},
and the assumed symmetries. We now have the machinery to
explore this conjecture in some detail.

It is clear that for every effective field theory we have
constructed there exists a unique S--matrix satisfying the above
properties; indeed proving that this was the case was the sole
purpose of much of the above discussion. Thus, what remains to be
shown is that, given a suitable S--matrix we can reconstruct a unique
effective field theory (or well--defined class of equivalent theories) which
yields this S--matrix when evaluated to all orders in perturbation
theory. Effective field theory and S--matrix theory would then be
equivalent.

By the cluster decomposition postulate the S--matrix is
equivalent to the full set of on--shell amputated connected Green's
functions on the boundary of the physical region. The $Z_2$ symmetry
means that only Green's functions with an even
number of external legs are nonvanishing, while Lorentz invariance
means that each Green's function depends only on the momentum
invariants $\{z_{ij}, i\neq j\}$ in the physical region. The unitarity and
analyticity postulates together with Lorentz invariance guarantee
the Landau and Cutkovsky rules for the position of the singularities
and the discontinuities across the cuts, given that we assume we have
only a single scalar particle of mass $m$. Furthermore, analyticity
allows us to analytically continue the amputated connected Green's functions
to unphysical regions in the variables $\{z_{ij},i\neq j\}$, and in
particular to the Euclidean region.
Since this analytic continuation is unique, the S--matrix is equivalent
to the full set of amputated connected on--shell Green's functions
\hbox{$\{\tilde{G}^c_{2n}(p_1,\ldots,p_{2n}); p_i^2=-m^2,
n=2,3,\ldots\}$}. We may also extend these Green's functions to
off--shell momenta, but this extension will not be unique since
we only know them at the isolated points $p_i^2=-m^2$. The S--matrix
is thus equivalent to the full set of connected amputated
Euclidean Green's functions \hbox{$\{\tilde{G}^c_{2n}(p_1,\ldots,p_{2n});
p_i^2>0, n=2,3,\ldots\}$} or in
other words to the generating functional $\tilde{W}[J]$,
only if we identify Green's functions which are equal when we set
$p_i^2=-m^2$. This is simply the off--shell redundancy discussed in
\S 4.2 above.

We next rewrite the non--local generating functional $\tilde{W}[J]$
in the form \ewpert, with $\Lambda$ infinitesimal and
$S_{\rm int}[\phi,0^+]$ quasi--local. This is possible because of
the essential singularity \ereglim\ in the regulating
function $K_\Lambda(p)$ at $\Lambda=0$. As all loop
momenta are suppressed, $\tilde{G}^c_{2m}$ and $V^r_{2m}(0)$ are both
analytic functions in the Euclidean region and share the same
singularities and discontinuities on the boundary of the physical
region; in fact from \exxiix\ $\tilde{G}^c_{2m}
=\delta^4\big(\sum_{i=1}^{2m}p_i\big)\sum_r g^r V^r_{2m}(0)$. In the
Euclidean region $V^r_{2m}(0)$ differs only infinitesimally from
$V^r_{2m}(0^+)$
since here the propagator $P_{0}(p)$ is free from singularities.
However $V^r_{2m}(0)$ and $V^r_{2m}(0^+)$
will differ on the boundary of the physical region; $V^r_{2m}(0^+)$ will
be analytic (and thus regular) since, as explained following \ereglim,
the singularities and discontinuities of $V^r_{2m}(0)$ arise precisely
from those diagrams in which an appropriate subset of internal lines
have gone on shell, and are thus no longer suppressed. Indeed it was
precisely by examining the singularity structure of individual Feynman
diagrams that the analytic structure of S--matrix elements was
unraveled\refs{\rlandau,\rui,\rsmatrix}. The construction of the
regular vertices $V^r_{2m}(0^+)$ is unique, since they may be obtained by
the analytic continuation of $V^r_{2m}(0)$ away from the Euclidean
region in which, in the limit $\Lambda\to 0$, they coincide.

Having obtained the stable\foot{If it were not stable, the S--matrix
would not exist. A `perturbative S--matrix' has no meaning within
S--matrix theory.} quasi--local effective action
$S_{\rm int}[\phi,0^+]$, we may evolve it upwards using the exact
renormalization group equation \eiix. To do this we must choose a
particular regulating function $K$; if we choose one satisfying the
properties following \kdef\ the evolution is well defined and
preserves the regularity of $S_{\rm int}[\phi,\Lambda]$. We stop the
evolution as soon as the some of the dimensionless vertices become
unnaturally large\nobreak\footsy{Or more properly when there exists no field
reparameterization such that none of the vertices are unnaturally
large; if we consider only one particular parameterization $\Lambda_0$
may be underestimated.} for Euclidean momenta of order $\Lambda$, violating
the bound of lemma~1, or if the action $S_{\rm int}[\phi,\Lambda]$
begins to develop unstable solutions as the quantum corrections are
removed.\foot{We discount the possibility that the action may
destabilize only momentarily, stabilizing again at higher scales;
unstable classical actions which are stabilized by quantum corrections
are of course possible, but only non--perturbatively.} This gives us
a naturalness scale $\Lambda_0$;
by lemma~2 $\Lambda_0\gsim\Lambda_R\sim m$. The
renormalization group thus defines a set of one--to--one mappings
labeled by $K$ from the amputated connected Green's functions to the
bare action $S[\phi,\Lambda_0]$.

Furthermore, as we explained in \S 4.2 and \S 4.3, the redundancy in
the bare interaction $S_{\rm int}[\phi,\Lambda_0]$ is in
one--to--one correspondence with that in the connected amputated Green's
functions. Thus if we construct equivalence classes out of all those
bare actions $S[\phi,\Lambda_0]$ which differ only in the form of
their regulating functions $K$ and in their redundant interactions (so
in other words we consider two actions equivalent if they are equal, up to
integrations by parts, when we put them on--shell by setting
$p^2=-m^2$), we have a one--to--one correspondence between each class
of stable equivalent effective field theories and the S--matrix from which they
have been reconstructed. This establishes Weinberg's equivalence
theorem, at least to all orders in perturbation theory, for the
simple case of a single scalar particle with a $Z_2$ global symmetry.

Of course the equivalent effective field theories will not all be
renormalizable; indeed since the S--matrix elements are dimensionless
functions of $p_i\cdot p_j/m^2$, most of the equivalent effective field
theories will have $\Lambda_0\sim m$, and thus be
`nonrenormalizable' (albeit perfectly consistent with fundamental
principles) in the sense explained in \S 3.3. Only a very small subset
of theories will have their S--matrix elements tuned in such a way
that $\Lambda_0\gg m$.
These `renormalizable' effective theories are very special in the sense that
they are both convergent and universal, as proven in \S 2.3 and
\S 3.1 above; S--matrix elements for processes involving energies
$E\lsim m\ll\Lambda_0$ may then be expressed to a high
level of accuracy in terms
of a finite number of physically relevant couplings. This is the
(only) reason for the general prevalence of such theories; the scale
of the next theory, which is after all what determines
$\Lambda_0$, is often much larger than the masses of the particles in
the current theory, which thus appears `renormalizable'. The relative
rarity in nature of theories involving scalar particles is
then simply a reflection of the unnaturalness and fine tuning problems of light
scalars\nobreak\refs{\rvx,\rxx}. If we were to insist (perhaps
unreasonably) on being able to calculate S--matrix elements precisely,
or at arbitrarily high energies,
in terms of only a finite number of parameters, then in this instance
at least the theory is essentially unique; it consists of a free massive
scalar, with all connected scattering amplitudes identically zero.

We conclude this section with a brief discussion of the possible ways
in which Weinberg's conjecture may fail. Clearly a given
effective field theory contains a lot of information which becomes
redundant when S--matrix elements are computed. To give this redundant
information physical meaning would require a theory of off--shell
particles, since as explained in \S 4.2 the redundancy in the
bare action is equivalent to that in the off--shell Green's functions.
There is, so far as we know, no experimental evidence that such a
theory is needed. Moreover there are many conceptual problems to be
overcome before it could be constructed. First, and perhaps foremost,
the Hilbert space for such a theory would no longer consist of a Fock
space of free $n$-particle states of definite energy and momentum,
since these are by construction on--shell. Instead we would have to
somehow include nonasymptotic mutually interacting off--shell particle
states, which approach the on--shell states asymptotically when
the interactions are turned off, but cannot be written as
linear superpositions of them. In other words we would have to drop
our implicit assumption that the Fock space is asymptotically
complete. We know of no consistent way in which this can be done; on
the contrary it was shown long ago \refs{\rpl,\rdirac}\ that in a
relativistic quantum theory the only experimental observables are
the momenta and internal quantum numbers of free particles.
It seems to us that this
consideration alone is enough to justify Heisenberg's proposition
\ref\rheis{W.~Heisenberg, \ZP\vyp{120}{1943}{513,673}.}
that only the S--matrix has an observable physical meaning.

If we suspend our disbelief for a moment, however, a Pandora's box is
opened. The parameterization of the theory, and indeed in some sense
the fields themselves, assumes a physical significance since the
equivalence theorem proven in \S 4.1 is no longer valid; field
redefinitions change the off--shell `physics'. Indeed the regulating
function itself now assumes physical significance, and only the rather
special actions of the form \enewclassact\ permit free particles which
propagate in the usual way. If the action is of the form \classact\
the existence
of propagating particles depends critically on the order $\sigma$ of the
essential singularity of the regulating function\rpu. If $\sigma$ is odd,
free particles which are off--shell in a timelike direction no longer
propagate but have wave--functions which grow exponentially, so it is
at best necessary to restrict the Hilbert space to states with spacelike
off--shellness\rpu. When $\sigma$ is even, off--shell
propagation is stable, and the on--shell states are approached
asymptotically. However all off--shell propagation is acausal, albeit
only at greater than zeroth order in $\hbar$ and by amounts suppressed by $\exp
-(l^2\Lambda_0^2)^{\sigma}$, where
$l^2$ is the distance from the light--cone. For the action
\enewclassact\ the acausality shows up not in the propagator, but in
the off--shell Green's functions, as can be shown in some detail  by
writing the results of \rv\ in position, rather than momentum, space.
It is also possible to extend the results in \rv\ to our much more general
form of effective action, and it seems as though it is impossible to
eliminate completely the acausal
behaviour using any action generated by the prescription in \S 5.2
(although the amount of acausality may be very significantly reduced
by increasing $\sigma$). Indeed it seems that off--shell
acausality is a necessary feature of any theory which is not strictly
local (as is discussed in \riv\ in the context of string field
theory); the only way to eliminate it completely is to take
$\Lambda_0$ to infinity, resulting in conventional local quantum field
theory with its axiomatic micro--causality.

To us, none of this is particularly relevant; since only on--shell
asymptotic free particle states are directly observable, the
notional acausality
associated with off--shell scattering amplitudes has no meaningful
physical interpretation.\foot{As stated by Dirac\rdirac; {\it
``Causality applies only to a system which is left undisturbed. If a
system is small, we cannot observe it without producing a serious
disturbance and hence we cannot expect to find any causal connection
between the results of our observations.''}} Rather by
relinquishing micro--causality as a fundamental assumption\foot{A
possibility noted at the close of ref.\rpu.} we have
been able to construct a wider class of quantum field theories which are
causal macroscopically, and indeed interpolate continuously between
conventional local quantum field theory and S--matrix theory.

\newsec{Extension to Other Theories.}

The proofs of boundedness, convergence and universality
described in this article can
easily be generalized to show that within perturbation theory all
quantum field theories, with fields of arbitrary spin\foot{The Lorentz
structure being preserved by the flow equations.},
can be defined perturbatively at energies much lower than the
naturalness scale $\Lambda_0$ by a set of low energy renormalization
conditions on the physically relevant coupling constants. If the
dimension of the space--time is sufficiently large, the number of
physically relevant couplings will be finite, and the theory
renormalizable.
Of course we have may also have more than one quantum field and choose to
have several small expansion parameters (as may be desirable if, for example,
the particles' self interactions are not the same strength as their mutual
interactions). As long as we choose the vertices in the interaction
Lagrangian so that they are of at least first order in one of the expansion
parameters and so that, for given order in each of these expansion
parameters, vertices involving a sufficiently large numbers of fields of each
type vanish, then the type of inductive proof presented in \S 2.2, \S 2.3
and \S 3.1 can be carried out successively in each of the expansion
parameters. So in order to prove renormalizability we simply
have to make sure that the boundary conditions on the flow equations
satisfy the same general naturalness and analyticity criteria as
described in \S 3.1.

This is not to say that the S--matrix for such theories will always be
unitary and causal however,
since unitarity may depend on the existence of a symmetry (for
example a gauge symmetry) to remove unphysical modes, and it may
be difficult to preserve this symmetry at each order in
perturbation theory under the renormalization group flow. Furthermore the
spectrum of the renormalized theory may differ from that expected in the bare
theory if hidden or nonlinearly realized symmetries are not respected by
the perturbative flow. We believe that such technical problems can be
overcome. However it may even then be difficult to obtain a wide
separation of the scales $\Lambda_0$ and $m$ if the propagator and
interaction are related by the symmetry; if this is the case
boundedness, convergence and universality will be of limited practical
value. We will postpone discussion of all such effective theories to later
articles in this series; here we will describe briefly the extension
of our results for effective scalar theory in four dimensional
space--time with a $Z_2$ global symmetry realized in
Wigner mode to other effective theories with both scalars and fermions
in space--times of arbitrary dimension, and with linearly realized global
symmetries which also leave the vacuum invariant.

Firstly, consider an effective theory with a single scalar field, but
without the $Z_2$ symmetry. The effective action \actgen\ (or in
momentum space \exxxi) will then be supplemented by all possible
interaction terms with an odd number of fields, and an expansion
parameter $g'$ say. There will also be a linear tadpole term
proportional to $\Lambda^3\phi$, but this is redundant since it may
always be eliminated by a simple field redefinition $\phi\to\phi
+c(\Lambda)\Lambda$, where $c(\Lambda)$ is a number; such a
redefinition does not belong to the class \eredef\efdef, but
guarantees that the field equation \eclassem\ is
satisfied for every $\Lambda$ (including eventually $\Lambda=0$), and
thus that after the redefinition the amputated connected Green's
functions are related to the S--matrix elements in the usual way.
Alternatively we could simply set the tadpole to zero by the
renormalization condition at $\Lambda=0$.\footsy{A more sophisticated
approach would be to modify the exact renormalization group
equation\eiix\ in such a way that once the linear term is eliminated
at $\Lambda_0$ it remains zero\ref\rellwanger
{U.~Ellwanger, \ZP\vyp{C38}{1993}{619}.}; this
results in an extra tadpole term
on the right hand side which must then, albeit rather easily, be
included in the bounding arguments.}

If we separate the interaction $S_{\rm int}\equiv S_{\rm int}^+
+S_{\rm int}^-$, with $S_{\rm int}^\pm$ even and odd respectively
under the $Z_2$ transformation $\phi\to -\phi$, the evolution equation
\eiix\ also separates:
\eqn\evolpm{\eqalign{
{\partial S_{\rm int}^+ \over \partial \Lambda}
&= {1 \over 2} \int {d^4 p \over (2 \pi )^4}
{\partial P_{\Lambda} \over\partial \Lambda}
\biggl[{\delta S_{\rm int}^+ \over \delta \phi_p}
{\delta S_{\rm int}^+ \over \delta \phi_{-p}}
+{\delta S_{\rm int}^- \over \delta \phi_p}
{\delta S_{\rm int}^- \over \delta \phi_{-p}}
- { \delta^2 S_{\rm int}^+ \over \delta\phi_p
\delta \phi_{-p}} \biggr],\cr
{\partial S_{\rm int}^- \over \partial \Lambda}
&= {1 \over 2} \int {d^4 p \over (2 \pi )^4}
{\partial P_{\Lambda} \over\partial \Lambda}
\biggl[2{\delta S_{\rm int}^- \over \delta \phi_p}
{\delta S_{\rm int}^+ \over \delta \phi_{-p}}
- { \delta^2 S_{\rm int}^- \over \delta\phi_p
\delta \phi_{-p}} \biggr].\cr}}
Thus if we were to take $g'\ll g$ at the scale $\Lambda_0$, the
resulting hierarchy is stable under the evolution; the even
vertex functions satisfy all the same bounds as in the $Z_2$ theory
obtained by setting $g'$ to zero, while the odd vertex function obey
similar bounds suppressed by an extra factor of $g'/g$. In this
sense the odd terms constitute a `soft'
breaking of the $Z_2$ symmetry. This means in particular that there
need be
no additional naturalness or fine tuning problem for the three point vertex,
despite the fact that its coupling has canonical dimension one: if we
were to take $g'\sim (m/\Lambda_0)$, the renormalized three--point
coupling will be of order $m$. The odd vertices will then behave under the
evolution as if they were all of one dimension higher than their
canonical dimension. This is rather fortunate since, as may be seen by
examining the effective potential, when $m\ll\Lambda_0$ the vacuum is
only stable if
\eqn\evacstab{g'\lsim (m/\Lambda_0)g^{1/2}.}

Scalar theories in spacetimes of dimension other than four exhibit
other interesting features. In three dimensions the relevant
interactions are $\{\phi^n; n=3,4,5,6\}$; an unnaturally large
$\phi^3$ coupling can only be avoided by taking $g'\lsim
(m/\Lambda_0)^{3/2}$, while to avoid an unnaturally large
$\phi^4$ coupling we also need $g\lsim m/\Lambda_0$. Again these
conditions are also sufficient to satisfy the stability condition
\evacstab\ (which holds for scalar theories in any dimension).

In two dimensions there are in general always an infinite number
of operators of a given canonical dimension because the field itself
is dimensionless; the canonically relevant operators are
$\{\phi^n;n=2,3,\cdots\}$ and
$\{\phi^{n-1}\partial^2\phi;n=2,3,\cdots\}$. The expansion in irrelevancy is in
this case simply a
derivative expansion (just as it would be for a nonlinear
sigma--model). It turns out that we are able to define a $\Lambda_0$
insensitive theory by setting only a finite number of low energy
renormalization conditions. However, this must be accompanied by
unnaturally small boundary conditions at $\Lambda_0$ for all the
relevant coupling constants not determined by these low energy
renormalization conditions. If we consider that the unnaturalness of an
infinite number of couplings is unacceptable then we must set low
energy renormalization conditions on all the relevant coupling
constants, i.e on an infinite number. To reduce the infinite number of
couplings to a finite subset in a natural way it is necessary to
impose some additional symmetry (for example
conformal invariance, or periodicity in the field).

On the other hand as we increase the space--time dimension beyond
four, the number
of interactions at each order in irrelevancy decreases rather rapidly,
since the dimension of the field goes as $\smallfrac{d}{2}-1$.
For $d=5$ or $6$,
$\phi^3$ is the only canonically relevant interaction term. Unlike the
conventional theory, which is unstable, the effective theory may be stable
provided the coupling is not too large (\evacstab\ again). In six
dimensions the coupling $g'$ is asymptotically free\foot
{A slight abuse of the
term since we only know how the coupling behaves at scales below
$\Lambda_0$; cf. the discussion of triviality in \S 5.3.}; at low scales it
therefore grows and the vacuum may eventually become unstable, causing
a phase transition. If the theory has the $Z_2$ symmetry, the most
relevant coupling at low scales is $g\Lambda_0^{-2}\phi^4$ so the
theory is almost free (and in the limit $\Lambda_0/m\to\infty$ would
become `trivial'). Such effective theories are actually rather
interesting since the presence of the scale $\Lambda_0$ is first
signaled by the observation of a very weak interaction.

The moral of this discussion is that what used to be called
`super--renormalizable' theories are if anything less natural
than `non--renormalizable' ones, since they require that more
couplings be fine tuned to ensure the separation of scales
$m\ll\Lambda_0$. Moreover if we (as proposed in \S 3.3)
describe an effective theory as renormalizable only if it can be
used to calculate amplitudes to a given level of precision in terms
of a finite number of physically relevant couplings, then the higher
the dimension of space--time the more renormalizable (i.e. predictive)
the theory becomes; `super--renormalizable' theories are less
renormalizable than `non--renormalizable' theories!

It is trivial to generalize all these results to theories which involve
more than one scalar field, perhaps with various linearly realized
global symmetries (for example linear sigma models), provided all the
particles have similar masses. Scalar particles with widely separated
mass scales will be considered in a later article\rbtdec.
\nobreak

Moving on to theories with fermionic fields, the flow equations respect global
Euclidean invariance and furthermore the statistics of the field has
no effect on the bounding arguments;
although the flow equation \eiix, which now becomes
\eqn\eflowferm{
{\partial S_{\rm int} \over \partial \Lambda}
= {1 \over 2} \int {d^4 p \over (2 \pi )^4}
{\partial P_{\Lambda} \over\partial \Lambda}
\tr\Bigg\lbrace (i\pslash +m)
\biggl[{\delta S_{\rm int}\over\delta\bar\psi_p}
{\delta S_{\rm int}\over \delta\psi_{-p}}
+ { \delta^2 S_{\rm int}\over\delta\bar\psi_p\delta\psi_{-p}}\biggr]
\Bigg\rbrace,}
has an extra minus sign in the second term due to the closed fermion
loop, this disappears in the process of taking bounds. The
power counting is different for fermions, since the inverse propagator
has dimension one, but this causes no real difficulties. Thus,
the proof of renormalizability immediately extends to theories
containing fermions.
In fact in two dimensions, the situation for fermionic theories is very
similar to that for scalars with $Z_2$ symmetry in four dimensions,
except that the four--fermion coupling now appears asymptotically
free at low energies.

Consider instead a purely fermionic theory in a space--time dimension $d>2$.
The fermion field has canonical dimension $\half(d-1)$, so the number
of physically relevant operators at a given order in irrelevancy is
significantly fewer than in a scalar theory. Indeed the only
canonically relevant operators are the
fermion kinetic and mass terms.
Unlike in the scalar theories there is no
fine tuning problem since corrections to the mass term are at most
logarithmic, as can be seen by writing the flow equation for the
two--point vertex for zero external momentum and using the usual
inductive argument. Although the bare mass is still
small (of order $m/\Lambda_0$) at the naturalness scale, this is not
unnatural if it is regarded as a soft breaking of global chiral symmetry.
 The interaction term of lowest
dimension is $\Lambda_0^{2-d}(\bar\psi\Gamma\psi)^2$, where
$\Gamma$ is a Dirac or Weyl matrix.\foot{Not all such interactions are
independent however, since they are related by Fierz transformations;
this is a new form of redundancy for fermionic theories.}
Purely fermionic theories are thus almost free
in the same sense as the scalar theories with $Z_2$ symmetry and $d>4$
discussed above, and as $\Lambda_0/m\to\infty$ they become trivial.
However for the fermionic theories not only is the onset
of new physics indicated by the presence of a new weak interaction,
but the structure of this interaction may be inferred from the form of
the interacting currents. Moreover, for fermions this can happen in
four dimensional space--time. Indeed it has --- the Fermi theory of
weak interactions was such an effective theory, and the $SU(2)\times
U(1)$ electroweak sector of the standard model was constructed
precisely so that at energies far below the electroweak scale the
effective Fermi theory was recovered.\foot{It would be an interesting, if
somewhat academic, exercise to develop the Fermi theory beyond tree
level, including further interactions due ultimately (i.e. in the more
fundamental electroweak theory) to box graphs and Higgs exchange,
and attempting to fix couplings using, besides low energy weak decays,
the more recent data from electron--neutrino scattering and atomic
parity violation.}

We can therefore see an important difference in four dimensions between a
theory containing only bosons and one containing only
fermions. In the former we can only gain insight into the scale of
the new physics by performing very careful scattering experiments
and looking for tiny discrepancies between
the measurements and the results predicted using the conventional
quantum field theory. In the latter, any deviation from the free
theory, i.e. any interaction, however weak, immediately gives us
information about the scale of the new physics. Of course, the same
would be true for a bosonic theory if one of the irrelevant operators led
to the violation of a symmetry (for example CP) respected by all the relevant
operators. However before we can apply our techniques to effective
electroweak theory we must first learn how to incorporate gauge
symmetries, spontaneously broken symmetries and nonlinear symmetries
into the flow equations.

Since $N=1$ global supersymmetry may be linearized by the introduction
of auxiliary fields, it is respected by the flow equations, and we may
construct renormalizable supersymmetric effective theories (the
incorporation of the regulating function is explained in ref.\ref
\rsusyKW{G.~Kleppe and R.~P.~Woodard, \PL\vyp{B253}{1991}{331}.}). It
is easy to see that the auxiliary fields remain auxiliary under the
renormalization group flow by using the same arguments as in
\S 5.1 and \S 5.2. The number of physically relevant coupling constants at
each order in irrelevancy is significantly reduced by the
supersymmetry, and there is of
course no fine tuning problem. Supersymmetry thus increases the
renormalizability of the theory. However at the naturalness scale we
still need an infinite number of couplings, the effective
supersymmetric field theory becoming equivalent (in the sense described in
\S 5.4) to supersymmetric S--matrix theory. The supersymmetry
may be broken softly by introducing small non--supersymmetric terms in the
same way as the $Z_2$ symmetry was broken above; the flow equations
take the form \evolpm\ if we take $S^+$ as the supersymmetric action
and $S_{\rm int}^-$ as the soft breaking. Clearly some of the
advantage gained from the supersymmetry is then lost, however, and in
particular the mass terms in the bosonic sector may become very large unless
$g'\ll (m/\Lambda_0)^2$.

\newsec{Renormalizable Effective Field Theory.}

In this article we have given a detailed discussion of the
perturbative renormalizability and general consistency of an effective $Z_2$
symmetric scalar quantum field theory, which can be straightforwardly extended
to include any theory of spin--zero and spin--half particles with
global symmetries realized in Wigner mode.

We formally define our effective quantum field theory through the
usual Euclidean functional integral, which in practice means that we use a
classical action at some scale $\Lambda$ to determine a set of Feynman
rules by which we may
compute Euclidean Green's functions (and thus by analytic continuation
S--matrix elements) order by order in perturbation theory. The action
is assumed to be consistent with the appropriate global symmetries, in
particular rotational invariance (which will guarantee that S--matrix
elements are Lorentz invariant) and any internal global symmetries.
The form of the classical action is further restricted by the
following conditions:
\item{1.} Stability: the field equations must have a unique
translationally invariant solution, to be identified as the vacuum
state, and stable propagating solutions, at most one for each field, which
may be identified as free particle states of mass $m$. This was ensured
perturbatively by the conditions imposed on the regulating function in
\S 1.1, and nonperturbatively by the construction of \S 5.2, by which
a stable classical action was derived by the classical evolution of a
manifestly stable primordial action.
\item{2.} Finiteness: the behaviour of the regulating function and
vertices at large Euclidean momenta must be such that all Feynman
diagrams are finite.
\item{3.} Analyticity: we assume that the Euclidean action (and thus both the
regulating function and the vertex functions) is real and, when
analytically continued, regular both in the fields and in their
momenta (see \S 1.1 and \S 3.1). When combined with condition 1. this
means that the only finite singularity in the analytic continuation of a
given Feynman diagram results from the simple pole in the propagator.
When combined with condition 2, analyticity
implies that the action has an essential singularity in Euclidean
momenta at the point at infinity.
Actually given analyticity condition 2 will be automatically satisfied if,
in a field representation in which
the propagator has a Lehmann representation, the Euclidean vertex functions
are uniformly bounded.
\item{4.} Naturalness: we assume that there is a particular
scale $\Lambda_0$ at
which none of the terms in the action is unnaturally large on the
scale of $\Lambda_0$ (although some may be unnaturally small, of
course). In fact, given conditions 2. and 3, this condition is almost content
free; $\Lambda_0$ is in practice given by the order of magnitude of the
largest vertex.

\noindent We also need to make a (rather general) technical assumption
on the ordering of the vertices in order to set up a perturbative
expansion (see \S 3.1).

Although clearly interelated, each of these conditions carries with it
a different physical implication. The stability condition is necessary
to ensure that a Fock space of stable free particle states can be
constructed, and thus that the quantum theory has a set of physical observables
(the momenta and internal quantum numbers of these states). The
finiteness condition is necessary to ensure that transition amplitudes
between these states are finite; the two conditions together are
necessary for the existence of the S--matrix. The analyticity
condition is then sufficient to ensure that the S--matrix is both
unitary and causal (\S 2.4 and \S 5.3). Analyticity is the closest
we get to a locality
assumption, which is why we refer to our analytic action as
`quasi--local'; in the words of Landau \rpl {\it ``the
principle of locality of interactions expresses itself in the analytic
properties of the fundamental quantities of the theory''}.
Stronger assumptions (such as insisting that the
action is a polynomial in momenta, which would enable the fields to
satisfy a
local commutativity condition) are, we believe, both unnecessary
and unphysical. It is thus not surprising that such strong locality
assumptions result in theories with infinities, and which can then
only be defined by some artificial analytic
continuation procedure (in for example the number of space--time
dimensions), or by introducing unphysical degrees of freedom
(Pauli--Villars fields) which violate unitarity except in the limit in
which they are removed, or indeed as a limiting case (either
$\Lambda_0/m\to\infty$ or, more interestingly, $\sigma\to\infty$)
of the quasi--local theories discussed here. In fact strictly
local nonasymptotically--free theories are almost certainly
noninteracting (`trivial'\rvi); as we
explained in \S 5.3 in an
effective quasi--local theory triviality is no longer an issue.

A precise
yet physical formulation of causality itself is necessarily problematic
simply because the position of a relativistic particle is not an
observable \rdirac; this is why we prefer to use the analyticity of
the classical action as a
fundamental assumption from which both unitarity and causality can be
derived. In this sense effective field theory is closer in
spirit to S--matrix theory than to conventional field theory; indeed in
 \S 5.4 we were able to show that they are in some sense
equivalent. However in effective field theory the analyticity
assumption is much more precise than in S--matrix theory, where the
complicated analytic structure of transition elements must in practice
be inferred from the
structure of the Feynman diagrams which contribute to them\rsmatrix.
The redundancy involved in the field theoretic description (as
expressed by the equivalence theorem of \S 4.1) seems to us a small
price to pay for this immense simplification.

The full power of effective field theory only becomes apparent however
when we take into account the naturalness condition, and in particular
consider theories where the particle mass $m\ll\Lambda_0$ (which for
scalar theories means that $m$ must be fine--tuned to unnaturally
small values --- the `naturalness problem'). The ensuing separation
of scales, as expressed in Wilson's exact renormalization group
equation \rii, forms the underlying basis for all the remaining
nontrivial `renormalizability' results proven in our
paper: boundedness (\S 2.1),
convergence (\S 2.2) and universality (\S 3.1) all follow
because the
physics at one scale is relatively independent of the physics at
another scale if the scales are well separated.\foot{Closely related
results will be proven in subsequent publications; in particular
infrared finiteness and Weinberg's theorem\rbtir, the decoupling
theorem for a theory with two scalar particles with widely separated
mass scales\rbtdec, and the
renormalizability of composite operator insertions and the operator
product expansion\rbtope.}

In the conventional formulation of perturbative quantum field
theory, in which we calculate Feynman diagrams with some ultraviolet
regularization and then subtract the ultraviolet divergences order by order,
all the different scales in a theory come into play at the same time,
and processes occurring at totally different scales
becomes hopelessly entangled. It is this confusion of different
scales that is ultimately responsible for the artificial and
unconvincing nature of the conventional discussion of renormalization.
Formulating the theory as an effective theory as described above\foot{
Since all possible terms in the effective Lagrangian are generated by
the evolution, it no longer makes sense to suppress irrelevant terms
at $\Lambda_0$.}, and
evolving (or `renormalizing') it using Wilson's exact renormalization group
equation, no infinities ever arise, and the separation of scales
becomes entirely self--evident because only an infinitesimal change of
scales is considered. Considerable technical simplification also comes from the
remarkably simple form of the exact renormalization group equation; without
this equation, and in particular using only the old BPHZ machinery\ri,
proving the renormalizability of the effective theory would have been
impossibly difficult.

In \S 1.2 we showed that the exact renormalization group makes it
possible to make completely arbitrary changes in the scale $\Lambda$
at which we define our effective quantum field theory without changing
the Green's functions. Since the flow equations preserve the
solutions to the equations of motion (\S 5.1) and the regularity of the
vertex functions, the existence and unitarity of the S--matrix is
guaranteed at all scales; the nonanalyticity of the Green's functions
in the physical region only appears at $\Lambda=0$, due again to the
essential singularity in the action (\S 2.4). Thus when
considering a physical processes at a certain energy we can simply
choose to use the quantum theory evolved to a convenient scale,
often a scale of the same order.

Using the exact renormalization group equations determining the flow
of the vertices between different scales the explicit but straightforward
bounding arguments of \S 2 and \S 3 enabled us to prove that to
all orders in perturbation theory the effective theory at low
scales, with a finite number of relevant coupling
constants fixed by appropriate renormalization conditions (we
considered both on and off--shell conditions), is bounded
(\S 2.2), convergent (\S 2.3) and universal (\S 3.1): it is
independent of the precise value of the naturalness scale $\Lambda_0$, and
of the form of both the regulating function and irrelevant vertices at
$\Lambda_0$, up to small corrections of order
$\Lambda_R/\Lambda_0$. Consequently, the theory is renormalizable in the
sense that we are able to calculate low energy Green's functions to
high accuracy using a finite number of parameters defined at
these low energies, without knowing any details of
the behaviour of the theory at high energies.
We can in particular set the irrelevant coupling constants to
zero (at least in perturbation theory) and let
$\Lambda_0/\Lambda_R \rightarrow \infty$ as we would
in the conventional formulation of local quantum field theory. It is thus
inevitable that, if the scale at which `new' physics appears is much
greater than the scale currently probed, then the
interactions of the observed particles will be well described
by a conventionally
`renormalizable' quantum field theory. Seen in this light the
considerable success of such theories is not so surprising;
it seems to us rather ironic that what was once considered to be
the main defect of quantum field theory, namely the renormalization of
the `infinities', turns out to be the main reason for its success!

If we want to consider processes at energies approaching the
naturalness scale $\Lambda_0$, or improve the
precision of the theory at a given scale,
we must necessarily increase the number of physically relevant parameters
determined experimentally (\S 3.2), which means that the theory
becomes less and
less predictive (and more and more like S--matrix theory). The
infinite number of parameters needed in the effective
theory at and above $\Lambda_0$ should not in itself be regarded as hopeless
however; S--matrix theory is still highly nontrivial. In fact the
situation is very similar to that in a classical theory: for example the
classical theory of the scattering of light--by--light is an effective
classical field theory; it requires an arbitrarily large number
of parameters for processes with energies approaching the
electro--production threshold; and these parameters are
in fact all determined in terms of
two parameters (the mass of the electron and the fine structure
constant) by the underlying quantum theory (QED). Similarly an
effective quantum field theory may be replaced at the naturalness
scale by a new effective quantum field theory containing either
new particles or new fields which do not themselves correspond to
asymptotic particle states, or it may be superceded
by some other theory embodying genuinely new physical principles.
Or it may not.

However, thus far we have only considered the physics of
scalar and spin half particles, with any linearly realized
global symmetries
being also symmetries of the vacuum. All theories of physical
interest\foot{With the sole exception of the effective theory of
very soft neutrino--neutrino scattering.}
involve particles of spin one which require local symmetries in order
to be simultaneously Lorentz invariant and unitary. Furthermore many
symmetries are spontaneously broken or realized nonlinearly. Thus the
theories discussed here can only really be considered as
toys. In future articles in this series we will consider the
renormalizability of effective QED.

\bigskip
\noindent{\bf Acknowledgements.}
\medskip
We would both like to thank Ian Aitchison and David Lancaster for
their patience during the
development of many of the ideas presented here. RDB would like to
thank Howard Georgi for a short discussion some years ago which helped
to motivate the present study, Poul Damgaard for some comments on
the final manuscript, and Tim Morris for bringing
ref.\refs{\rWH,\rWerice} to his
attention. We would also like to thank the Royal
Society and the SERC for financial support and RST would like to thank
Jesus College, Oxford and the Leathersellers' Company for a graduate
scholarship.

\vfill\eject

\appendix{A}{}

As explained in \S 2.2, if we have a real function of $m$
Euclidean momenta, $f(p_1,...,p_{m})$, we define the norm of
$f(p_1,...,p_{m})$ to be
\eqn\Ni{\Vert f(p_1,...,p_{m})\VertL \equiv
\max_{\{p_1,\ldots,p_m\}}\Big[\prod_{i=1}^{m}K_{\Lambda}(p_i)^{1/4}
\vert f(p_1,...,p_{m})\vert\Big],}
where the maximum is taken over all bounded Euclidean momenta, and
$K_{\Lambda}(x)$ is the regulating function described in \S 1.1.
This definition is indeed that of a norm because (dropping the subscript):
\item{(1)}$\Vert f\Vert \geq 0$, and $\Vert f\Vert = 0$ if and only if
$f=0$. The first two statements are trivial; the last is true because
if $\Vert f\Vert =0$, then since $K_{\Lambda}(x)$ has no zeros
$f(p_1,...,p_{2m})=0$ for all $p_i$.
\item{(2)} $\Vert a f\Vert = \vert a \vert\Vert f\Vert$, where $a$ is
any complex number. This is trivial.
\item{(3)} The triangle inequality; $\Vert f+g\Vert\leq\Vert f\Vert
+ \Vert g\Vert$, where $g(p_1,...,p_{m})$ is another function of the
same $m$ momenta. Again the proof is straightforward; $\vert
f+g\vert\leq\vert f\vert +\vert g\vert$, $K_{\Lambda}(x)$ is
strictly positive, and then $\max ({f}+{g})
\leq\max{f}+\max{g}$ whenever ${f}$ and ${g}$
are real and positive.

In addition to the above, we also have the Cauchy-Schwarz
inequality; $\Vert fg\Vert\leq\Vert f\Vert\Vert g\Vert$, where the
product $f(p_1,\ldots,p_{m})g(q_1,\ldots,q_{n})$ is considered as
a function of $(p_1,\ldots,p_{m},q_1,\ldots,q_{n})$ (though the $p_i$
and the $q_i$ are not necessarily completely independent of one
another). Again the result is easily proven;
$\vert fg\vert =\vert f\vert\vert g\vert$, and $\max ({f}{g})
\leq\max{f}\max{g}$.

These results are then used throughout the paper to prove the various
bounds on the norms of the vertex functions and Green's functions.

\appendix{B}{}

In the proof of boundedness in \S 2.2, and in subsequent proofs,
we needed to know that if $f(p_1\ldots p_{m})$ represents a
vertex function, or the $j_{th}$ momentum derivative of a
vertex function (in which case its momentum and Euclidean indices
are implicit), then if
\eqn\eAi{\Vert \partial_{p_i^{\mu}}f(p_1 \ldots p_{m}) \VertL
\leq c\Lambda^{-n},}
and
\eqn\eAii{\Vert \partial_{p_j^{\mu}}\partial_{p_k^{\nu}}
f(p_1 \ldots p_{m})\VertL
\leq c\Lambda^{-n-1},}
for all $p_j^{\mu}$ and $p_k^{\nu}$, for all components of
$f(p_1\ldots p_{m})$ and for all $\Lambda \in [\Lambda_R,
\Lambda_0]$, where $c$ is a constant, then we can also say that
\eqn\eAiii{\biggl\Vert \sum_{\mu =1}^{4}\sum_{i=1}^{m}
(p-q)^{\mu}_i\int_0^1
d\rho\,\partial_{k_i^{\mu}}f(k_1 \ldots k_{m}) \biggr\Vert_{\Lambda}
\leq c\Lambda^{-n+1},}
where $q_i^2 \sim \Lambda_R^2$ for all $i$, $k_i\equiv q_i +\rho r_i$,
$r_i\equiv p_i - q_i$. In fact it is sufficient to show that
\eqn\eAiv{\Vert r^{\mu}_j\partial_{k_j^{\mu}}
f(k_1 \ldots k_{m}) \VertL
\leq c\Lambda^{-n+1},}
for each Euclidean component $\mu$ of each momentum $p_j^\mu$, and given
$\rho\in [0,1]$. To simplify the notation we henceforth drop both
indices $i$ and $\mu$, writing $\bar{p}$ for the single real number
$p_j^\mu$, etc.

In the remainder of this appendix we will prove \eAiv. For simplicity
we take the the regulating function $K(x)=e^{-x}$, denote
$\partial_{\bp}f(p_1,\ldots,p_n)$ by $F(p_i)$, and make the definition
\eqn\eAxx{C\equiv c\prod_{i\neq j}e^{p_i^2 /4\Lambda^2}
e^{{(p_{j}^2-\bp^2)}/4\Lambda^2}
.}
So \eAi\ and \eAii\ become $\vert F(p_k)\vert \leq C
\Lambda^{-n}e^{\bp^2 /4\Lambda^2}$ and $\vert { \partial \over
\partial p_{j}^{\nu}} F(p_i)
\vert \leq C \Lambda^{-n-1}e^{\bp^2/4\Lambda^2}$ respectively.
It will also be useful to define
\eqn\eAv{\Phi (q_i,r_i)\equiv F(p_i)- F(q_i),}
where $q_i^2 \sim \Lambda_R^2$. Then, since $e^{q_i^2/\Lambda^2}
\lsim 1$ and $e^{m^2/\Lambda^2}\lsim 1$, \eAi\ can be rewritten as,
\eqn\eAvi{\vert F(q_i)\vert\leq C\Lambda^{-n},\qquad
\vert\Phi(q_i,r_i)\vert\leq C\Lambda^{-n}
e^{\bp^2 /4\Lambda^2}.}
while \eAii\ yields
\eqn\eAiix{ \Bigl\vert {\partial \over \partial p_{k}^{\nu}}
\Phi (q_i,r_i)\Bigr\vert \leq C\Lambda^{-(n+1)}
e^{\bp^2 /4\Lambda^2},}
or, for the special case $\nu =\mu$ and $k=j$, and using the fact
that the exponential is convex,
\eqn\eAix{ \Bigl\vert { \partial \over \partial (\rho\br )}
\Phi(q_i,\rho r_i) \Bigr\vert \leq C\Lambda^{-(n+1)}
e^{\bp^2 /4\Lambda^2}.}

Multiplying both sides of \eAix\ by $\bp -\bq\equiv\br$, taking the modulus
of both sides, integrating with respect to
$\br$ between $0$ and $\br$ and taking the modulus again, we obtain
$$
\Biggl\vert\int^{\br}_0 \Bigl\vert\br' {\partial \over
\partial (\rho\br')} \Phi (q_i,\rho r_i') \Bigr\vert \,
d\br'\Biggr\vert \leq \quad
C \Lambda^{-(n+1)} \int_0^{\br}\br e^{(\br'+\bq)^2
/4\Lambda^2}\, dr'.
$$
Also, we know that for any function
the modulus of the integral of the modulus is greater than or equal to the
modulus of the integral, so on the left-hand side of our inequality we may
remove the modulus sign inside the integral while still satisfying the
inequality. Now letting $\br'\rightarrow \br'-\bq$ on the right--hand
side, and using the fact that
$$\Bigl\vert\int_{\bq}^{\bp}xe^{x^2/4\Lambda^2}\, dx\Bigl\vert
=2\pi^{1/2}\bigl\vert\Lambda^2(e^{{\bp^2 /4\Lambda^2}}
-e^{{\bq^2/4\Lambda^2}})\bigr\vert
\leq \Lambda^2e^{{\bp^2 /4\Lambda^2}},$$
where we ignore constants of order unity in the final inequality,
we find that
\eqn\eAx{\Biggl\vert\int^{\br}_0 \br' {\partial \over
\partial (\rho\br')} \Phi(q_i,\rho r_i') \,d\br'\Biggr\vert
\leq C\Lambda^{-(n+1)} \Biggl(\Lambda^2e^{{p^2 /4\Lambda^2}}
+\vert\bq\vert \Bigl\vert\int_{\bq}^{\bp}
e^{\br^2/4\Lambda^2} \, d\br \Bigr\vert \Biggr) .}

Integrating the integral on the left--hand side of \eAx\ by parts,
taking the modulus, and then using the inequality \eAx\ we can thus
show that
\eqnn\eAxi
$$\eqalignno{
\vert\br\Phi(q_i,\rho r_i) \vert
& = \Biggl\vert \int^{\br}_0
\Phi(q_i,\rho r_i') \, d\br' + \int^{\br}_0 \br
{\partial\over\partial (\rho\br')} \Phi(q_i,\rho r_i') \,
d\br'\Biggr\vert\cr
&\leq \Bigl\vert \int^{\br}_0
\Phi(q_i,\rho r_i') \, d\br'\Bigr\vert
+\rho C\Lambda^{-(n+1)} \Biggl(\Lambda^2e^{{p^2 /4\Lambda^2}}
+\Lambda\Bigl\vert \int_{\bq}^{\bp} e^{\br^2
/4\Lambda^2}\, d\br\Bigr\vert\Biggr).&\eAxi\cr}
$$
The first term on the right may be bounded with the aid of
equation \eAvi;
\eqn\eAxii{\Bigl\vert \int^{\br}_0 \Phi(q_i,\rho r_i')\,
d\br' \Bigr\vert
\leq \Bigl\vert\int^{\br}_0 \vert \Phi(q_i,\rho r_i')
\vert \, d\br'\Bigr\vert
\leq C\Lambda^{-n} \Bigl\vert
\int^{\bp}_{\bq} e^{\br^2 /4\Lambda^2}\, d\br\Bigr\vert.}
Therefore, two of the terms on the right-hand side of \eAxi\ involve the
modulus of the integral
\eqn\eAxiii{\int^{\bp}_{\bq} e^{x^2 /4\Lambda^2}\, dx
= \int^{\bp}_{0}e^{x^2 /4\Lambda^2}\, dx
+ \int^{0}_{\bq}e^{x^2 /4\Lambda^2}\, dx.}
In order to bound the modulus of \eAxiii\ we
use the fact that whatever the value of $y$,
\eqn\eAxiv{\biggl\vert\int^{y}_0 e^{x^2 /4\Lambda^2}dx\biggr\vert
\leq \biggl\vert\int^{y}_0 {x\over \Lambda}e^{x^2 /4\Lambda^2}
dx\biggl\vert +\biggl\vert\int^{\Lambda}_0 e^{x^2 /4\Lambda^2}
dx\biggl\vert
= \Lambda ( 2 (e^{y^2 /4\Lambda^2} -2) + e^{{1/4}}).}
Therefore, using the explicit result for the second term on the right-hand side
of \eAxi\ and also the inequality \eAxiv, \eAxi\ may be written as
\eqn\eAxv{\vert\br\Phi(q_i,\rho r_i)\vert
\leq C\Lambda^{-n+1}e^{\bp^2 /4\Lambda^2}.}

Finally, using \eAv\ and \eAvi, we may write
\eqn\eAxvii{ \vert \br F(q_i + \rho r_i)\vert
\leq C\Lambda^{-n} \vert \br\vert  \, + \, C\Lambda^{-n+1}
e^{-\bp^2 /4\Lambda^2}.}
Multiplying both sides by $\prod_i e^{-(p_i^2 + m^2)/4\Lambda^2}$
we have
\eqn\eAxxi{ \Vert \br F(q_i + \rho r_i)\VertL
\leq c\Lambda^{-n} .}
The first term last factor on the right is easily seen to be
$\leq \Lambda$, so
\eqn\eL{ \Vert \br F(q_i + \rho r_i)\Vert\leq
c\Lambda^{-n+1},}
which is the required result \eAiv.

\appendix{C}{}

In this appendix we show that the `nonlocal regularization'
constructed by Kleppe and Woodard \rv\ is equivalent to the classical
renormalization group evolution of the primordial action
\eSinfty\ described in \S 5.2. They begin with a local action of the same
form as \eSinfty, introduce an auxiliary field $\psi$, and define
the auxiliary action coupling $\phi$ and $\psi$ by
\eqn\eAA{{\cal S}^{\rm nlr}[\phi ,\psi ;\Lambda ]
= \half (\phi , P\inv _{\Lambda} \phi )
-\half (\psi , L\inv _{\Lambda} \psi ) + S_{\rm int}^\infty[\phi + \psi].}
where $P_{\Lambda}=K_{\Lambda} P_\infty$ is the inverse
propagator just as in \classact\exxix, while $L_\Lambda$ is the regular
`propagator' \eldef.
The auxiliary field satisfies a classical equation of motion,
which gives $\psi[\phi;\Lambda]$. The regularized action is
then taken to be ${S}^{\rm nlr}[\phi ;\Lambda_0]=
{\cal S}^{\rm nlr}\big[\phi ,\psi[\phi ;{\Lambda_0}];{\Lambda_0}\big]$.

It is not difficult to see from the algorithm following \eSinfty\
that this construction will produce the same regularized action as
that using the classical renormalization group equation. In order to
prove this equivalence we derive the exact renormalization group
equation satisfied by the interacting part of ${S}^{\rm nlr}$,
\eqn\eBa{S^{\rm nlr}_{\rm int}[\phi;\Lambda]
= -\half(\psi[\phi;\Lambda], L\inv _{\Lambda}\psi[\phi;\Lambda])
+ S^\infty_{\rm int}\big[\phi + \psi[\phi;\Lambda]\big].}
Differentiating both sides of \eBa\ with respect to $\Lambda$, we
obtain
\eqn\eBB{\eqalign{
{\partial S^{\rm nlr}_{\rm int}[\phi;\Lambda]\over\partial\Lambda}
&= - \biggl({\partial\psi[\phi;\Lambda]\over\partial\Lambda},
L\inv _{\Lambda}\psi[\phi;\Lambda]\biggr)
-{1\over 2}\biggl(\psi[\phi;\Lambda],{\partial L\inv _{\Lambda}
\over\partial\Lambda}\psi[\phi;\Lambda]\biggr) \cr
& \hskip 3in + \biggl({\partial\psi[\phi;\Lambda]\over\partial\Lambda},
{\delta S^\infty_{\rm int}\big[\phi +\psi[\phi;\Lambda]\big]
\over\delta\psi}\biggr)\cr
&= - {1 \over 2} \biggl(\psi[\phi;\Lambda],
{\partial L\inv _{\Lambda}\over\partial\Lambda}\psi[\phi;\Lambda]\biggr),
\cr}}
the second equality following from the equation of motion for $\psi$.
Using in addition the classical equation of motion obtained
by functional differentiation of \eAA\ with respect to $\phi$, the
solution for $\psi[\phi;\Lambda]$ may be written in the form (see
Theorem A.1 of ref.\rv)
\eqn\eBE{\psi[\phi;\Lambda]
= -\biggl({K_{\Lambda}-1\over K_{\Lambda}}\biggr)\phi
+ L_{\Lambda}{\delta S^{\rm nlr}[\phi;\Lambda]\over\delta\phi}
= L_{\Lambda} {\delta S^{\rm nlr}_{\rm int}[\phi;\Lambda]\over\delta\phi}.}
Substituting this into \eBB, and noting that (see \eldef)
${\partial L_{\Lambda}\over\partial\Lambda}
= {\partial P_{\Lambda}\over\partial\Lambda}$ we obtain
\eqn\eBF{
{\partial S^{\rm nlr}_{\rm int}[\phi;\Lambda]\over\partial\Lambda}
= {1 \over 2}\biggl(
{\delta S^{\rm nlr}_{\rm int}[\phi;\Lambda]\over\delta\phi},
{\partial P_{\Lambda}\over\partial\Lambda}
{\delta S^{\rm nlr}_{\rm int}[\phi;\Lambda]\over\delta\phi}
\biggr);}
the interaction part of the nonlocally regularized action satisfies the
same classical exact renormalization group equation as the
interaction part of the regularized action $S[\phi,\Lambda]$.
When $\Lambda \rightarrow \infty$, $K_{\Lambda}\rightarrow 1$ and
$L_{\Lambda}\rightarrow 0$, so from \eBE\ we see that
$\psi[\phi;\Lambda]\rightarrow 0$.
Thus, in this limit $S^{\rm nlr}_{\rm int}[\phi;\infty]
=S_{\rm int}^{\infty}[\phi]$; the interaction part of the nonlocally
regularized action becomes the local interaction. This completes the proof.

\footatend\vfill\supereject\immediate\closeout\rfile\writestoppt
\baselineskip=14pt\centerline{{\bf References}}\bigskip{\frenchspacing%
\parindent=20pt\escapechar=` \input refs.tmp\vfill\eject}\nonfrenchspacing
\vfill\eject\immediate\closeout\ffile{\parindent40pt
\baselineskip14pt\centerline{{\bf Figure Captions}}\nobreak\medskip
\escapechar=` \input figs.tmp\vfill\eject}
\end